%% file: qrw-V3_12.10.21.tex
\documentclass[12pt]{article}

\usepackage{amsfonts,amsmath,amssymb}
\usepackage{latexsym}

\usepackage{epsfig}
\usepackage{hyperref}
\usepackage{color}

\usepackage{epsfig}
    \def\sepand{\rule{14cm}{0pt}\and}

\usepackage[T1]{fontenc} 

\input eqmacros
\input totalmacroe

\graphicspath{{pictures/}}

\renewcommand{\imath}{\mathrm{i}}
\newcommand{\emath}{\mathrm{e}}

\begin{document}

\hfill

\vspace{1cm}

\begin{center}

{\Large \bf Quantum walk on a comb \\ with infinite teeth
}

\medskip
\vspace{1.5 truecm} 


\vspace{1.5 truecm}

{\bf Fran\c cois David}

\vspace{0.4 truecm}

Institut de Physique Th\'eorique,\\
Universit\'e Paris-Saclay, CNRS, CEA, Institut de physique th\'eorique\\
91191, Gif-sur-Yvette, France

\vspace{1.3 truecm}

{\bf Thordur Jonsson}
\vspace{0.4 truecm}
\vspace{0.4 truecm}

Division of Mathematics

The Science Institute, University of Iceland

Dunhaga 3, 107 Reykjavik, Iceland

\end{center}
 \textwidth 150mm
  \textheight 215mm
   \setlength{\unitlength}{0.01in}
    \def\sepand{\rule{14cm}{0pt}\and}



\topmargin 0pt
\oddsidemargin 5mm
\headheight 0pt
\topskip 0mm

\addtolength{\baselineskip}{0.5\baselineskip}

\pagestyle{empty}

\hfill

\noindent {\bf Abstract.}  We study continuous time quantum walk on a comb with infinite teeth
and show that the return probability to the starting point decays with time $t$ as $t^{-1}$.    We analyse the diffusion along the spine and into the teeth
and show that the walk can escape into the teeth with a finite probability and goes to infinity along the spine
with a finite probability.  
The walk along the spine and into the teeth behaves qualitatively as a quantum walk on a line.
This behaviour is quite different from that of classical random walk on the comb.

\newpage

\tableofcontents

\newpage
\pagestyle{plain}

\section{Introduction}
Quantum walks have been extensively studied since late last century, the main motivation coming 
from quantum computation and the search for efficient algorithms.   A pedagogical intoduction to the subject 
can be found in
\cite{kempe} and a comprehensive overview is given in \cite{salvador}.  One of the characteristic 
features of quantum walks is that they tend to move around faster and sometime very much faster than classical random walks.

In the past the relation between classical random walks and the geometry of the underlying space  (which often is a graph) have been
studied intensively, see e.g.\  \cite{georw,georw2}.  In particular, geometric properties are often reflected in the heat
kernel.  The purpose of this paper is to investigate a simple graph and compare the behaviour of quantum 
random walk to that of classical random walk.  There are already many results in this direction, in particular on 
$\bbZ^d$ and various finite graphs, see \cite{salvador} and \cite{grimmet}.  

There are two classes of quantum walks.  The first is the so-called coined quantum walks which are analogous to discrete time classical random walk where the walker flips a coin at each timestep to decide
which vertex to go to next.   In the quantum case the coin is an extra quantum degree of freedom which can complicate 
the analysis.  The other class is continuous time quantum walk where no coin is needed
and the time development is given (in continuous time) by the one parameter unitary group generated by the 
Laplacian of the underlying graph.  In many cases 
the behaviour of coined walks and the continuous time walk are qualitatively similar but the relation 
between the two is not trivial, see \cite{childs}.   In this paper
we are exclusively concerned with continuous time random walks.

In the next section we define continous time quantum walk on a graph 
following \cite{fahri1, fahri2}.    We study such a walk on a 
comb which is the integer line (the spine) with an integer half line (a tooth) 
attached at each vertex of the spine.    The comb is an interesting example of a nonhomogeneous
graph where one can make detailed calculations and classical random walk on a comb is well understood \cite{ben,havlin,bertacchi,bertacchi2,DJW}.  
 
In Section 3 we find the spectrum of the Laplacian on the comb and the corresponding eigenfunctions.
We find that there are two classes of eigenfunction.  The first class are nonlocalized functions which 
oscillate along the spine and in the teeth.
Then there are eigenfunctions which are localized on the spine and decay exponentially in the teeth.

Knowing the spectrum of the Hamiltonian and its eigenfunctions allows us to express the propagator as
a contour integral in the complex plane and the large time behaviour can be analysed by steepest descent methods.
This is the content of Section 4.
The probability that the walk
 is back at the starting point situated on the spine after time $t$ has a powerlaw decay with an
exponent $-1$, while the corresponding exponent for classical random walk is $-3/4$.  
The motion of the quantum walker is ballistic in the sense that if the wave function is concentrated at
one vertex at time $0$, then the wave front moves with a velocity of order 1 into the teeth, and moves with a different velocity of order 1 along the spine.
We calculate these velocities and the respective probabilities that the walk disappears into the teeth or along the spine as time goes to infinity.

A few technical details are relegated to the Appendices.  There we also discuss some numerical results
on the behaviour of the wave function for large time.

\section{Continuous time quantum walk}

Let $G$ be a graph with vertex set $V$ and edge set $E$.   The graph may be finite or infinite but we
are mainly interested in the infinite case.   If $a\in V$ we let 
$\sigma_a$ denote the number of nearest neighbours of $a$.
Let $M$ be a matrix indexed by $V$ with matrix elements
\beq{matrix}
M_{ab}=\left\{ \begin{array}{ll}\ \ \sigma_a ~& {\rm if}~ a=b\\
-1~ & {\rm if}~(a,b)\in E\\
\ \ 0 ~ & ~{\rm otherwise}\end{array}\right.
\eeq
If we consider classical continuous time random walk on $G$ and $p_a(t)$ is the probability to be 
at the vertex $a$ at time $t$ then
\beq{classical}
\frac{dp_a(t)}{dt}=-\sum_{b\in V} M_{ab}\, p_b(t).
\eeq

The quantum walk on $G$ is defined by introducing the Hilbert space 
$\cH =L^2(V)$, i.e.\ square 
summable complex valued functions defined on the vertex set of $G$ with the usual inner product.  
We use Dirac notation and for $a\in V$ let $|a\kt$ denote the function which takes the value $1$ at $a$ and is $0$ elsewhere.
These functions make up an orthonormal basis for $\cH$ and we define a Hamiltonian operator $H$ by
\beq{Hamiltonian}
\br a|H|b\kt = M_{ab}.
\eeq
The state of the quantum walk at a given time is given by an element of $\cH$ which changes with time
according to the Schr\"odinger equation with Hamiltonian $H$.  If the state of the quantum walk at time $t=0$ is 
$|\psi_0\kt$, assumed to be normalized, then its state at time $t$ is
\beq{timeevolution}
|\psi_t\kt =e^{-\imath tH}|\psi_0\kt.
\eeq
If the walk starts out at vertex $a$ at time $t=0$ (i.e.\ $|\psi_0\kt =|a\kt$) then the probability amplitude to find 
the walk at $b$ at time $t$ is 
\beq{amplitude}
A_t(a,b)=\br b|e^{-\imath tH}|a\kt
\eeq
and the probability to find the walk at $b$ at time $t$ is
\beq{probability}
P_t(a,b)=|A_t(a,b)|^2.
\eeq
It is easy to see from \rf{classical} that the corresponding probability for classical random walk is
\beq{classical2}
P^{\rm class}_t(a,b)= \br b| e^{-tH}|a\kt .
\eeq
The classical spectral dimension $d_s$ of the graph is defined by the decay with $t$ 
of the return probability to the starting point,
\beq{specdim}
P^{\rm class}_t(a,a)\sim t^{-d_s/2},~~t\to\infty ,
\eeq
and it is not hard to see that if $d_s$ exists it is independent of $a$.   Analogously we define the
{\it quantum spectral dimension} $d_{qs}$ (if it exists) by
\beq{qspecdim}
P_t(a,a)\sim t^{-d_{qs}/2},~~t\to\infty .
\eeq
The spectrum of $H$ lies on the nonnegative real axis so the classical 
spectral dimension is determined by the density of states of $H$ close to 0.  
More precisely, let $\{  |E,\lambda \kt\}_{E,\lambda}$ be the normalized 
eigenkets of $H$ with eigenvalue $E$ where the index $\lambda$ takes care of multiplicity.  Then we have
\beq{calc}
P^{\rm class}_t(a,a)=\int_0^\infty dE \int d\lambda \, |\br a|E, \lambda\kt|^2 e^{-tE}.
\eeq
Assuming that 
\beq{density}
\int d\lambda \, |\br a|E, \lambda\kt|^2 \sim E^{\gamma}
\eeq
for $E$ close to zero we find easily that
\beq{calc2}
P^{\rm class}_t(a,a)\sim t^{-1-\gamma}
\eeq
for $t\to\infty$ so
$d_s = 2(1+\gamma ) $.
By an analogous argument the absolute value of the 
amplitude $A_t(a,a)$ decays no faster than  $ t^{-1-\gamma}$ for large $t$ and 
we conclude that
\beq{specdimerel}
d_{qs}\leq 2\,d_s.
\eeq
We get an equality in \rf{specdimerel} if the decay of $A_t(a,a)$ 
is determined by the edge of the spectrum of $H$ and this is the case for regular 
graphs like $\bbZ^d$ \cite{luck,grimmet} .
For the comb we find that localized states with high energy
dominate the return probability to the starting point and we conjecture that 
this will hold quite generally for graphs with localized eigenstates for the 
Hamiltonian.
   
We consider continuous time quantum walk on an infinite comb.
The infinite comb is a graph $C$ with vertex set $V= \{ (n,j): n\in\bbZ,\, j\in \bbZ_0^+\}$,
where $\bbZ_0^+$ denotes the nonnegative integers.   The
edge set $E$ is defined by stating which vertices are nearest neighbours, i.e.\ connected by an edge.
If $j>0$ the neighbours of $(n,j)$ are $(n,j+1)$ and $(n,j-1)$.  The neighbours of 
$(n,0)$ are $(n,1)$, $(n-1,0)$ and $(n+1,0)$.  This is an infinite linear graph with a discrete 
half line attached at each vertex.
The infinite linear graph is often referred to as the {\it spine} and the half lines we call {\it teeth}.
 
 Classical random walk on $C$ has been studied 
in great detail, see e.g.\ \cite{ben,bertacchi,bertacchi2,DJW}.  The main results are that the spectral dimension is $3/2$ and the diffusion along the spine
is anomalous in the sense that the average distance squared $\br x^2\kt_t$ travelled along the spine after time $t$ scales as $t^{\oh}$.

We use the Hilbert space $\cH =L^2(V)$ and let
$|n,j\kt$ denote the function which is equal to 1 on the vertex $(n,j)$ and 0 elsewhere.
The time development of a quantum walk on $C$ is given by the unitary operator
\beq{1}
U(t)=e^{-\imath tH}
\eeq
where the Hamiltonian $H$ is minus the Laplacian on $C$ (in agreement with \rf{Hamiltonian}), given in Dirac notation by
$$
H  = -\sum_{n=-\infty}^\infty \sum_{j=0}^\infty  [  \delta_{j,0}(|n+1,0\kt\br n,0 |+|n-1, 0\kt\br n,0| +
|n,1\kt\br n,0| -3|n,0\kt\br n,0|)  +
$$
$$
(1-\delta_{j,0}) (|n,j+1\kt\br n,j| + |n,j-1\kt\br n,j| -2|n,j\kt\br n,j| ) ] .
$$
If a quantum walker is at the vertex $(n_1,j_1)$ at time $t=0$ then the probability amplitude that the walker
is located at $(n_2,j_2)$ at time $t$ is given by
\beq{4}
A_t(n_1,j_1;n_2,j_2)=\br n_2,j_2|e^{-\imath tH} |n_1,j_1\kt.
\eeq
The probability that the walker is at $(n_2,j_2)$ at time $t$, given that he is at $(n_1,j_1)$ at time $t=0$, is
\beq{5}
P_t(n_1,j_1;n_2,j_2)=| A_t(n_1,j_1;n_2,j_2)|^2.
\eeq
In order to estimate this probability we begin by finding the eigenvalues and eigenfunctions of the Hamiltonian.

\section{Diagonalising $H$}
The eigenfunctions of $H$ can be taken to be Bloch waves in the $n$ variable since the Hamiltonian is invariant 
under translations along the spine.  We therefore begin by making the Ansatz
\beq{6}
\phi_{\alpha,\theta}(n,j)=\br n,j|\theta,\alpha\kt=Ae^{\imath \alpha n+\imath \theta j}+Be^{\imath \alpha n-\imath \theta j}
\eeq
for the eigenfunctions,
where
$H|\alpha,\theta\kt =E|\alpha,\theta\kt$, $\alpha\in [0,2\pi)$,  $\theta\in [0,\pi )$ and $A,B$ are constants.  
For $j > 0$ the function $\phi_{\alpha, \theta}(n,j)$
satisfies the equation
\beq{7}
2\phi_{\alpha, \theta}(n,j)-\phi_{\alpha, \theta}(n,j+1)-\phi_{\alpha, \theta}(n,j-1)=E\phi_{\alpha, \theta}(n,j)
\eeq
which shows that 
\beq{7a}
E=2-2\cos\theta .
\eeq
  On the spine, i.e.\ when $j=0$, the equation takes the form
\beq{8}
3\phi_{\alpha, \theta}(n,0)-\phi_{\alpha, \theta}(n-1,0)- \phi_{\alpha, \theta}(n+1,0)- \phi_{\alpha, \theta}(n,1) 
=E \phi_{\alpha, \theta}(n,0).
\eeq
Using \rf{7a} 
it is straightforward to check that the ratio between the coefficients $A$ and $B$ is given by
\beq{9}
{A\over B}= -{1-2\cos\alpha+e^{\imath \theta} \over 1-2\cos\alpha +e^{-\imath \theta}}.
\eeq
The eigenvalues of $H$ we have found are 
the same as on the discrete line $\bbZ$ but the eigenvalues are infinitely degenerate,
$\alpha$ being the degeneracy index.   

In order to use the eigenfunctions to compute the probability amplitudes \rf{4} we need to normalize the eigenfunctions correctly.
Let us take
\beq{10}
A=y+e^{i\theta},~~B=-(y+e^{-i\theta})
\eeq
where $y=1-2\cos\alpha $.  We will show in Appendix A that in this case
\beq{11}
\br\theta' ,\alpha'|\theta ,\alpha\kt =N(\alpha ,\theta )\delta (\alpha -\alpha')\delta (\theta -\theta'),
\eeq
where $N(\alpha ,\theta )=4\pi^2 (y^2 +2y\cos\theta +1)$. 

There is a second class of solutions $\{ |\gamma,\alpha \kt\} $
to the eigenvalue problem for $H$.  They are of the form
																		
\beq{n1}
\br n,j|\gamma ,\alpha\kt =c\,e^{\imath \alpha n} (-1)^je^{-\gamma j},
\eeq
where $\alpha\in [0,2\pi)$, $\gamma >0$ and $c$ is a normalization factor.  We will see that $\gamma$ is in fact uniquely determined by $\alpha$.
The equation for $\psi_\alpha (n,j)= \br n,j|\gamma,\alpha\kt$ at   $j>0$ implies that
\beq{n2}
E=2+2\cosh\gamma.
\eeq
The equation at $j=0$ and arbitrary $n$ gives
\beq{n3}
3-2\cos\alpha +e^{-\gamma } =E.
\eeq
It follows that
\beq{n4}
1-2\cos\alpha =e^\gamma
\eeq
and $\alpha$ lies in the interval $(\pi /2,3\pi /2)$.   We see also that for these solutions
 $4<E<16/3$.

If we choose $c$ such that 
\beq{n5}
\sum_{n,j}\overline{\psi}_\alpha (n,j)\psi_{\alpha'}(n,j)=\delta (\alpha -\alpha'),
\eeq
then a simple calculation shows that
\beq{n6}
|c|^2= \frac{1}{2\pi}(1-e^{-2\gamma}).
\eeq
We will prove in Appendix \ref{aCompleteness} that the set of functions $\{ \phi_{\alpha ,\theta}\}\cup \{ \psi_\alpha\}$ is complete. 

\section{The probability amplitudes}

In this section we study the probability amplitudes \rf{4} and the corresponding probabilities \rf{5}.  
We first 
calculate the probability amplitude to be back at the starting point after time $t$
in the limit $t\to\infty$.
We then study the scaling properties of the amplitudes as  $t$ and $n$ and/or $j$ become large.
The main result is that the motion is ballistic along the teeth and the spine.  This means, looking at the spine,
that the amplitude oscillates with $n$ up to a value $n=v_ct$ with an amplitude of order $t^{-\oh}$.  For $n>v_ct$ 
it decays exponentially.  Similarly, looking in tooth $n$, the amplitude oscillates in $j$ with an amplitude of order 
$t^{-\oh}$ up to a critical value $j=u_ct$ after which it decays exponentially.  We show that the probability that the 
walk is in the $n$th tooth in the limit $t\to\infty$ is nonzero and of order $n^{-1/4}$.  

The probability that the walk is at 
a finite distance from the spine as $t\to\infty$ is also nonzero.  The probabilities that the walk escapes to infinity along the spine and into the teeth sum to 1 as they should.  This is different from the classical case where the walk 
cannot escape into the teeth.  But this is expected in the quantum case 
since the quantum walk on the discrete half-line is not recurrent.

If we scale both $j$ and $n$ with $t$ we always find that the amplitude decays exponentially  This is discussed in 
subsection 4.5 and Appendix C.

\subsection{An integral representation for the amplitude}

Here we aim to derive a formula for the amplitude
$A_t(0,0;n,j)$ which allows us to 
analyse the large $t$ behaviour and the scaling behaviour when $n$ and/or $j$
become large.  

\subsubsection{Representation in the $z=e^{-i \theta}$ complex plane}

The starting point is
\bea
A_t(0,0;n,j) &  = & \int_0^{2\pi} d\alpha \int_0^\pi d\theta \, \frac{\br n,j|\theta ,\alpha\kt\br \theta ,\alpha | 0,0\kt }{N(\alpha ,\theta )}\,e^{-\imath tE(\theta )}\nonumber
\\&&+\int_{\pi /2}^{3\pi /2}d\alpha \, \br n,j|\gamma ,\alpha\kt\br \gamma ,\alpha |0,0\kt\,e^{-\imath tE(\gamma)},\label{amp}
\eea 
using the completeness of the eigenfunctions of the Hamiltonian.
The normalization constant 
$N(\alpha ,\theta )=4\pi^2 (1-2\cos\alpha +e^{i\theta })(1-2\cos\alpha+e^{-i\theta})$ 
so the first integral above can be written
\bea
&&-{\imath \over 2\pi^2}
\int_0^{2\pi} d\alpha \int_0^\pi d\theta \,\sin\theta\left[ \frac{e^{\imath \theta j}}{1-2\cos\alpha +e^{-\imath \theta}} -
\frac{e^{-\imath \theta j}}{1-2\cos\alpha +e^{\imath \theta}}\right]\,e^{\imath \alpha n} e^{-\imath tE(\theta )} \nonumber \\
&&= -{\imath \over 2\pi^2}
\int_0^{2\pi} d\alpha \int_0^{2\pi} d\theta \,\sin\theta \, \frac{e^{\imath \alpha n +\imath \theta j}}{1-2\cos\alpha +e^{-\imath \theta}}\,e^{-\imath tE(\theta )} .\label{integral}
\eea
Putting $w=e^{\imath \alpha}$ and $z=e^{-\imath \theta}$ we have
\beq{x1}
\int_0^{2\pi} d\alpha \, \frac{e^{\imath \alpha n }}{1-2\cos\alpha +e^{-\imath \theta}}=\imath \oint_{|w|=1}\frac{w^n}{ w^2-(1+z)w+1}\,dw .
\eeq
The denominator in the $w$-integral has simple zeroes at $w=w_-$ and $w=w_+$ with
\beq{x2}
w_\pm(z)= {1+z\over 2}\pm\oh\sqrt{(z+3)(z-1)}.
\eeq
We define the square root 
\beq{xx2}
S(z)= \sqrt{(z+3)(z-1)}
\eeq
such that it is positive for $z>1$ and analytic in the complex plane except for a cut along the real axis from $-3$ to $1$.
Then $w_-$ is inside the unit circle (and $w_+$ is outside) except for $z=\pm 1$.  Evaluating  the $w$-integral \rf{x1} we find
\beq{x3}
{2\pi\, w_-^n \over \sqrt{(z+3)(z-1)}}\quad \text{if}\ n\ge 0\ ,\quad {2\pi\, w_+^n \over \sqrt{(z+3)(z-1)}}\quad \text{if}\ n< 0.
\eeq 
It follows that \rf{integral} equals an integral over $z$ along the unit circle, going from $\arg(z)=-\pi$ to 
$\arg(z)=+\pi$:
\beq{x4}
{1\over 2\pi \imath }\oint_{|z|=1}{1-z^{-2}\over \sqrt{(z+3)(z-1)}}\,w_-^{|n|}\,z^{-j}\,e^{-\imath tE}\,dz.
\eeq
Now consider the 2nd integral on the right hand side of \rf{amp}.
Explicitly it is
\beq{x5}
{1\over 2\pi}\int_{\pi/2}^{3\pi/2} d\alpha \,(1-e^{-2\gamma}) \,e^{\imath \alpha n}(-1)^j e^{-\gamma j} e^{-\imath tE},
\eeq
where $1-2\cos\alpha =e^{\gamma}$  and $E=2+2\cosh\gamma$.  Now put $z=-e^{\gamma}$ and change the variable of integration in
\rf{x5} to $z$.  When $\alpha$ goes from $\pi/2$ to $\pi$, then $z$ decreases from $-1$ to $-3$ and increases back to $-1$ as $\alpha$ goes from 
$\pi$ to $3\pi/2$.    We have 
\beq{x6}
2\sin\alpha\,d\alpha = -dz
\eeq
and
\beq{x7}
\sin\alpha= \eta\oh \sqrt{3-2z-z^2}
\eeq
with $\eta=+1$ for $\alpha\in (\pi/2 ,\pi)$ and $\eta =-1$ for $\alpha\in (\pi ,3\pi/2)$.
Furthermore,
\beq{x8}
e^{\imath \alpha}={z+1\over 2}+ \eta \imath \oh \sqrt{(z+3)(1-z)}.
\eeq
It follows that \rf{x5} equals
\bea
\label{x8.5}
&&{-1\over 2\pi} \int_{-1}^{-3} {dz\,(1-z^{-2}) \over \sqrt{(z+3)(1-z)}  } \left( {z+1\over 2}+ \oh \imath \sqrt{(z+3)(1-z)}\right)^n\,z^{-j}e^{-\imath tE}\nonumber\\
&&+ {1\over 2\pi} \int_{-3}^{-1} {dz\,(1-z^{-2}) \over \sqrt{(z+3)(1-z)}  } \left( {z+1\over 2} - \oh \imath \sqrt{(z+3)(1-z)}\right)^n\,z^{-j}e^{-\imath tE}.
\eea
These integrals can be viewed as a closed contour integral from -1 to -3 and back to -1 where the first half is along the upper edge of the
cut of the square root and the second half is along the lower edge of the cut.  
\begin{figure}[h]
\begin{center}
\includegraphics[width=4in]{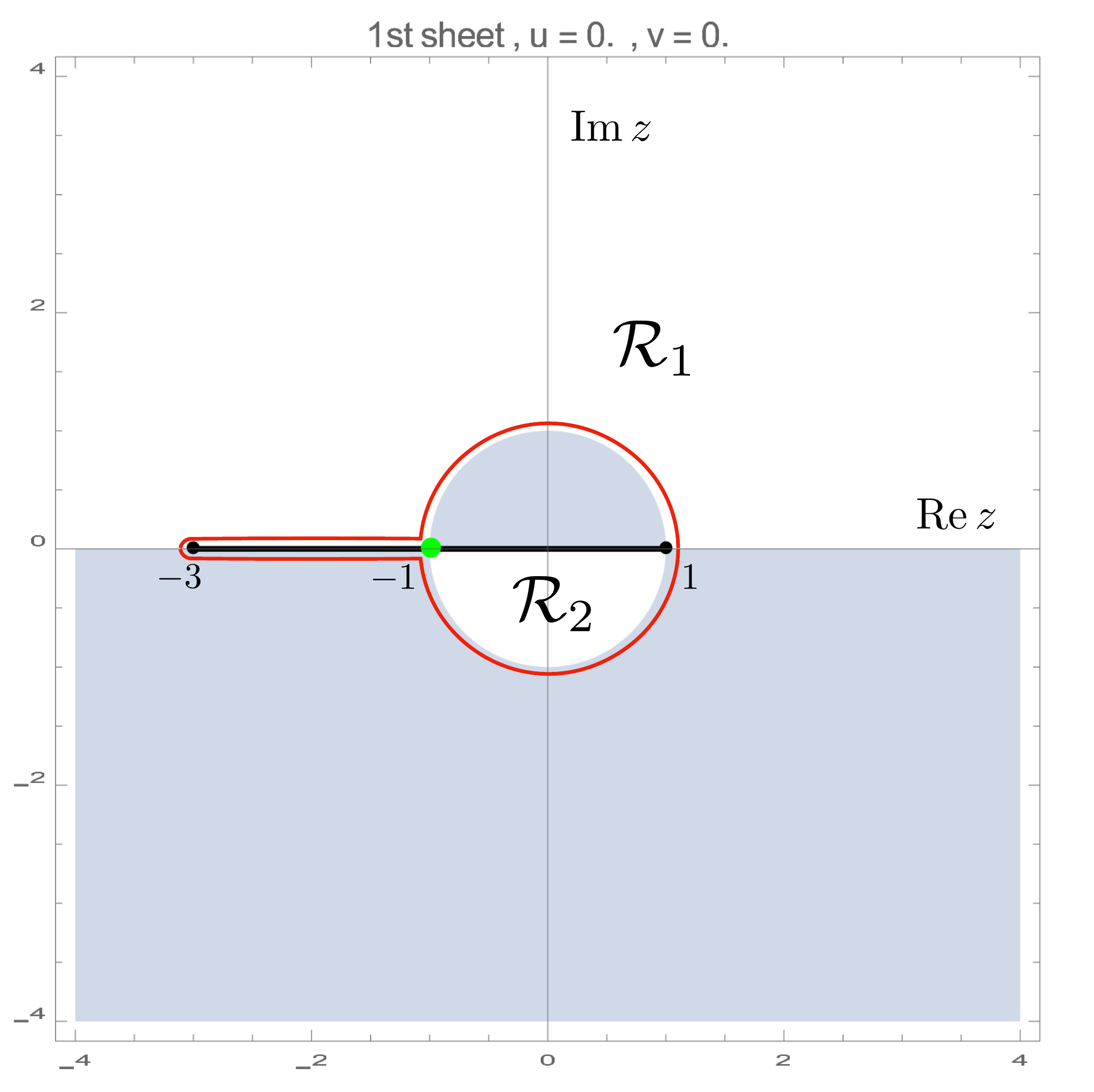}
\caption{The integration contour in the $z$ complex plane (red) of the integral \rf{x9} for $A_t(0,0;n,j)$. There is a square-root cut along $[-3,1]$ (black line) and the first sheet is represented. The white (resp.\ grey) regions are the ``safe'' (resp.\ ``dangerous'') regions where the real part of the potential $V_0(z)$ given by \ref{x9.4 } is negative (resp.\ positive). Here $V_0(z)$ is purely imaginary along the original path of integration. The points 
$z=\pm 1$ are extrema of $V_0(z)$ (saddle points). The points $z=-3$ and $z=1$ are branch points.}.
\label{fContourZ}
\end{center}
\end{figure}
When $n>0$ this is enough, but we can use the symmetry $\alpha\to\pi-\alpha$ in  the integral \rf{x5} 
to replace $n$ by $-n$ in the calculation when $n<0$.
Combining \rf{x8.5} with the integral \rf{x4} we see that the total amplitude can be written as a closed contour integral
\beq{x9}
A_t(0,0;n,j)={1\over 2\pi \imath }\oint_{\Gamma}{1-z^{-2}\over \sqrt{(z+3)(z-1)}}\,w_-^{|n|}\,z^{-j}\,e^{-\imath t(2-z-1/z)}\,dz,
\eeq
where $\Gamma$ is the contour in the complex plane depicted on Fig~\ref{fContourZ}.
Using the fact that the integrand is analytic in $z$ we see that  $\Gamma$ can be replaced by
any c.c.w. contour in the complex plane around the cut $[-3,1]$ for the square root, and staying away from the pole at $z=0$ of the ``potential''
\begin{equation}
\label{x9.4 }
V_0(z)=-\imath \, E= \imath \,(z+1/z-2)
\end{equation}
which gives an essential singularity in the exponential $e^{-\imath  t E}=e^{t V_0(z)}$ in \rf{x9}. The other essential singularity is of course at $z=\infty$.
The large time behaviour ($t\to\infty$) of the amplitude will be studied through the representation \rf{x9} by the complex saddle point method.

\subsubsection{Representation in the $w_+$ complex plane}

Changing variable in the $z$-integral to $w=w_+(z)$ gives us an alternative representation of the
amplitude.  This removes the square root and will be used in Section 4.4.2 where we give details.

\subsection{Return probability and quantum spectral dimension}
\label{sRetProb}  
\subsubsection{Large $t$ steepest descent analysis}
\label{ssRetProbSD}

We begin by considering the case $n=j=0$ which gives the amplitude $A_t(0,0;0,0)$ 
that a quantum walk 
returns to the starting point after time $t$ and therefore yields the return probability and the quantum spectral dimension.

\begin{figure}[h]
\begin{center}
\includegraphics[width=4in]{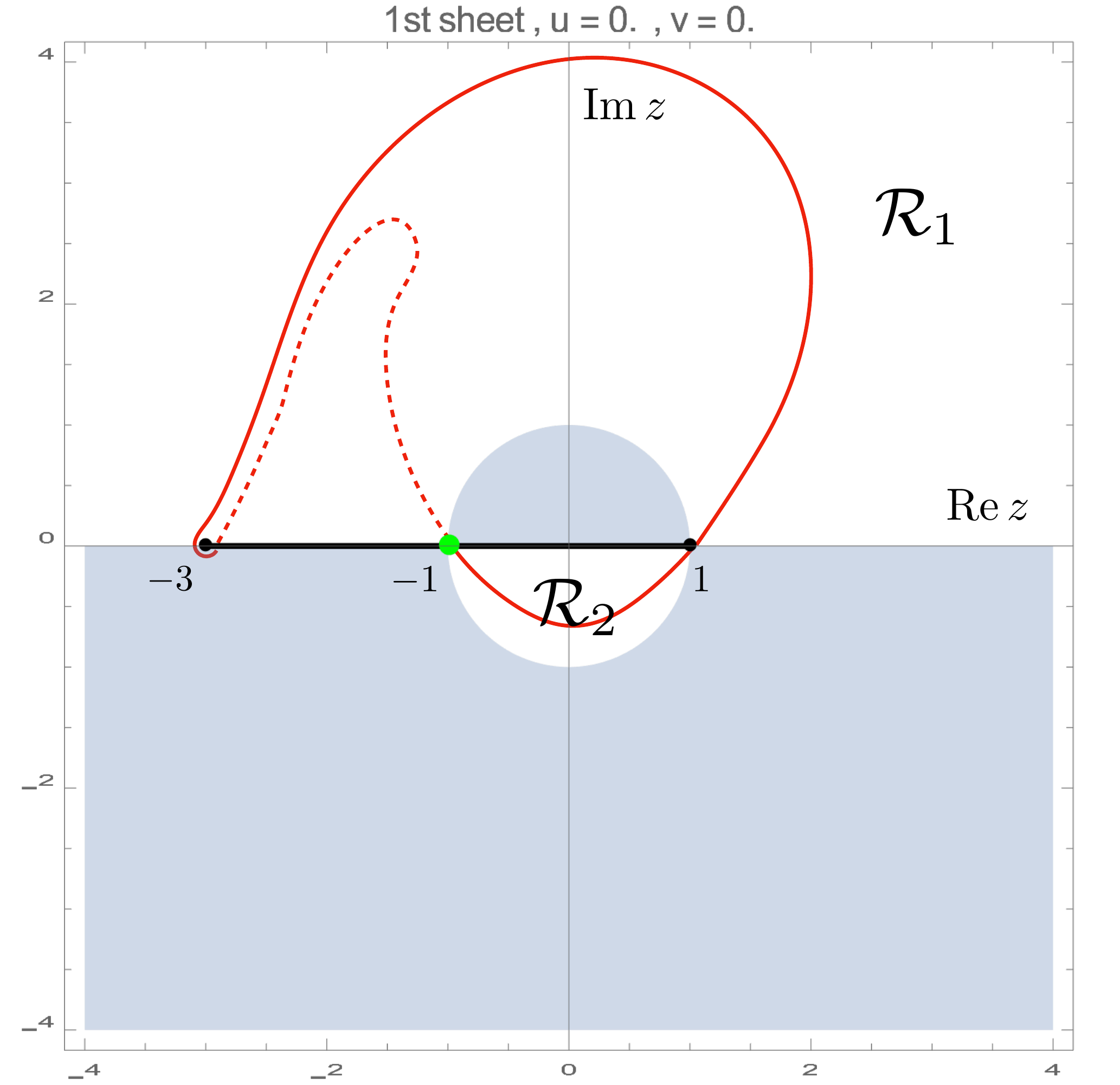}
\caption{The deformed integration contour $\mathcal{C}$ (red). It is obtained by deforming 
the original contour of Fig.~\ref{fContourZ}, picks the saddle points at  
$z=1$, $z=-1$ and the branch point at $z=-3$, and  stays in the white ``safe regions''.  The dashed part of the path is in the 2nd sheet. }
\label{fContourZ2}
\end{center}
\end{figure}

We note that the function $V_0(z)$ has a negative real part in the region
$\mathcal{R}_1$ where ${\rm Im}\,z>0$ and $|z|>1$ and also in the region $\mathcal{R}_2$ where ${\rm Im}\,z<0$ and 
$|z|<1$.  As an integration contour in \rf{x9} we choose a path $\mathcal{C}$ that starts at $z=1$ and moves to
$z=-3$ in the region $\mathcal{R}_1$.  From $-3$ we enter the region $\mathcal{R}_1$ in the 2nd sheet of the Riemann 
surface of $S$ and proceed to the point $z=-1$.  From -1 we move into the region $\mathcal{R}_2$ in the
first sheet and close the contour by going to $z=1$ inside $\mathcal{R}_2$, see Fig.~\ref{fContourZ2}

With the choice of integration contour described above 
the integrand in \rf{x9} goes exponentially to 0 as $t\to\infty$ except at the saddle points
$z=\pm 1$ and at the endpoint of the cut $z=-3$.   In order to see how the amplitude decays as
 $t\to\infty$
it therefore suffices to consider the contribution to the integral from small regions around these 
3 points, using steepest descent analysis.  Near the saddle point $z=1$ we can write 
$z=1+ u$ (with contour $u=e^{\imath \pi/4}v$, $v$ real) so that $V_0(z)=\imath  u^2+\mathrm{O}(u^3)$ and its dominant contribution in \rf{x9} gives a term of order $\mathrm{O}(t^{-3/4})$:
\begin{equation}
\label{sp1contrib}
\int_{-e^{-\imath {\pi\over 4}}\infty}^{+e^{-\imath {\pi\over 4}}\infty} {du\over 2\imath \pi}{2u\over\sqrt{4 u}} e^{\imath  t u^2}=e^{-3\imath \pi\over 8}{\sqrt{2}\over 4\pi} \Gamma(3/4)\, t^{-3/4}.
\end{equation}
Near the saddle point $z=-1$ we write $z=1+ u$ (with contour $u=e^{-\imath \pi/4}v$, $v$ real) so that $V_0(z)=- 4 \imath -\imath  u^2 -\imath  u^3+\mathrm{O}(u^4)$.  We shall need to expand to the next order as compared to the first saddle point.
The dominant and first subdominant contributions give an integral, where the dominant contribution, expected to be $\mathrm{O}(t^{-1})$ is odd and integrates to zero. The subleading contributes at 
order $\mathrm{O}(t^{-3/2})$:
\begin{equation}
\label{sp2contrib}
\int_{-e^{\imath {\pi\over 4}}\infty}^{+e^{\imath {\pi\over 4}}\infty} {du\over 2\imath \pi}    \imath  (-u -3/2 u^2 + \imath  t u^4) e^{\imath  t u^2}\,e^{-4 \imath  t}=    {3\over 4 \sqrt{\pi}} e^{\imath \pi/4} e^{-4 \imath  t}\, t^{-3/2}.
\end{equation}
At the branch point at $z=-3$ we write $z=-3+u$ so that $V_0(z)= -\imath \,16/3 + \imath \,8/9\, u 
+ \mathrm{O}(u^2)$. The dominant term in the integral is found to be of order $\mathrm{O}(t^{-1/2})$, and given by
\begin{equation}
\label{brpcontrib}
\oint {du\over 2\imath  \pi} {{8\over 9}\over \sqrt{-4 u}} e^{-{16\over 3} \imath t}e^{\imath {8\over 9} t u} 
= \sqrt{{9\over 8 \pi}} e^{-\imath {3\pi\over 4}} e^{-{16\over 3}\imath t}\, t^{-1/2}
\end{equation}
with the integral contour over $u$ going from $\imath \infty$ to zero in the first sheet, and back from zero to $\imath \infty$ in the second sheet.

Comparing the contributions \rf{sp1contrib}, \rf{sp2contrib} and \rf{brpcontrib}, one sees that the contribution \rf{brpcontrib} of the branch point at $z=-3$ dominates the large time asymptotics of the 
the probability amplitude  for returning to $(0,0)$. The probability of return after a time $t$ is therefore
\begin{equation}
\label{RetProbF}
P_t(0,0;0,0)=|A_t(0,0;0,0)|^2\sim {9\over 8\pi}\,t^{-1}.
\end{equation}

\subsubsection{Discussion}
The large time decay of the return probability \rf{RetProbF} as $t^{-1}$ means that the quantum spectral dimension of the comb is $d_{qs}=2$. Therefore this is an example where the quantum spectral dimension is not twice the classical spectrum dimension $d_{qs}\neq 2d_{s}$ since for the comb $d_s=3/2$ \cite{DJW}.  It satisfies however the inequality $d_{qs}\le 2 d_s$. 
  
  We expect that a similar large time scaling as $t^{-1}$ holds for the return probability starting from 
  any vertex on the comb.  
 
 The quantum spectral dimension for the comb is the same as for the quantum walk on the discrete line $\mathbb{Z}$. This is perhaps not completely unexpected since at $z=-3$ the
states with energy $E>4$ are dominating and these states are essentially supported on the
spine of the comb which is classically a one-dimensional object 
for which $d_{qs}$ is also 2 \cite{luck}. 
%
%
%
%
%
%

\subsection{Propagation into the teeth}

\subsubsection{Principle: propagation into the first tooth}
We study the behaviour of the wave function which propagates at a given velocity $u$ along a tooth by performing  the following rescaling.
We consider the tooth at the origin by setting $n=0$ and writing $j=ut+\tilde{\jmath}$ where 
$\tilde{\jmath}\in\bbZ$ and $u>0$ are fixed and we 
let $t$ increase along a sequence such that $tu\in\bbZ_+$.   We let
\beq{x10}
V_u(z)=\imath (z+z^{-1}-2) -u\log z.
\eeq
The amplitude reads
\beq{AmplU}
A_t(0,0;0,j)={1\over 2\pi \imath }\oint_{\Gamma}{1-z^{-2}\over \sqrt{(z+3)(z-1)}}\,
z^{-\tilde \jmath}\,e^{t V_u(z)}\,dz.
\eeq
The saddle points which give the dominant contribution to \rf{AmplU} in the large $t$ limit, $u$ and 
$\tilde\jmath$ being fixed and of order $\mathrm{O}(1)$, are given by
\beq{x11}
V'_u(z)=\imath (1-z^{-2})-uz^{-1}=0
\eeq
which has the solutions
\beq{x12}
z=z_\pm \equiv -{\imath u\over 2}\pm\oh\sqrt{4-u^2}.
\eeq
There is a critical value for the velocity
\begin{equation}
\label{ }
u_c=2.
\end{equation}
The behaviour of the amplitude is different depending on whether $0<u<2$, $u=2$ or $u>2$.

\subsubsection{Velocity $u<2$}
\label{sssu<2}
\begin{figure}[h]
\begin{center}
\includegraphics[width=4.in]{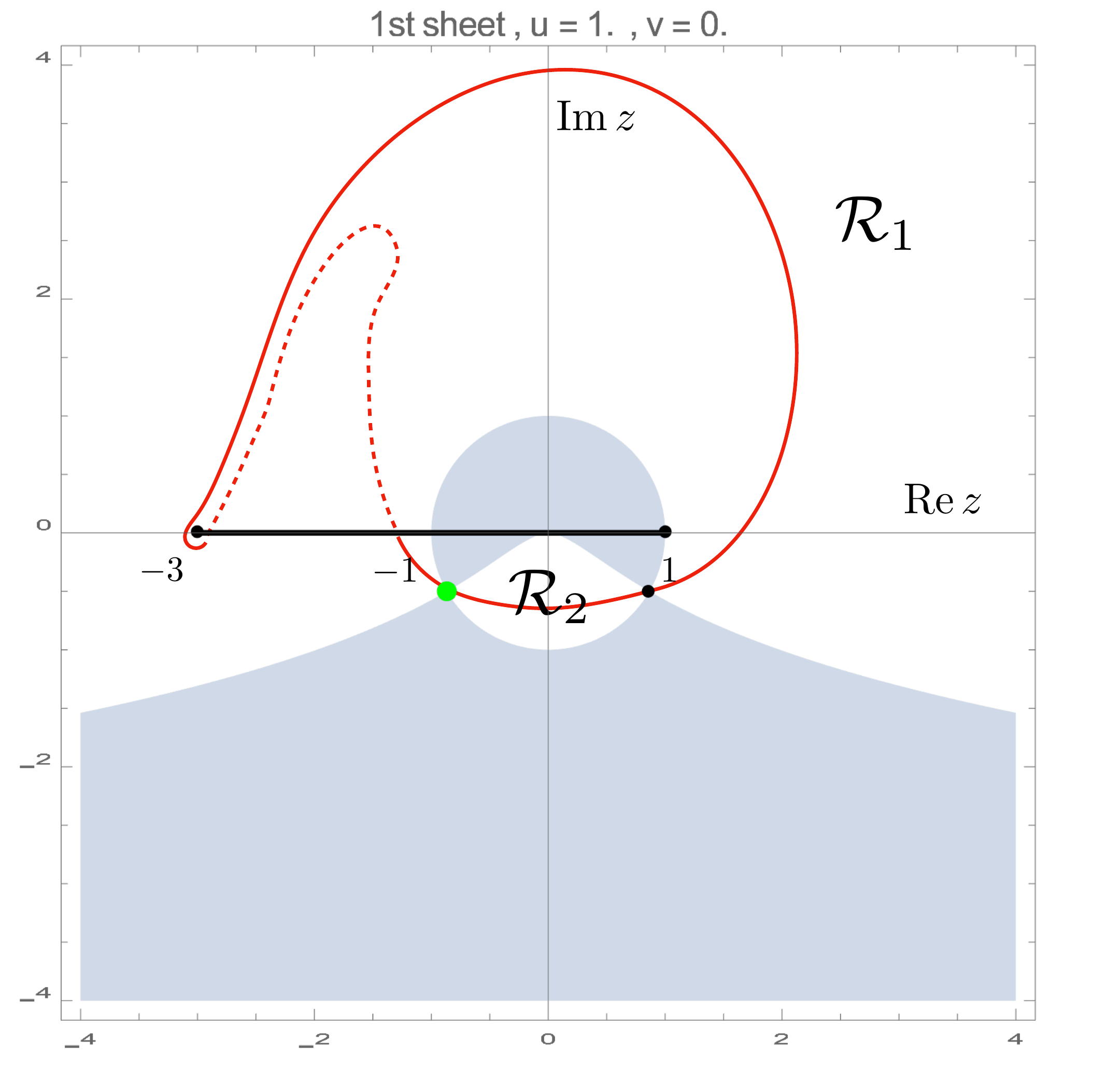}
\caption{The deformed integration contour which has to be taken in the integral \ref{AmplU} in the case $0<u<2$. It picks the two saddle points $z_+$ and $z_-$ in the lower half plane, 
marked green and black. In the ``safe'' white regions $\mathrm{Re}(V(z))<\mathrm{Re}(V(z_+))=\mathrm{Re}(V(z_-))$, in the ``dangerous'' gray regions $\mathrm{Re}(V(z))>\mathrm{Re}(V(z_+))=\mathrm{Re}(V(z_-))$. Again the dashed part of the contour lies in the second sheet.}
\label{DeformedC}
\end{center}
\end{figure}

We begin by considering the case $u<2$ in which case the saddle points lie on the lower half of the unit circle and can be parametrized by an angle $\varphi$
\beq{x13}
z_\pm=\pm e^{\mp \imath \varphi},~~~\varphi =\arcsin {u\over 2}.
\eeq  
Now we can deform the integration contour without entering the 2nd sheet, see Fig.\ \ref{DeformedC},
and the integrand decays exponentially with $t$  except at the saddle points, where the potential is purely imaginary:
\beq{x14}
V(z_+)=\imath (2\cos\varphi +u\varphi)
\eeq
and
\beq{x15}
V(z_-)=-\imath (2\cos\varphi +u(\varphi -\pi))=\overline{V(z_+)}+\imath u\pi.
\eeq
In particular, the contribution of the branch cut at  $z=-3$ is now exponentially small at targe $t$, since $\mathrm{Re}(V_u(-3))=-u \log(3)<0$, and it can be neglected.
Let 
\beq{x16}
I(z)={1\over 2\pi \imath } {1-z^{-2}\over \sqrt{(z+3)(z-1)}}.
\eeq
Then the dominant contribution to the contour integral coming from the two saddle points is
\beq{x17}
{\sqrt{2\pi}\over\sqrt{t}}e^{-2\imath t}\left[ {I(z_+)z_+^{-\tilde{\jmath}} e^{tV(z_+)}\over \sqrt{-V''(z_+)}} + (z_+\mapsto z_-)\right].
\eeq
Going through the saddle point along the path of steepest descent fixes the sign of the 
square root of $-V''(z_\pm)$.

We can write
\beq{x18}
z_+^{-\tilde{\jmath}} e^{tV(z_+)}=e^{\imath \varphi j}e^{2\imath t\cos\varphi}
\eeq
and
\beq{x19}
z_-^{-\tilde{\jmath}} e^{tV(z_-)}=(-1)^j e^{-\imath \varphi j}e^{-2\imath t\cos\varphi}.
\eeq
Furthermore, \beq{x20}
V''(z_+)=2\imath e^{3\imath \varphi}+ue^{2\imath \varphi}
\eeq
and $V''(z_-)=\overline{V''(z_+)}$.   
Our final expression for the amplitude is therefore
\beq{x21}
A_t(0,0;0,j) ={\sqrt{2\pi}\over\sqrt{t}}e^{-2\imath t}\left[ {I(z_+)e^{\imath \varphi j}e^{2\imath t\cos\varphi}
\over \sqrt{-V''(z_+)}} +{I(z_-)(-1)^j
e^{-\imath \varphi j}e^{-2\imath t\cos\varphi}
\over \sqrt{-V''(z_-)}}\right]+O(t^{-3/2}).
\eeq
The amplitude is a superposition of two periodic function in $j$, with respective 
wave numbers $\varphi$ and $\varphi+\pi$, and amplitude oscillating with $t$.
The probability for being at $j=ut+\tilde\jmath$ at time $t$ is
\begin{equation}
\label{AtoP}
P_t(0,0;0,u t+\tilde\jmath;t)=|A_t(0,0; 0, ut+\tilde\jmath )|^2
\end{equation}
which oscillates with $t$ and decreases as $t^{-1}$.
For large $t$, averaging the probability \rf{AtoP} with respect to $\tilde\jmath$ on an interval 
 $\tilde \jmath\in [-\ell/2,\ell/2]$, with $1\ll\ell\ll t$ (for instance $\ell=\sqrt{t}$) we get a coarse grained asymptotic probability profile for $j = u t$
\begin{equation}
\label{PCorGr}
P_{\mathrm{cg}}(0,ut; t)= 2\pi\left({I(z_+)|^2\over |V''(z_+)|}+ {|I(z_-)|^2\over |V''(z_-)|}\right){1\over t}={c(u)\over t},
\end{equation}
since the cross terms vanish when we average, and
\begin{equation}
\label{CofUExpl}
c(u)={1\over 2\pi}{u^2\over\sqrt{4-u^2}}\left({1\over\sqrt{26+6 u^2 +5\sqrt{4-u^2}}}+{1\over\sqrt{26+6 u^2 -5\sqrt{4-u^2}}}\right). 
\end{equation}
Furthermore, we have absolute bounds on the asymptotics for the local probability for $0<u<2$
\begin{equation}
\label{Pbounds}
{d_-(u)\over t} \le
P_t(0,0;0,u t+\tilde\jmath )\le {d_+(u)\over t}.
\end{equation}
with
\begin{equation}
\label{DofUExpl}
d_\pm(u) = {1\over 2\pi}{u^2\over \sqrt{4-u^2}}\left(
{1\over(26+6 u^2 -5\sqrt{4-u^2})^{1/4}}\pm {1\over (26+6 u^2 +5\sqrt{4-u^2})^{1/4}} \right)^2 .
\end{equation}

\subsubsection{Velocity u=2}
\label{ssucrit}
At $u=2$ the two saddle points merge since  
$z\equiv z_+=z_-=-\imath $, $\varphi =\pi/2$ and $V(z_+)=\imath \pi$.   In this case $V''(z_+)=0$ and we must expand the potential to third order at $z=-\imath $ to get the large time behaviour of the amplitude.
The natural scaling for $j$ and the integration variable $z$ are
\begin{equation}
\label{Ueq2Scal}
j=2 t + \hat \jmath\, t^{1/3}
\ ,\quad z=-\imath + \hat z\, t^{-1/3}
\end{equation}
so the leading oscillating part in \rf{x9} near the saddle point is 
\begin{equation}
\label{ }
z^{-j} \emath^{\imath t(z+1/z-2)}=\emath ^{- 2 \imath t} \emath^{\imath {\pi\over 2} j} \emath^{-{\imath\over 3}(\hat z^3+ 3 \hat \jmath \hat z)}(1+\mathrm{O}(t^{-1/3})).
\end{equation}
Integrating \rf{x9} over $z$, in the large $t$ limit, by the steepest descent method, the integration contour picks the $z=-\imath$ saddle point as the dominant contribution, and the integration gives an Airy function for the amplitude
\begin{equation}
\label{AiryFrontA}
A_t(0,0;0,j)= c_2 \,\emath ^{- 2 \imath t} \emath ^{\imath {\pi\over 2} j}\,   t^{-1/3}\,\mathrm{Ai}(\hat \jmath)(1+\mathrm{O}(t^{-1/3}))
\end{equation}
with a constant $c_2=-\sqrt{\imath-1/2}$. The probability to be at $j= 2 t + t^{1/3}\hat\jmath$ at time $t$
is therefore to leading order
\begin{equation}
\label{AiryFrontP}
P_t(0,0; 0, 2 t + t^{1/3}\hat\jmath ) = \oh\sqrt{5}\, t^{-2/3}\, \mathrm{Ai}(\hat\jmath)^2.
\end{equation}

\subsubsection{Velocity u>2}
For $u>2$ the saddle points move away from the unit circle and are pure imaginary:
\beq{x22}
z_\pm= -\imath \left( {u\over 2}\pm \oh\sqrt{u^2-4}\right).
\eeq
They can be parametrized by $\psi>0$ as
\begin{equation}
\label{z2psi}
z_+= - \imath  e^\psi\,,\ z_+= - \imath  e^{-\psi} \,,\ \psi=\mathrm{acosh}(u/2)
\end{equation}
so that 
\begin{equation}
\label{VzpmU>2}
V(z_\pm)=\mp\, 2(\psi\cosh(\psi) -\sinh(\psi))+ \imath (\pi\cosh(\psi)-2).
\end{equation}
The potential $V$ has a real part at the saddle points. The real part of $V(z_+)$ is negative and the real part of $V(z_-)$ is positive. 
The steepest descent path goes through $z_+$, not  through $z_-$, so we pick a single contribution, giving an amplitude which decays exponentially with $t$ for $u>2$.
The steepest descent calculation gives the large $t$ scaling form for the amplitude at $j=u t +\tilde\jmath$ with $\tilde\jmath=\mathrm{O}(1)$
\begin{equation}
\label{Asymu>ucv0}
A_t(0,0;0,u t +\tilde\jmath)= e(u)\, \emath^{- 2\imath\,t} \emath^{\imath {\pi\over 2} (ut+\tilde\jmath) }\ t^{-1/2}\ \emath^{- t \varpi(u)}\ \emath^{-\tilde\jmath\,\chi (u) }\ (1+\mathrm{O}(t^{-1/2}))
\end{equation}
where
\begin{equation}
\label{ }
\begin{split}
\varpi(u)&=-\sqrt{u^2-4}+u\ \log\left({u+\sqrt{u^2-4}\over 2}\right)\ ,\quad
\chi(u)= \log\left({u+\sqrt{u^2-4}\over 2}\right)\\
\end{split}
\end{equation}
and for completeness
\begin{equation}
\label{ }
e(u)={1\over \imath \sqrt{2\pi}}{ 4 u\over (u^2-4)^{{1\over 4}} \sqrt{-(u+\sqrt{u^2-4}- 2\imath)(u+\sqrt{u^2-4}+6\imath)}}.
\end{equation}

\subsubsection{Propagation into the $n$-th tooth}
The large time propagation along the $n$-th tooth, with $n$ of order $\mathrm{O}(1)$, 
can of course be studied by the same steepest descent method as we used for the tooth at $n=0$. 
One must take into account the additional $w_-(z)^{|n|}$ term in the integral 
representation \rf{x9}, with $w_-(z)$ given by \rf{x2}. 
The asymptotics for the amplitude \rf{x9} become (for $u<2$)
\begin{equation}
\label{OscAmpV0}
A_t(0,0;n,j) ={\sqrt{2\pi}\over\sqrt{t}}e^{-2\imath t}\left[ {I(z_+)w_-(z_+)^{|n|}e^{\imath \varphi j}e^{2\imath t\cos\varphi}
\over \sqrt{-V''(z_+)}} +{I(z_-)(-1)^jw_-(z_-)^{|n|}
e^{-\imath \varphi j}e^{-2\imath t\cos\varphi}
\over \sqrt{-V''(z_-)}}\right]
\end{equation}
and the asymptotics for the coarse grained probability at $j= u t$ is, by the same argument as before,
\begin{equation}
\label{PCorGrN}
P_{\mathrm{cg}}(n,ut; t)=2\pi\left({|I(z_+)|^2 {|w_-(z_+)|}^{2|n|}\over |V''(z_+)|}+ {|I(z_-)|^2  {|w_-(z_-)|}^{2|n|}\over |V''(z_-)|}\right){1\over t}={c(u,n)\over t}
\end{equation}
where
\begin{equation}
\label{CofUExplN}
c(u,n)={1\over 2\pi}{u^2\over\sqrt{4-u^2}}\left({|w_-(z_-(u))|^{2|n|}\over\sqrt{26+6 u^2 +5\sqrt{4-u^2}}}+{|w_-(z_+(u))|^{2|n|}\over\sqrt{26+6 u^2 -5\sqrt{4-u^2}}}\right).
\end{equation}

\subsubsection{Discussion}
The above results are qualitatively similar to the behaviour of the  amplitude for a quantum walk on the discrete line \cite{luck}. 
The wave function expands linearily with time, with a front moving with constant velocity. Here the front velocity is $u_{\mathtt{tooth}}=2$, to be compared with the front velocity in both directions on the discrete line, which is $u_{\mathtt{line}}=\pm 1$. The shape of the front is an Airy function given by \rf{AiryFrontA} and  \rf{AiryFrontP}, as for the line. 
Ahead of the front is an exponentially decaying 
evanescent wave. Behind the front, the bulk of the wave function is a biperiodic function with some more complicated structure given by the interplay between the two saddle points. 
The coarse grained probability density function has an explicit form, self similar with the time $t$ as it expands, given by \rf{PCorGr}, \rf {CofUExpl}. 
One can check that the spatial extent of the transition region at the front is of order $t^{1/3}$.

The propagation into the $n$-th tooth, is very similar, but with a density function depending on $n$ (see \rf{CofUExplN}), and decreasing exponentially with $|n|$, since 
$w_-(z_\pm)|<1$.   This is what we expect intuitively since the wave function 
has to move $n$ steps along the spine before starting to propagate into the $n$th tooth. 

\subsubsection{Probability to be in the teeth at large $t$}
We first calculate 
\beq{probteeth}
P_T(n)\equiv \lim_{t\to\infty} \sum_{j=0}^\infty P_t(0,0; n,j ),
\eeq
the probability that the walk ends up in the $n$th tooth.
In the large $t$ limit we can replace $P_t(0,0; n,j )$ by the coarse grained probability \rf{PCorGrN} and the sum
over $j$ is replaced by an integral over $u$ from $0$ to $2$.  The values of $u>2$ give a vanishing contribution in the limit $t\to\infty$.  Using \rf{PCorGrN} we find
\beq{IntRepPT}
P_T(n)=\int_0^2 c(u,n)\, du.
\eeq
We change the integration variable to $\vp$, where $u=2\sin\vp$,  cf.\ \rf{x13}. 
 After some calculations we find
\beq{IntRepTT2}
P_T(n)={1\over\pi}\int_0^\pi d\vp\, {(1-\cos^2\vp)|w_-(e^{-i\vp})|^{2|n|}\over
\sqrt{(5+3\cos\vp)(1-\cos\vp)}}.
\eeq
For large $|n|$ the above ingegral is dominated by the region around $\vp=0$ where $w_-$ equals $1$.
It is not hard to check that $|w_-(e^{-i\vp})|^2=1-\sqrt{2\vp}+O(\vp )$ for small $\vp$ so
$P_T(n)\sim n^{-4}$ as $n\to\infty$.
The total probabilty of ending up in a tooth is given by
\begin{eqnarray}
P_{\rm Teeth} & = & \sum_{n=-\infty}^{\infty} P_T(n) \nonumber \\
&=& {1\over\pi} \int_0^\pi d\vp\, {(1-\cos^2\vp)\over
\sqrt{(5+3\cos\vp)(1-\cos\vp)}} {1+|w_-(e^{-i\vp})|^{2}\over 1-|w_-(e^{-i\vp})|^{2}}\,.\label{TotProbTeeth}
\end{eqnarray}
The function $w_-$ depends on $\vp$ in a complicated way so the integral is hard to evaluate analytically but numerically we find $P_{\rm Teeth}= 0.63159137\ldots$

There is an alternative way to calculate $P_T(n)$ which we now explain and 
does not rely on using the coarse grained probability density.   The probability of being in tooth $n$ at time $t$ is
\beq{ProbToothnt}
p_n(t)= \sum_{j=0}^\infty |A_t(0,0;n,j)|^2
\eeq
where $A_t(0,0;n,j)$ is given by \rf{x9}.  We can carry out the sum over $j$ in \rf{ProbToothnt} and find
(for $n\geq 0$)
\beq{ProbToothnt2}
p_n(t)=\oint dz_1\oint d\bar{z_2} \,I(z_1)I(\bar{z_2}){w_-(z_1)^nw_-(\bar{z_2})^n\over
1-z_1\bar{z_2}}e^{-\imath tE(z_1)+\imath E(\bar{z_2})},
\eeq
where the integration contours in the $z_1$- and $z_2$-planes are as is explained in subsection 4.1.1. 
We fix $\bar{z_2}$ in the minimal contour $\Gamma_0$ depicted in Fig.\ \ref{fContourZ}.  We then integrate $z_1$ along a contour $\Gamma_1$ enclosing the minimal contour.   Next we deform $\Gamma_1$ to the contour $\Gamma_2$ depicted in Fig.\ \ref{fContourZ2}.   In the deformation process we may pick up a contribution from the 
pole at $z_1=(\bar{z_2})^{-1}$.  The integral over $\Gamma_2$ goes to 0 as $t\to\infty$ since the integrand decays exponentially in $t$ except at 3 points which give contributions which have a power law decay in $t$ as $t\to\infty$ by the same arguments as in subsection 4.2.1.  We only pick up a contribution from the pole when 
$\bar{z_2}$ is on the unit circle in the upper half plane.  Using that $E(z)=E(z^{-1})$ we see that the pole 
contribution is independent of $t$ and
\beq{Limit}
\lim_{t\to\infty}p_n(t)=P_T(n)=2\pi\imath\int dz\, I(z)I(z^{-1})w_-^n(z)w_-^n(z^{-1}),
\eeq
where the integration is along the unit circle in the upper half plane and we have renamed the integration variable.   Changing the integration variable to $\vp$, where $z=e^{\imath\vp}$, we recover the result 
\rf{IntRepTT2} after some calculations.

\subsection{Propagation along the spine}
\subsubsection{Principle}
We now study the propagation along the spine.
We first consider the case $j=0$ and scale $n=vt+\tilde{n}$ where $v$ is fixed (and chosen to be positive) 
and $\tilde{n}\in\bbZ$ is $\mathrm{O}(1)$
as $t\to\infty$.  
The amplitude reads
\begin{equation}
\label{Ainwplane}
A_t(0,0;n,0)={1\over 2\pi \imath }\oint_{\Gamma}{1-z^{-2}\over \sqrt{(z+3)(z-1)}}\,w_+^{-\tilde n}\,e^{t\, W_v(z)}\,dz
\end{equation}
with $w_+(z)= w_-(z)^{-1}$ given by \rf{x2}, and the potential is
\beq{x24}
W_v(z)=\imath (z+z^{-1}-2)-v\log (w_+(z)).
\eeq
The large $t$ limit can be studied by the steepest descent method. 
The saddle point equation is
\beq{x25}
\imath(1-z^{-2})-{v\over \sqrt{(z+3)(z-1)}}=0.
\eeq
One can check that there is a critical velocity
\begin{equation}
\label{vcrit}
v_c= {3\sqrt{3}\over 4}
\end{equation}
such that when  $0<v<v_c$ the saddle point equation \rf{x25} has four real solutions, and two complex.  Two of them lie in the interval $(-3,-1)$
and two in $(-1,1)$.  At $v=v_c$ the two saddle points in $(-3,-1)$ merge and they become complex for $v>v_c$ while the solutions in $(-1,1)$ stay real.

\subsubsection{Another integral representation ($w$\,-\,plane)}
In the case $u=0$, $v>0$, as well as in the general case $u>0$, $v>0$ that we shall discuss later, it is convenient to make a change of
variable and replace $z$ by $w=w_+(z)$. The mapping $z\leftrightarrow w$  reads explicitly
\begin{equation}
\label{ztowmapping}
w(z)= {1\over 2} \left(z+1+\sqrt{(z+3)(z-1)}\right),\quad z(w)=w+ 1/w -1
\end{equation}
and is such that  
\beq{x26}
\frac{w'(z)}{w(z)} = {1\over \sqrt{(z+3)(z-1)}}.
\eeq
\begin{figure}[h!]
\begin{center}
\includegraphics[width=3.5in]{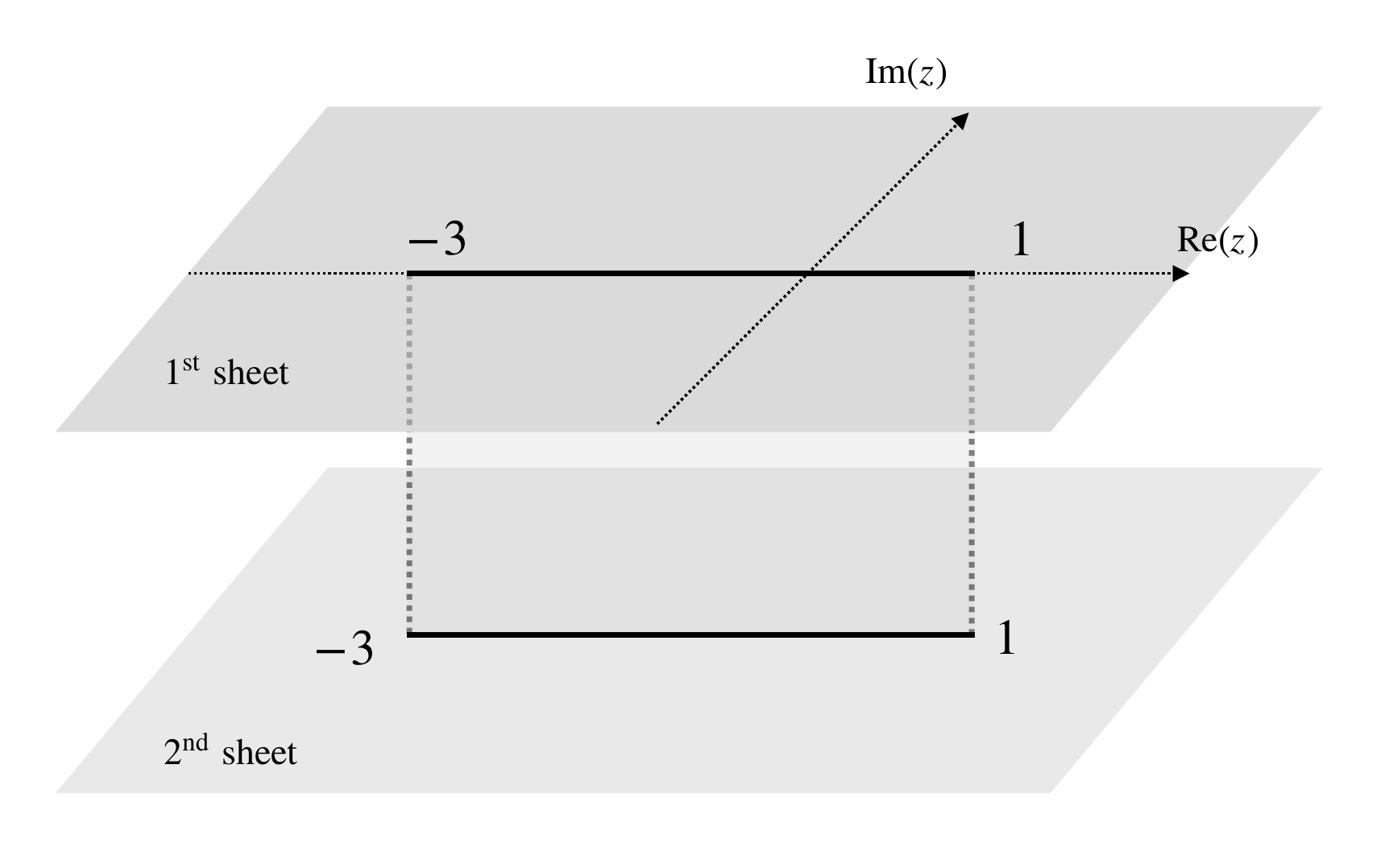} \\ 
$\big\downarrow$ \\
\bigskip \includegraphics[width=4in]{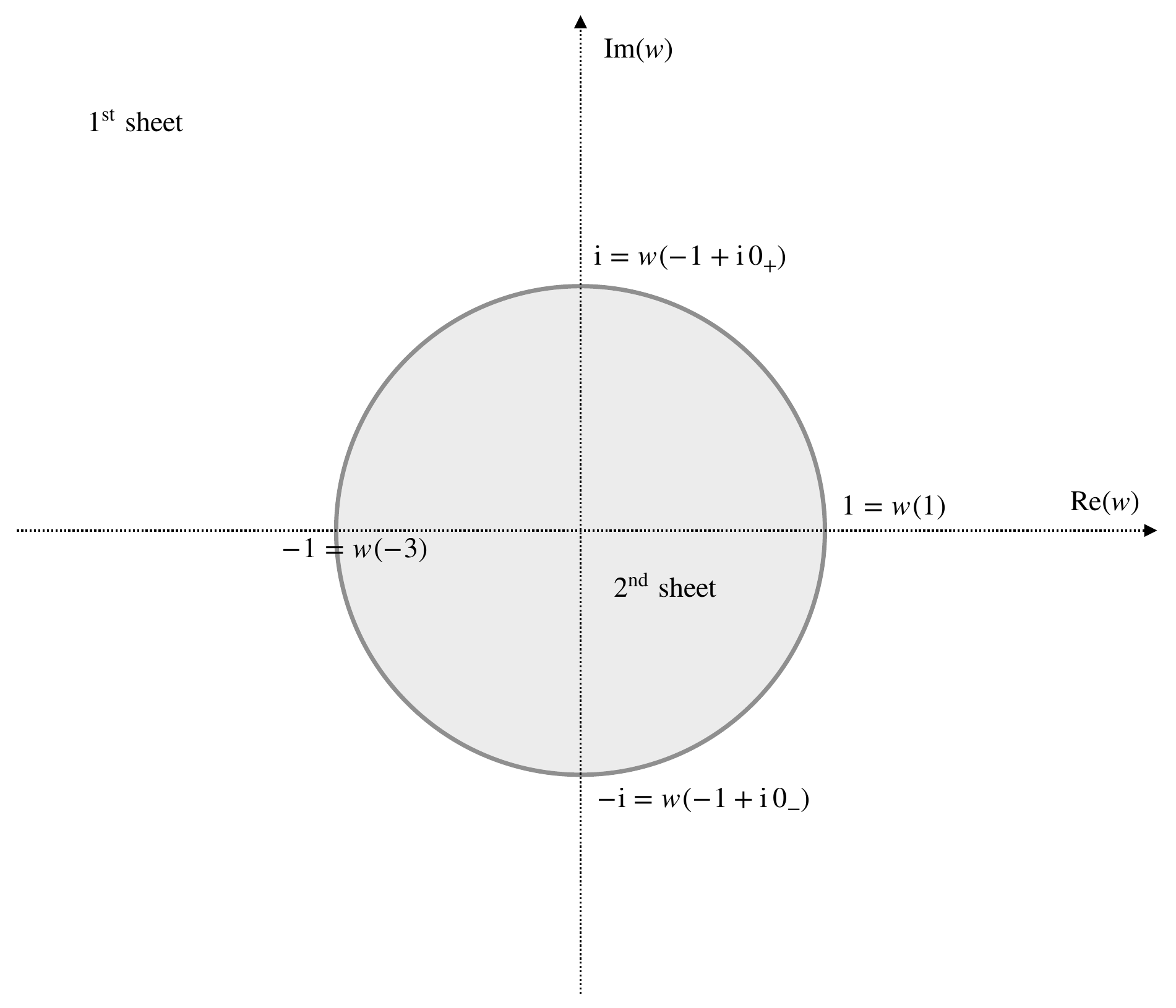}
\caption{The mapping $z\rightarrow w$ from the double sheeted $z$ plane with a branch cut along $[-3,1]$ (upper picture) and the complex $w$ plane (lower picture).}
\label{ztow}
\end{center}
\end{figure}
It is represented on Fig.~\ref{ztow}.
This mapping sends the double sheeted $z$ complex plane, with a cut along $[-3,1]$,
 onto the complex $w$ plane, with the first sheet mapped onto the exterior of the unit circle $|w|>1$ 
 by $w(z)=w_+(z)$, the cut onto the unit circle $|w|=1$, and the second sheet mapped onto the interior of the unit circle $|w|<1$ by $w(z)=w_-(z)$. 
The branch points of the cut are mapped onto $w(1)=1$, $w(-3)=-1$.
The other interesting points in the first and second sheets are mapped respectively onto
$w(-1)= \imath$, $w(0)=e^{\pi /3}$, $w(\infty)= \infty$ (first sheet) and onto
$w(-1)= -\imath$, $w(0)=e^{-\pi /3}$, $w(\infty)= 0$ (second sheet).

The integral representation of the amplitude becomes, for $n=v\,t+\tilde n$ and $j=0$,
\begin{equation}
\label{IntRepWu}
A_t(0,0;n,0)={1\over 2\pi \imath }\oint_{\Gamma} dw\ 
{(w+w^{-1}) (w+w^{-1}-2)
\over {(w+w^{-1}-1)}^{3}}\,w^{-\tilde n}\,\emath^{t\,\mathbb{W}(w;0,v)}
\end{equation}
with $\mathbb{W}$ the potential in the $w$ variable (defined in the general case $u\neq 0$ and $v\neq 0$, which will be studied later)
\begin{equation}
\label{WPotZgen}
\mathbb{W}(w;u,v)=\imath(z(w)+z(w)^{-1}-2 )- u\,\log(z(w))- v\,\log(w)
\end{equation}
and $\Gamma$ is a c.c.w contour encircling the unit circle $|w|=1$ once.

Using this representation, it is much easier to discuss how the contour $\Gamma$ must be deformed in the $w$ complex plane when using the steepest descent method, and which saddle points are relevant for  the large $t$ asymptotics of the amplitudes.
The saddle point equation $\mathbb{W}'(w;0,v)=0$ is now an algebraic equation of degree six:
\begin{equation}
\label{SPEquV}
{\imath (w-1)^3 (w+1)(w^2+1)\over w(w^2-w+1)^2} - v =0.
\end{equation}
We study this equation and the potential $\mathbb{W}(w)$ by a combination of analytic and numerical methods and summarize the results in the next sections.
 
\subsubsection{Velocity $v=0$}
\begin{figure}[h!]
\begin{center}
\includegraphics[width=2.5in]{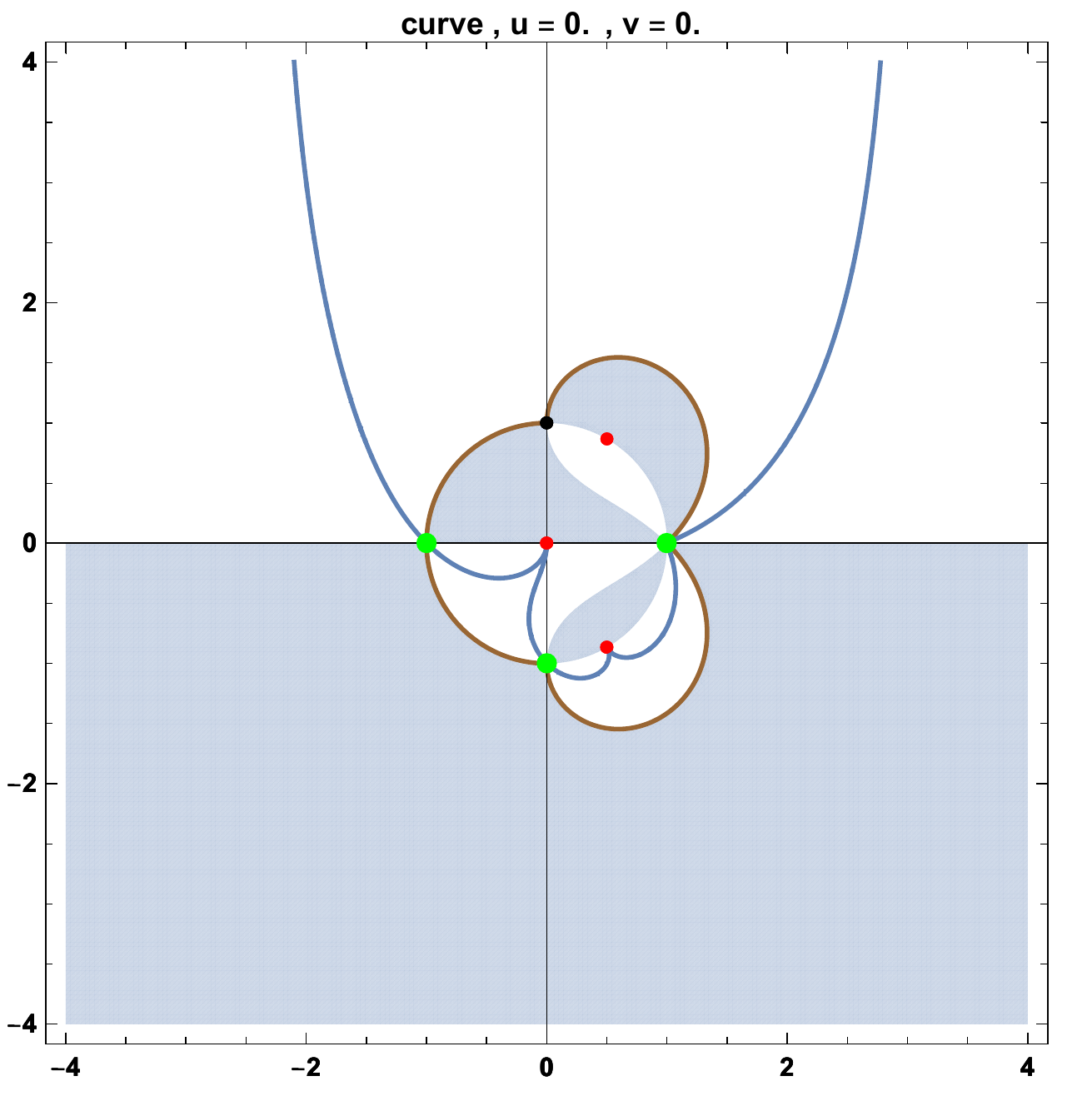}
\caption{The steepest descent path (blue) for the integral \ref{IntRepWu} in the $w$ plane for $v=0$, $u=0$. The original integration path is in brown and corresponds to the red contour in Fig.\ \rf{fContourZ}. 
Again in the ``safe'' white regions the real part of the potential is negative $\mathrm{Re}\,\mathbb{W}(w) <0$, in the ``dangerous'' gray regions $\mathrm{Re}\,\mathbb{W}(w) >0$. The saddle points (extrema of $\mathbb{W}(w)$ are represented by black dots (when irrelevant) and green dots (when picked by the steepest descent contour). The red dots are the singular points where $\mathbb{W}(w)$ diverges (has a pole).}
\label{u=0_v=0}
\end{center}
\end{figure}

Fig.~\ref{u=0_v=0} shows the saddle points and the steepest descent integration path in the case $u=v=0$ already studied for the return probability in Section~\ref{sRetProb}, using the $z$-plane representation.
In the $w$-plane, there are four saddle points: $w=1$ which is a triple point ($\mathbb{W}'=\mathbb{W}''=\mathbb{W}'''=0$), and $w=\imath$, $-1$ and $-\imath$ which are simple points ($\mathbb{W}'=0.\, \mathbb{W}''\neq 0$).
The red points are the poles of $\mathbb{W}$, at $w=0$, $\emath^{\imath\pi/3}$,  $\emath^{-\imath\pi/3}$ (there is a fourth pole at infinity $w=\infty$).
The white regions are the domains where the real part of the potential is negative ($\mathrm{Re}(\mathbb{W})<0$), while the gray regions  are the domains where the real part of the potential is positive ($\mathrm{Re}(\mathbb{W})>0$). The brown curve is the original integration curve considered in the $z$-plane, the two rightmost lobes correspond 
to the integration over the unit circle, the leftmost half circle to the integration around the cut from $z=-1$ to $z=-3$ and back.
The steepest descent method consists in deforming the integration contour in the region where the real part of the potential $\mathbb{W}$ is maximally negative, until one reaches a steepest descent contour where the variation of the integrand is real, i.e.\ such that the differential $d\mathbb{W}$ is real ($\mathrm{Im}(dz\,\mathbb{W}'(z))=0$).
The steepest descent path goes through some saddle points, starting and ending in valleys, i.e.\ along directions where $\mathrm{Re}(\mathbb{W})$ goes to $-\infty$.

It is clear from Fig.~\ref{u=0_v=0} that when $u=v=0$, the original contour must be deformed in the white regions only. The steepest descent path is depicted in blue. It goes through the three saddles points at $w=-1$, $w=-\imath$ and $w=1$ (depicted in green), and does not go through the saddle point at $w=\imath$.
One notes that $\mathrm{Re}(\mathbb{W})=0$ at these three saddle points. Therefore the three saddle points contribute by some power of $t$ times an oscillatory term in the large $t$ limit. These are the three terms discussed in the $z$-plane representation in sect.~\ref{ssRetProbSD}. The analysis of this section can of course be repeated here. The dominant (least decreasing power) term comes from the $w=-1$ saddle point.

\subsubsection{Velocity $v<v_c$}
\begin{figure}[h!]
\begin{center}
\includegraphics[height=2.5in]{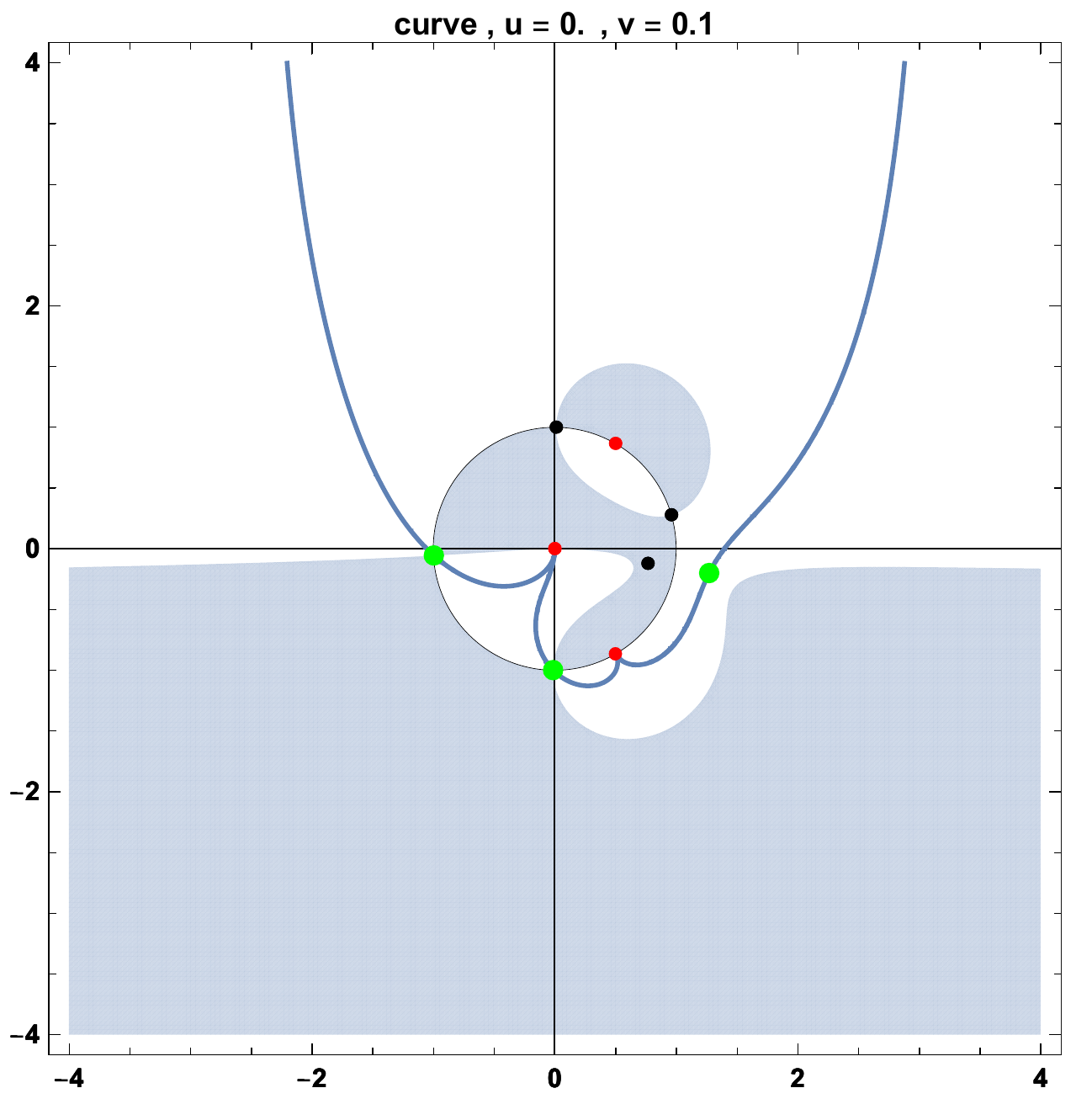}
\includegraphics[height=2.5in]{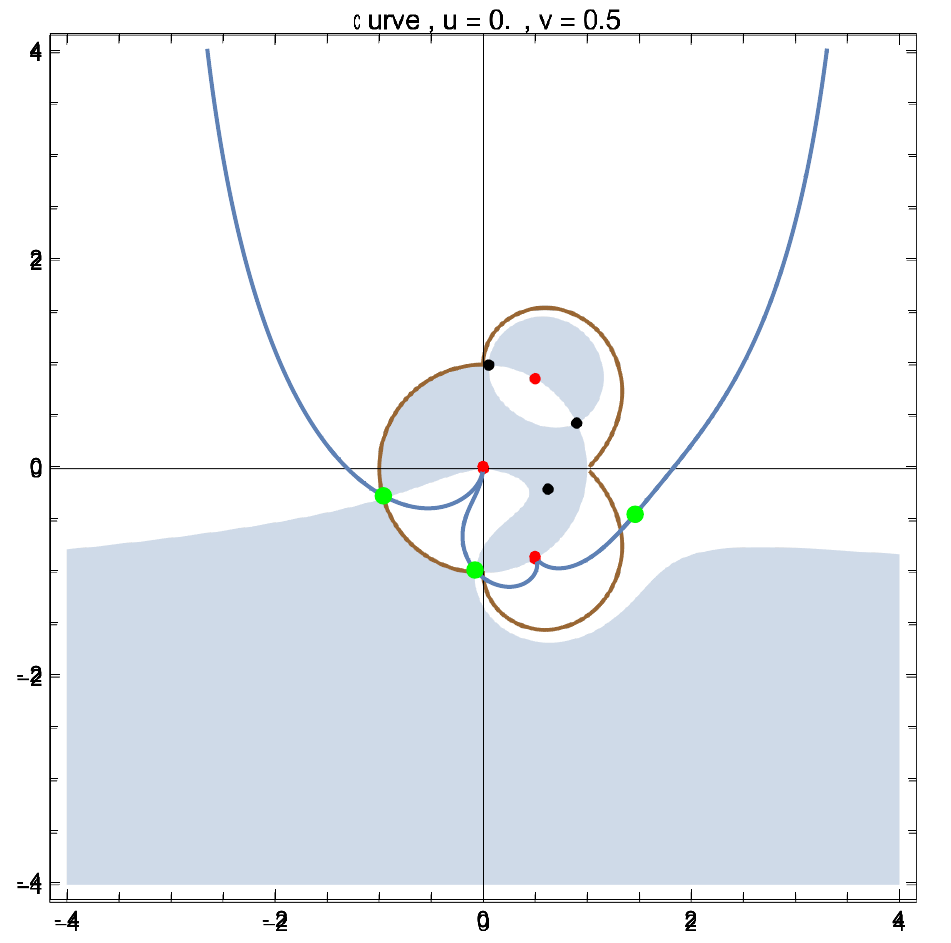}
\includegraphics[height=2.5in]{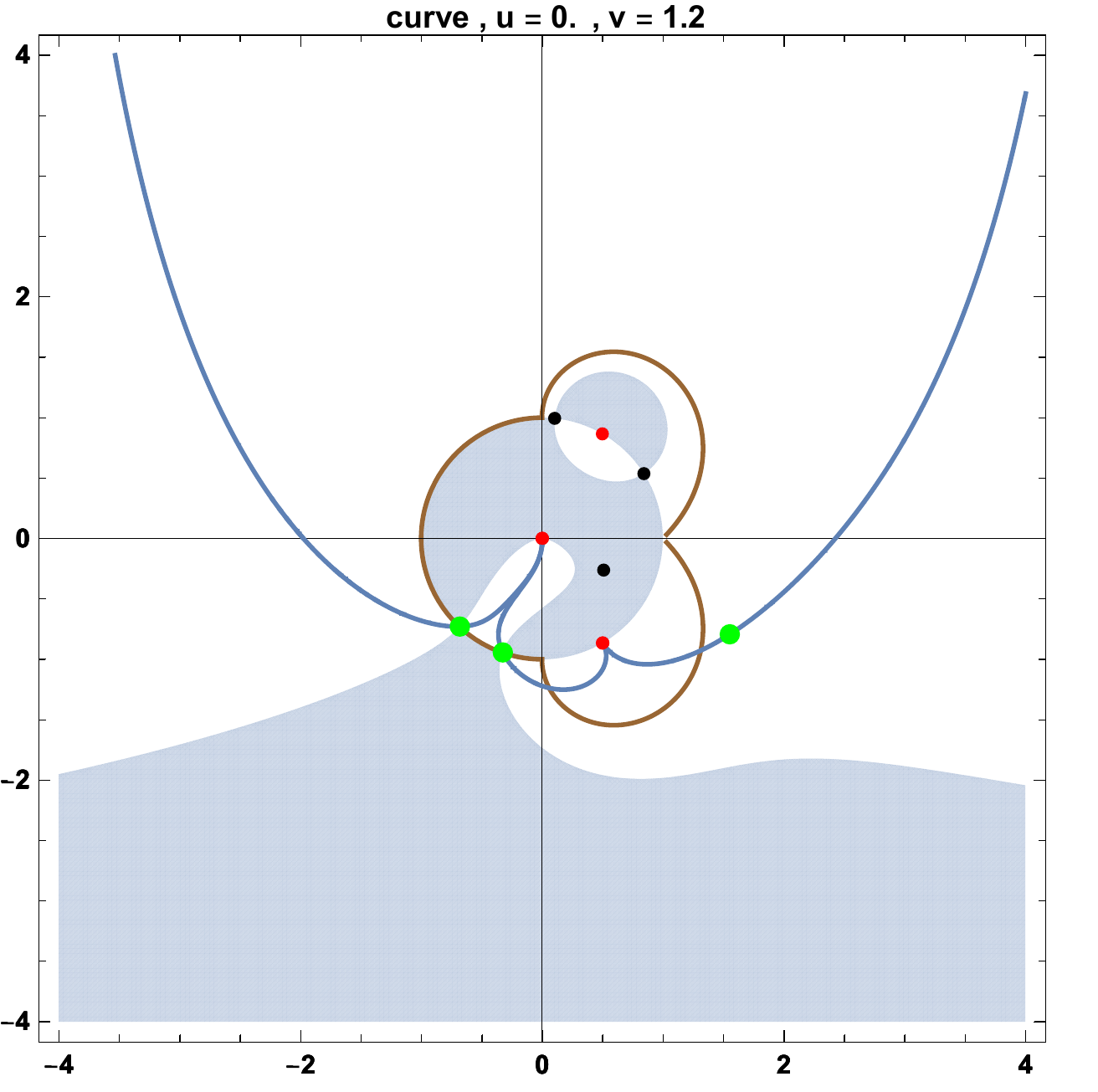}
\caption{ Three examples of steepest descent path (blue) in the $w$ plane for the integral \ref{IntRepWu}, for $u=0$ and three values of $v=0.1$, $0.5$ and $1.2$, in the interval  $0<v<v_c$. The representation conventions are the same as those of Fig.~\ref{u=0_v=0} for the $u=v=0$ case.}
\label{u=0_v<vc}
\end{center}
\end{figure}

Fig.~\ref{u=0_v<vc} shows how the saddle points and the steepest descent integration path evolve when $v$ increases, but stays below $v_c$.
The two leftmost saddle points (in green) move along the unit circle and come closer as $v$ increases. The fact that they are on the unit circle means that the potential $\mathbb{W}$ given by \rf{WPotZgen} is 
purely imaginary at these two saddle points. 
The rightmost triple saddle points splits into three separate simple saddle points. Only one of them (in green) is picked by the steepest descent path, and it is the one with $\mathrm{Re}(\mathbb{W})<0$. 
Therefore is will give a subdominant (exponentially decaying with $t$) term.
The two other ones have respectively $\mathrm{Re}(\mathbb{W})=0$ and $\mathrm{Re}(\mathbb{W})>0$ but are not relevant, since they do not lie on the steepest descent contour.

Through the steepest descent method, the two relevant saddle points $w_1$ and $w_2$ 
give for the amplitude to move 
along the spine, with $n=v\, t+\tilde n$, large $t$ asymptotics of the  form given in 
\rf{OscAmpU0}. It is a universal decaying power $t^{-1/2}$, multiplied by 
a sum of two oscillatory terms, whose amplitudes $a_k$ and wave numbers 
$\Phi_k$, $k=1,2$,  depend on the 
velocity $v$. The $\Phi$'s are nothing but the imaginary part of the potential $\mathbb{W}$ at the saddle points 
$w_1$ and $w_2$. The amplitudes are obtained from the second derivatives $\mathbb{W}''$ and the integrand in \rf{Ainwplane} at the two saddle points:
\begin{equation}
\label{OscAmpU0}
A_t(0,0;n,0) =
{1\over\sqrt{t}}\left({a_1(v)\,w_1(v)^{-\tilde n}\,\emath^{\imath t \Phi_1(v)} + a_2(v)\,w_2(v)^{-\tilde n}\,\emath^{\imath t \Phi_2(v)} }\right) +\mathrm{O}(t^{-1}).
\end{equation}

The two relevant saddle points are on the left half of the unit circle, i.e.\ $\mathrm{Re}(w)<0$, $|w|=1$, and thus correspond in the $z$ plane to points on the real interval $-3< z<-1$. They are associated to eigenmodes of the original Hamiltonian of the form \rf{n1}, which are localized on the spine ($n$ direction) and decay exponentially along the teeth ($j$ direction). Therefore, the part of the wavefuntion which propagates along the spine in the $n\to\pm\infty$ direction stays localized close to the spine and one does not find 
a probability flux along the teeth in the $j\to\infty$ direction for large $n$.

\subsubsection{Velocity $v=v_c$}
\label{ssvcrit}
\begin{figure}[h!]
\begin{center}
\includegraphics[height=2.5in]{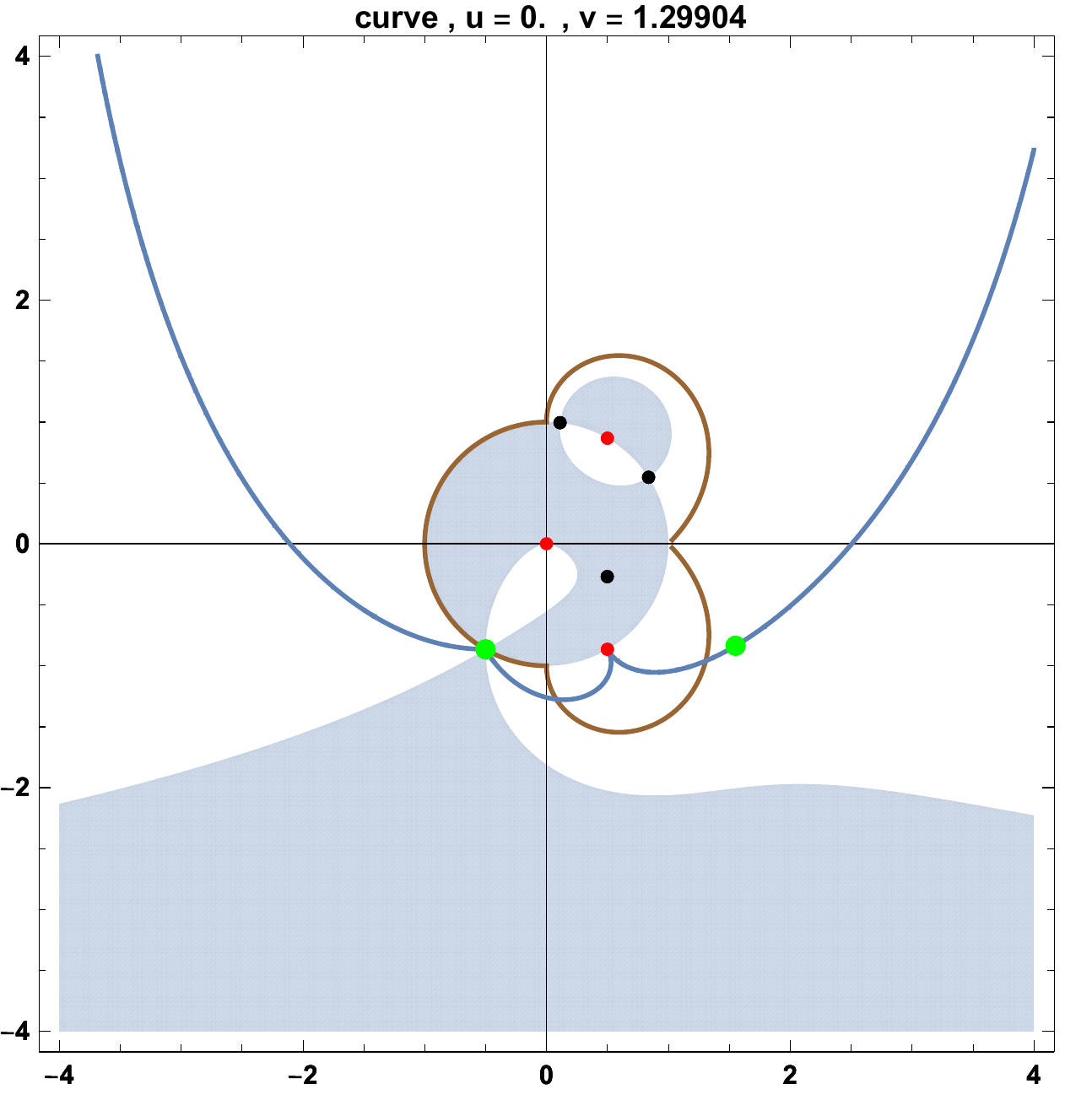}
\caption{ The steepest descent path (blue) in the $w$ plane at the critical velocity $v=v_c$, $u=0$. Same conventions as for the two previous figures.}
\label{u=0_v=vc}
\end{center}
\end{figure}

At the critical velocity  $v_c$ \rf{vcrit} 
the two relevant saddle points merge at $w_c=\emath^{-2\imath\pi/3}$ which is now a double saddle point. 
This is depicted on Fig.~\ref{u=0_v=vc} and $w_c$ is now the single relevant saddle point. 
Note that the steepest descent path changes discontinously, and does not go to the valley at $w=0$.
The subdominant rightmost saddle point stays subdominant since for this one $\mathrm{Re}(\mathbb{W})<0$.

A steepest descent analysis, similar to the one performed in subsection 
\ref{ssucrit} for the critical tooth velocity $u_c=2$ and $v=0$, shows that the correct scaling to study the asymptotics of the amplitude at the critical spine velocity is
\begin{equation}
\label{ }
n=v_c\, t + \hat n\, t^{1/3}
\end{equation}
and that the amplitude has an Airy profile in the $\hat n$ variable similar to the one obtained in \rf{AiryFrontA} for the propagation along a tooth at the critical velocity.

\subsubsection{Velocity $v>v_c$}
\label{ssvgrtvcrit}

\begin{figure}[h!]
\begin{center}
\includegraphics[height=2.5in]{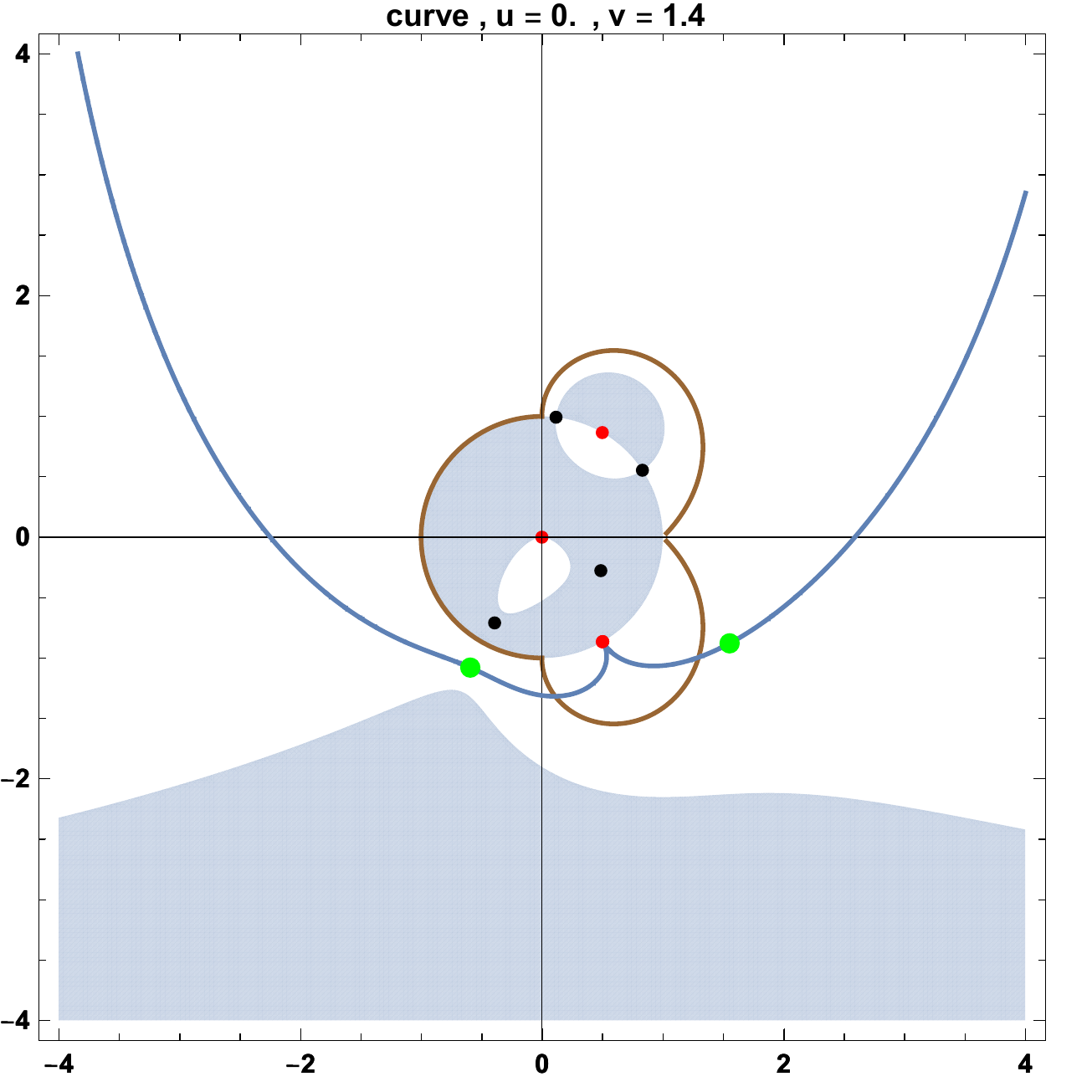}
\includegraphics[height=2.5in]{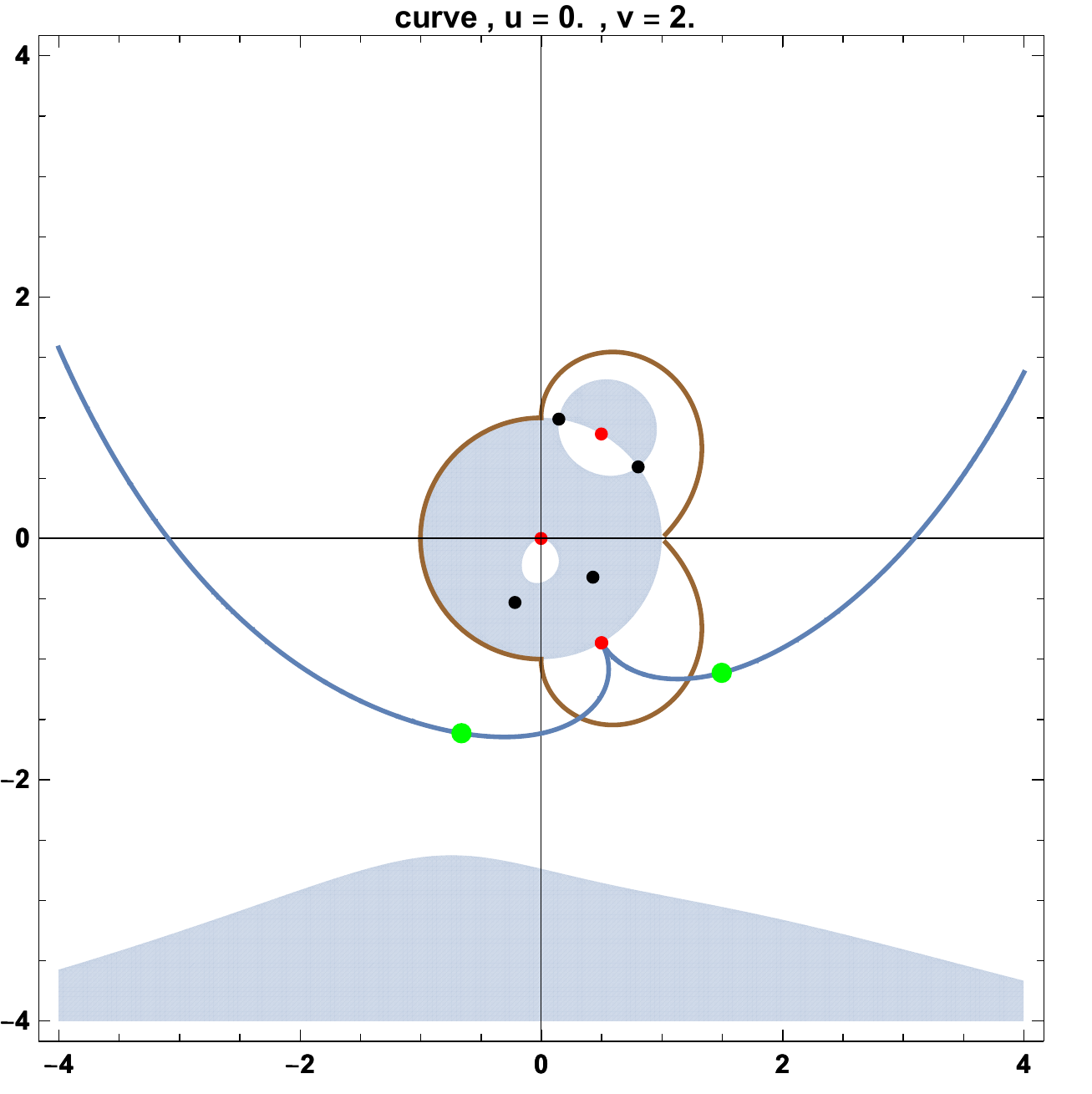}
\caption{Two examples of steepest descent path (blue) in the $w$ plane for $u=0$ and two different values of
$v>v_c$,  $v=1.4$ and $2$, same conventions as in the prevous figures.}
\label{u=0_v>vc}
\end{center}
\end{figure}

When $v$ is greater than $v_\mathrm{c}$, the double saddle point splits into two simple saddle points.
One ($w_{1'}$ in black) is such that $\mathrm{Re}(\mathbb{W})>0$.
The other one ($w_{2'}$ in green) is such that $\mathrm{Re}(\mathbb{W})<0$. It is this last one which is picked by the steepest descent path.
The second rightmost saddle point ($w_3$ in green) which was subdominant at $v=v_{\mathrm{c}}$ stays subdominant.
This is depicted on Fig.~\ref{u=0_v>vc}.

Steepest descent analysis now shows that the amplitude decays exponentially:
\begin{equation}
\label{OscAmpU0}
A_t(0,0;n,0) =
{1\over\sqrt{t}}\,a(v)\,w_{2'}(v)^{-\tilde n}\,\emath^{-t\,\chi(v)}\,\emath^{\imath\,t\,\phi(v)} 
\end{equation}
with $\mathbb{W}(w_{2'})=-\chi+\imath\phi$ and $\chi(v)$ increases with $v$.

\subsubsection{Discussion}
We have considered the case $v>0$ but the case $v<0$ is similar and symmetric.
The global picture which emerges from this analysis is the following:
A part of the wave packet moves along the spine and goes to infinity in the $n\to\pm\infty$ directions.
The evolution of the amplitude along the spine is qualitatively similar to that for the quantum walk on the discrete line $\mathbb{Z}$, or along a tooth. 
A front with an Airy profile moves at a constant velocity $v_c =\sqrt{27/16}$, which is 
different from the case of the line, where $v_c=1$, and of the tooth (half-line) where $v_c=2$. It is followed by a wave function (with a biperiodic fine structure) which contains the bulk of the quantum probability amplitude.

\subsubsection{Probability to be close to the spine at large $t$}

Here we calculate the probability to be at a finite distance from the spine in the $t\to\infty$ limit.    We need to calculate the amplitude $A_t(0,0;n,j)$ with $n=vt+\tilde{n}$, where $\tilde{n}$ and $j$ are $O(1)$ as $t\to\infty$
and we assume $n>0$ without loss of generality.
Since now we have a factor $z^{-j}$ in the integral representation for the amplitude it is easier to go back to the $z$-representation.  We can assume that $v<v_c$ since the $v>v_c$ gives an exponentially decaying contribution.   The saddle point equation in the $z$ variable reads
\beq{SaddlePz}
\imath (1-z^{-2}) -v{w_+' (z)\over w_+(z)}=0.
\eeq
There are two solutions $z_1$ and $z_2$, $z_1<z_2$, which lie in 
the interval $(-3,-1)$ along the cut in the lower half plane
corresponding to the two saddle points $w_1$ and $w_2$ in the $w$-plane discussed in subsection 4.4.4.
The saddle point approximation gives
\beq{SaddApprox}
A_t(0,0;n,j)={\sqrt{2\pi}\over \sqrt{t}}\left[ {I(z_1) e^{t\mathbb{W}(z_1)}\over w_+(z_1)^nz_1^j \sqrt{\mathbb{W}''(z_1)}}+ (z_1\mapsto z_2) \right],
\eeq
cf.\ \rf{x17}.
We note that the potential $\mathbb{W}$ is pure imaginary at the saddle points and $|w_+(z_k)|=1$, $k=1,2$.
The absolute value squared of the amplitude gives the probability to be at the vertex $(n,j)$ at time $t$.
 Averaging this probability over $\tilde{n}$, the cross terms in the probability average to $0$, and we get
 the coarse grained probability to be at $n=vt$ in the same way as in subsection 4.3.5:
 \beq{averagedProb}
 Q(vt,j;t)={2\pi \over  t}\left[ {|I(z_1)|^2\over |\mathbb{W}''(z_1)| |z_1|^{2j}} +(z_1\mapsto z_2) \right].
 \eeq
 Here $w_+$ and $e^{t\mathbb{W}}$ have vanished since they have absolute value 1 at the saddle points.
 Converting a sum over $n$ to an integral over $v$ we find that the probability to be at a distance $j$ from the spine in the large $t$ limit is given by
 \beq{ProbDistj}
P_S(j)={2\pi}\int_0^{v_c}dv\,\left[ {|I(z_1)|^2\over |\mathbb{W}''(z_1)| |z_1|^{2j}} +(z_1\mapsto z_2) \right].
\eeq
We split the integral in two parts and make a change of variable $v\to z_1=z$ for $-3<z<-2$ and
$v\to z_2=z$ for $-2<z<-1$.  Then we find
\beq{ProbDistj2}
P_S(j)={2\pi}\int_{-3}^{-1}dz\, \left|{dv\over dz}\right|  {|I(z)|^2\over |\mathbb{W}''(z)| |z|^{2j}}.
\eeq
By a simple calculation we find that
\beq{simpleCalc}
{dv\over dz}=w_+ {dz\over dw_+}.
\eeq
Furthermore, $|w_+(z)|=1$ for $-3<z<-1$ and 
\beq{derivative}
\left| {dw_+\over dz}\right| = {1\over \sqrt{(z+3)(1-z)}}
\eeq
so
\beq{ProbDistj3}
P_S(j)={1\over 2\pi}\int_{-3}^{-1}dz\,{(1-z^{-2})^2\over z^{2j}\sqrt{(z+3)(1-z)}},
\eeq
where we have used that $I(z)=(2 \imath \pi)^{-1}(1-z^{-2})((z+3)(1-z))^{-1/2}$.  
The total probabiity of being a finite distance away from the spine in the $t\to\infty$ limit is then
\beq{TotProbSpin}
P_{\rm Spine}=\sum_{j=0}^\infty P_S(j)={1\over 2\pi}\int_{-3}^{-1}dz\,{1-z^{-2}\over \sqrt{(z+3)(1-z)}}.
\eeq
This integral can be evaluated analytically and the numerical value is $0.368469\ldots$ so
$P_{\rm Spine}+P_{\rm Teeth}=1$ as expected from the unitarity of the time development.

As in subsection 4.3.7, the probability 
$P_S(j)$ can be caculated directly without making use of the coarse grained probability.  We outline the argument.   The probability to be at a distance $j$ from the spine at time $t$ is given by
\beq{ProbtjSpine}
q_t(j)=\sum_{n=-\infty}^\infty  |A_t(0,0; n,j)|^2.
\eeq
Noting that $|A_t(0,0;n,j)|=|A_t(0,0;-n,j)|$ we can do the sum over $n$ and find
\beq{SumOvern}
q_t(j)=-{1\over (2\pi )^2}\oint dz_1\oint d\bar{z_2}\, I(z_1)I(\bar{z_2}) (z_1\bar{z_2})^{-j} 
e^{-itE(z_1)+itE(\bar{z_2})} {1+w(z_1)w(\bar{z_2})\over 1 -w(z_1)w(\bar{z_2})},
\eeq
where $w(z) \equiv w_+(z)$ and the integration contours are as before.  We now make a change of variables
$z_i\to w_1=w(z_1)$ and $z_2\to w_2=w(z_2)$.  We have
\beq{ChangeofVar}
{dw\over w}={dz\over \sqrt{(z+3)(z-1)}}
\eeq
so
\beq{SumOvern2}
q_t(j)=-{1\over (2\pi )^2}\oint {dw_1\over w_1}\oint {d\bar{w_2}\over \bar{w_2}}\, {(1-z_1^{-2})(1-\bar{z_2}^{-2})
\over
(z_1\bar{z_2})^{j}}  e^{-itE(z_1)+itE(\bar{z_2})} {1+w_1 \bar{w_2}\over 1 -w_1 \bar{w_2}},
\eeq
where the integration contours for $w_1$ and $\bar{w_2}$ 
are the ones corresponding to the ones in the $z_1$ and $z_2$ planes as explained in subsection 4.4.2.  We fix $\bar{w_2}$ on the original contour and deform the $w_1$ integration contour to the speepest descent path.  Then we pick up poles when we deform through the poles which occur at
$w_1=\bar{w_2}^{-1}$.  The steepest descent integral tends to 0 as $t\to\infty$ by the same argument as before.
Viewing $E$ as a function of $w$ 
we see that $E(w)=E(w^{-1})$ since $z$ does not change as $w\mapsto w^{-1}$.  We conclude that the pole
contribution is $t$ independent.   By inspection we see that 
$\bar{w_2}$ must be located on the unit circle in the upper half plane between $-1$ and $\imath$
in order for the deformation to hit the poles.  Hence,
\beq{TotProbSpine2}
P_S(j)=\lim_{t\to\infty} q_t(j)={\imath\over \pi}\int_\gamma {dw\over w}(1-z^{-2})^2 z^{-2j},
\eeq
where $\gamma$ is the part of the unit circle between $-1$ and $\imath$ and we have renamed the integration variable.  Changing the integration variable back from $w$ to $z$ we find the integral
\rf{ProbDistj3}.

\subsection{Propagation in the bulk:  $u>0$, $v>0$}
\begin{figure}[h!]
\begin{center}
\includegraphics[height=2.5in]{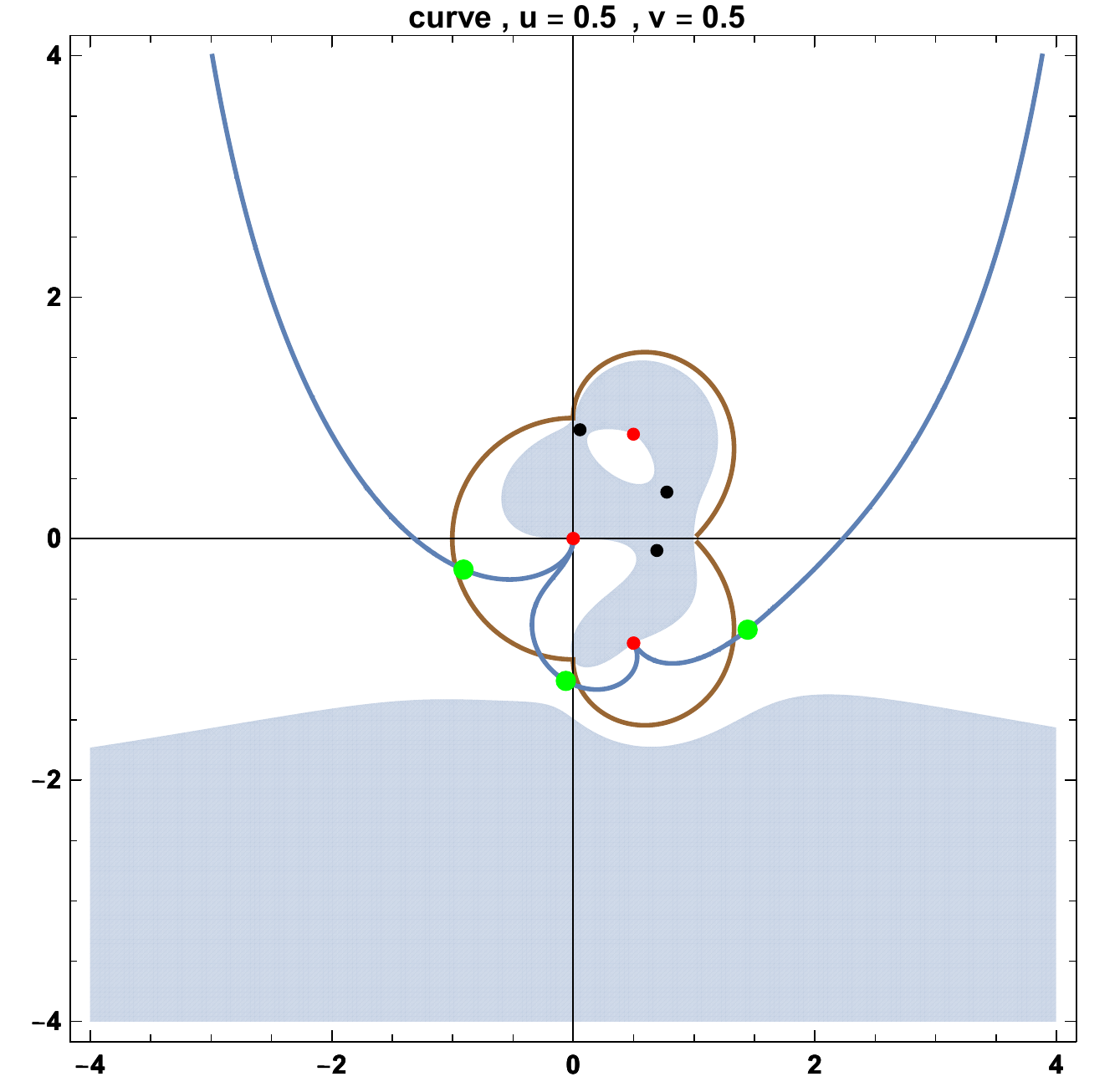}
\caption{An example of the steepest descent path in the $w$ plane for the general case $v>0$, $u>0$.}
\label{u>0_v>0}
\end{center}
\end{figure}

The saddle point equation $\mathbb{W}'(w)=0$ reduces to an algebraic equation of degree six:
\begin{equation}
\label{SPEquUV}
{\imath (w-1)^3 (w+1)(w^2+1)\over w(w^2-w+1)^2} -u\, {(w+1)(w-1)\over (w^2-w+1)}- v =0.
\end{equation}
The equation can be studied numerically, and this gives the general features of the six saddle points and 
the steepest descent path for the amplitude in the complex $w$ plane. 
An example in given on Fig.~\ref{u>0_v>0} for the values $u=.5$, $v=.5$, but the features are generic.

The main result is that as soon as $u$ and $v$ are nonzero (and positive), there are generically six distinct saddle points. Three of them are in the ``dangerous'' $\mathrm{Re}(\mathbb{W})>0$ region (in gray) and three of them in the ``allowed'' $\mathrm{Re}(\mathbb{W})<0$ region (in white). The original integration path (in brown) must be deformed in the white region only. In the example depicted in 
Fig.~\ref{u>0_v>0}, it picks the tree allowed saddle points (in green), which have a strictly negative $\mathrm{Re}(\mathbb{W})$. At large times the asymptotics of the  amplitude in the bulk of the comb for the $(j,n)$ variables, 
\begin{equation}
\label{ }
j=ut\ \quad n= vt \qquad\text{with}\quad u>0\ ,\quad v>0,
\end{equation}
has an exponential decay of a generic form similar to \rf{Asymu>ucv0} ($u>u_c$, $v=0$) or  \rf{OscAmpU0} ($u=0$, $v>v_c$) :
\begin{equation}
\label{OscAmpU0x}
A_t(0,0;n,j) =
{1\over\sqrt{t}}\,a(u,v)\,\emath^{-t\,\chi(u,v)}\,\emath^{\imath\,t\,\phi(u,v)} ,
\end{equation}
with $\chi=-\mathrm{Re}(\mathbb{W}(w_{\star}))$ and $\phi= \mathrm{Im}(\mathbb{W}(w_{\star}))$,  
where $w_\star$ is the dominant saddle point, i.e.\ the one with 
the least negative  $\mathrm{Re}(\mathbb{W})$. 
In the case of Fig.~\ref{u>0_v>0}, it turns out that $w_\star$ is the second saddle point (in c.c.w. order) with the largest $\mathrm{Im}(w)$, i.e. the saddle point which is the continuous deformation of the saddle poit at $w=-\imath$ when $u=v=0$.

This is general, and means that there is no propagation in the bulk of the comb.
The quantum walks propagates only along the teeth which are close to the initial point (in our case 
$n=j=0$), and along the spine in both directions. 

The behaviour of the saddle points and 
the steepest descent path as a function of $u$ and $v$ is somewhat involved and it is an interesting mathematical problem to study it. 
Depending on the values of $u$ and $v$, the steepest descent path may pick one, two or the three of the allowed saddle points. This is however not very relevant for the physics, where it is enough to know that there is an exponential decay of the amplitude at large $t$. We shall discuss and illustrate the different cases in Appendix~\ref{aSteepestPathUVPlane}

\subsection{Arbitrary starting point}
Here we show that all the previous results of this 
section generalize with minor modifications to the case when the quantum walk starts 
at an arbitrary place in a tooth rather than on the spine.   Without loss of generality we can assume that the walk starts
at the vertex $(0,j_0)$, $j_0>0$ at time $t=0$.

The amplitude to be on site $(n,j)$ at time $t$, starting at time zero from site $(0,j_0)$ 
is given by
\begin{equation}
\label{Aj0}
 \begin{split}
A_t(0,j_0;n,j) &  =  \int_0^{2\pi} d\alpha \int_0^\pi d\theta \, \frac{\br n,j|\theta ,\alpha\kt\br \theta ,\alpha | 0,j_0\kt }{N(\alpha ,\theta )}\,e^{-\imath t E(\theta )}
\\&+\int_{\pi /2}^{3\pi /2}d\alpha \, \br n,j|\gamma ,\alpha\kt\br \gamma ,\alpha |0,j_0\kt\,e^{-\imath t E(\gamma)} .
\end{split}
\end{equation}
Let us begin by considering the first integral over $\alpha$ and $\theta$   (the contribution of the extended states).
From (\ref{6}) we see that 
%
\begin{equation}
\label{ampli1}
\begin{split}
&\hskip10.em\frac{\br n,j|\theta ,\alpha\kt\br \theta ,\alpha | 0,j_0\kt }{N(\alpha ,\theta )}
\\
&=\ 
{1\over 4\pi^2} {\left(
\left(y+\emath^{\imath\theta}\right) \emath^{\imath (n\alpha+ j\theta)}
{-}
\left(y+\emath^{-\imath\theta}\right) \emath^{\imath (n\alpha- j\theta)}
\right) 
\left(
\left(y+\emath^{-\imath\theta}\right) \emath^{-\imath  j_0\theta}
{-}
\left(y+\emath^{\imath\theta}\right) \emath^{\imath j_0\theta}
\right)
\over
\left(y+\emath^{\imath\theta}\right) \left(y+\emath^{-\imath\theta}\right)
}
\\
&= \ {1\over 4 \pi^2} \emath^{\imath n\alpha}
\left(
\emath^{\imath\theta(j-j_0)}+\emath^{-\imath\theta(j-j_0)}-{y+\emath^{\imath\theta}\over y+\emath^{-\imath\theta}}\emath^{\imath\theta(j+j_0)}
-{y+\emath^{-\imath\theta}\over y+\emath^{\imath\theta}}\emath^{-\imath\theta(j+j_0)}
\right)
\\
&=\ {1\over 4 \pi^2} \emath^{\imath n\alpha}
\Big(\emath^{\imath\theta(j-j_0)}+\emath^{-\imath\theta(j-j_0)}-\emath^{\imath\theta(j+j_0)}-\emath^{-\imath\theta(j+j_0)}
\\
&\hskip6em  - {2\imath \sin\theta\over y+\emath^{-\imath\theta}}\ \emath^{\imath\theta(j+j_0)} + {2\imath \sin\theta\over y+\emath^{\imath\theta}}\ \emath^{-\imath\theta(j+j_0)}\Big)
\\
&= \ {1\over 4 \pi^2} \emath^{\imath n\alpha}
\left (4 \sin (\theta j) \sin(\theta j_0)- {2\imath \sin\theta\over y+\emath^{-\imath\theta}}\ \emath^{\imath\theta(j+j_0)} + {2\imath \sin\theta\over y+\emath^{\imath\theta}}\ \emath^{-\imath\theta(j+j_0)}\right)
\end{split}
\end{equation}
with  $y=1-2\cos\alpha $ as before.

Inserting this in (\ref{Aj0}) and integrating over $\alpha$, the first term gives a nonzero result only if $n=0$,
and we obtain
\begin{equation}
\label{ }
\delta_{n,0}\,A^0_t(j_0;j)\ , \ \text{with}\quad A^0_t(j_0;j)=\ {1\over 2\pi} \int_0^\pi \!\!d\theta\ \emath^{-\imath t E(\theta)}\, 4 \sin (\theta j) \sin(\theta j_0).
\end{equation}
Using the symmetry $E(\theta)=E(-\theta) $ this can be rewritten as
\begin{equation}
\label{zerotooth}
A^0_t(j_0;j)=
{1\over \imath\pi} \int_{-\pi}^\pi \!\!d\theta\ \emath^{-\imath t E(\theta)}\,  \emath^{\imath \theta j} \sin(\theta j_0).
\end{equation}
Integrating over $\alpha$ in the second and third terms in (\ref{ampli1}) 
can be done exactly as in Section 4.1.1 using (\ref{x1}) and (\ref{x2}). 
Using again the symmetry $E(\theta)=E(-\theta)$ we end up with a second contribution from the extended states, given by
\begin{equation}
\label{ }
A_t^{\scriptscriptstyle{\mathrm{ext}}}(0,j_0;n,j)={1\over 2\pi} \int_{-\pi}^\pi d\theta\ \emath^{-\imath t E(\theta)}\, {(-2 \imath \sin\theta)\over\sqrt{(\emath^{-\imath\theta}+3)(\emath^{-\imath\theta}-1)}}\ {w_-(\theta)}^{|n|}\ 
\emath^{\imath\theta(j+j_0)}
\end{equation}
with
\begin{equation}
\label{ }
w_-(\theta)= {1+\emath^{-\imath\theta}\over 2}-{1\over 2}\sqrt{(\emath^{-\imath\theta}+3)(\emath^{-\imath\theta}-1)}.
\end{equation}
The contribution of the localised states \rf{n1} is computed by the same method. One has
\begin{equation}
\label{ }
\br n,j|\gamma ,\alpha\kt\br \gamma ,\alpha |0,j_0\kt = {1\over 2\pi} \left(1-\emath^{-2\gamma}\right)\emath^{\imath\alpha n} (-1)^{j+j_0} \emath^{-\gamma(j+j_0)}
\end{equation}
with  $1-2\cos\alpha=\emath^\gamma$.
Inserting this into the second integral in (\ref{Aj0}) we get the contribution of the localized states
to the amplitude:
\begin{equation}
\label{ }
A_t^{\scriptscriptstyle{\mathrm{loc}}}(0,j_0;n,j)={1\over 2\pi}\int_{\pi/2}^{3\pi/2} d\alpha\, (1-\emath^{-2\gamma})\emath^{\imath\alpha n}\, (-1)^{j+j_0}\,\emath^{-\gamma(j+j_0)}\,\emath^{-\imath t E
(\gamma)}.
\end{equation}
Again, performing the change of variables
\begin{equation}
\label{ }
z=\emath^{-\imath\theta}\quad,\qquad z=-\emath^\gamma
\end{equation}
allows us to combine 
$A_t^{\scriptscriptstyle{\mathrm{ext}}}$ and $A_t^{\scriptscriptstyle{\mathrm{loc}}}$ into a single ``regular'' contour integral
\begin{equation}
\label{ }
A_t^{\scriptscriptstyle{\mathrm{ext}}}(0,j_0;n,j)+A_t^{\scriptscriptstyle{\mathrm{loc}}}(0,j_0;n,j)=A_t^{\scriptscriptstyle{\mathrm{reg}}}(0,j_0;n,j)
\end{equation}
given by
\begin{equation}
\label{Iint}
A_t^{\scriptscriptstyle{\mathrm{reg}}}(0,j_0;n,j)
={1\over 2\imath\pi}\oint_\Gamma {1-z^{-2}\over\sqrt{(z+3)(z-1)}}\, w_-^{|n|}\, z^{-(j+j_0)}\, \emath^{-\imath t(2-z-z^{-1})}\, dz.
\end{equation}
The integral \rf{zerotooth} can be written
\begin{equation}
\label{I0int}
A_t^0(j_0;j)= {1\over 2\imath\pi}\oint_\Gamma\  4\,z^{-j-1} (z^{-{j_0}}-z^{j_0}) \, \emath^{-\imath t(2-z-z^{-1})}\, dz.
\end{equation}
In both integrals we can take the same c.c.w. contour $\Gamma$ encircling the cut at $[-3,1]$, but the 
integrand in (\ref{I0int}) only has essential singularities at $z=0$ and $z=\infty$.
The final form for the amplitude is then
\begin{equation}
\label{Atj0final}
A_t(0,j_0;n,j)= \delta_{n,0}\ A_t^0(j_0;j) + A_t^{\scriptscriptstyle{\mathrm{reg}}}(0,j_0;n,j)
\end{equation}
The differences with the case $j_0=0$ studied previously are: (i) the $z^{-{j_0}}$ factor inside the integral for the ``regular term'', when compared to (\ref{x9}), which corresponds simply to a shift $j\to j+j_0$ in (\ref{x9}) and  
(ii) the special term $A_t^0$ for the initial tooth at $n=0$.
Note that when $j_0=0$, $A_t^0=0$ and we recover (\ref{9}).

We can now study the large time asymptotics by the same methods as before.
Indeed, the saddle point analysis is similar, with the same saddle points which are the extrema of the same potential.

The probability of return to the starting point is given by $P_t(j_0)=\left | A_t(0,j_0;0, j_0)\right | ^2$. For the ``regular term'' $A_t^{\scriptscriptstyle{\mathrm{reg}}}$ the large $t$ behaviour is governed by the two saddle points of the potential
$V_0(z)=2-z-1/z$ at $z=1$ and $z=-1$, and by the discontinuities at $z=1$ and $z=-3$. As for the $j_0=0$ case, the leading term is given by singularity at $z=-3$ and we find that
\begin{equation}
\label{ }
\left | A_t^{\scriptscriptstyle{\mathrm{reg}}}(0,j_0;0,j_0)\right | \propto t^{-1/2}.
\end{equation}
For the special term $A_t^0$, there is no cut and only the saddle points contribute. However
\begin{equation}
\label{ }
A_t^0(j_0;j_0)={1\over 2\imath\pi}\oint_\Gamma dz\ z^{-1}\ \left( z^{-2 j_0}-1\right)\ \emath^{-\imath t (2-z-1/z)}
\end{equation}
and the factor $(z^{-2 j_0}-1)$ vanishes at $z=\pm 1$ since $j_0$ is an integer. This implies that
\begin{equation}
\label{ }
\left | A_t^0(j_0;j_0)\right | = \mathrm{O}\left(t^{-3/2}\right)
\end{equation}
and is subdominant. Thus the quantum spectral dimension is still $d_{\mathrm{qs}}=2$ as expected.

The propagation into the teeth and along the spine can be studied in the same way. 
Choosing a tooth velocity $u$ and a spine velocity $v$, and rescaling
\begin{equation}
\label{ }
j=t\, u + \tilde \jmath \quad,\qquad n=t\, v+ \tilde n\, ,
\end{equation}
the large $t$ behavior of the amplitude is governed by the saddle points of the same potential as for $j_0=0$:
\begin{equation}
\label{ }
V_{u,v}(z)=\imath(z+1/z-2)-u\log z -v \log w_-(z).
\end{equation}
Depending on the value of real part of $V_{u,v}$ at a saddle point $z_c$, the large $t$ behavior is evanescent (if $\mathrm{Re}(V_{u,v})<0$), oscillatory with a power-like decay (if $\mathrm{Re}(V_{u,v})=0$). The conclusions are similar to those of Sec.\ 4.3 and 4.4.
Along a tooth which is at a finite distance for the origin ($v=0$, i.e. $\tilde n = n$), the propagation is oscillatory for $u<u_c=2$, and evanescent for $u>2$, so at large $t$ the wave function is localized in the interval $j\in [0, 2 t]$, with an Airy-like front expanding at velocity $u_c=2$.
Along the spine ($u=0$, i.e.\ $\tilde\jmath=j$), the wave function is localized at a finite distance from the spine.
The propagation is oscillatory for $v<v_c=3\sqrt{3}/4$, and evanescent for $v>v_c$, so at large $t$ the wave function is localized in the interval $n\in [0, v_c t]$, with a front expanding at velocity $v_c$. Away from these two othogonal directions, the wave function is evanescent.

It is interesting to study in more detail the behaviour of the wavefunction along 
the initial tooth $n=0$, since it is on this tooth that the amplitude has the additional $A_t^0$ term.
As before we set $j=t\, u + \tilde \jmath$. The general amplitude $A_t(0,j_0;n,j=ut+\tilde\jmath)$ is
\begin{equation}
\label{At0j0form}
\begin{split}
{1\over 2\imath \pi} \oint_\Gamma {dz\over z}\, \emath^{-t (\imath(2-z-1/z)+ u \log z)}\, z^{-\tilde\jmath}\ \omega_-(z)^{|n|} 
\left( \delta_{n,0}\,(z^{j_0}-z^{-j_0}) + {(z-z^{-1})\, z^{-j_0}\over \sqrt{(z+3)(z-1)}} \right).
\end{split}
\end{equation}
As before, the two saddle points for $0<u<2$ are
\begin{equation}
\label{zpmagain}
z_+ =\emath^{-\imath\phi}\ ,\quad z_-=-\emath^{\imath\phi}\ ,\quad \sin\phi=u/2,
\end{equation}
see Sec.\ 4.3.1 and 4.3.2.
We obtain by a saddle point analysis large $t$ asymptotics similar to (\ref{x21}), but now with  
functions $I(z_+,j_0)$ and $I(z_-,j_0)$ (obtained from the rightmost term in (\ref{At0j0form}) which 
depend explicitly on $j_0$.

\begin{figure}[h!]
\begin{center}
\includegraphics[width=2in]{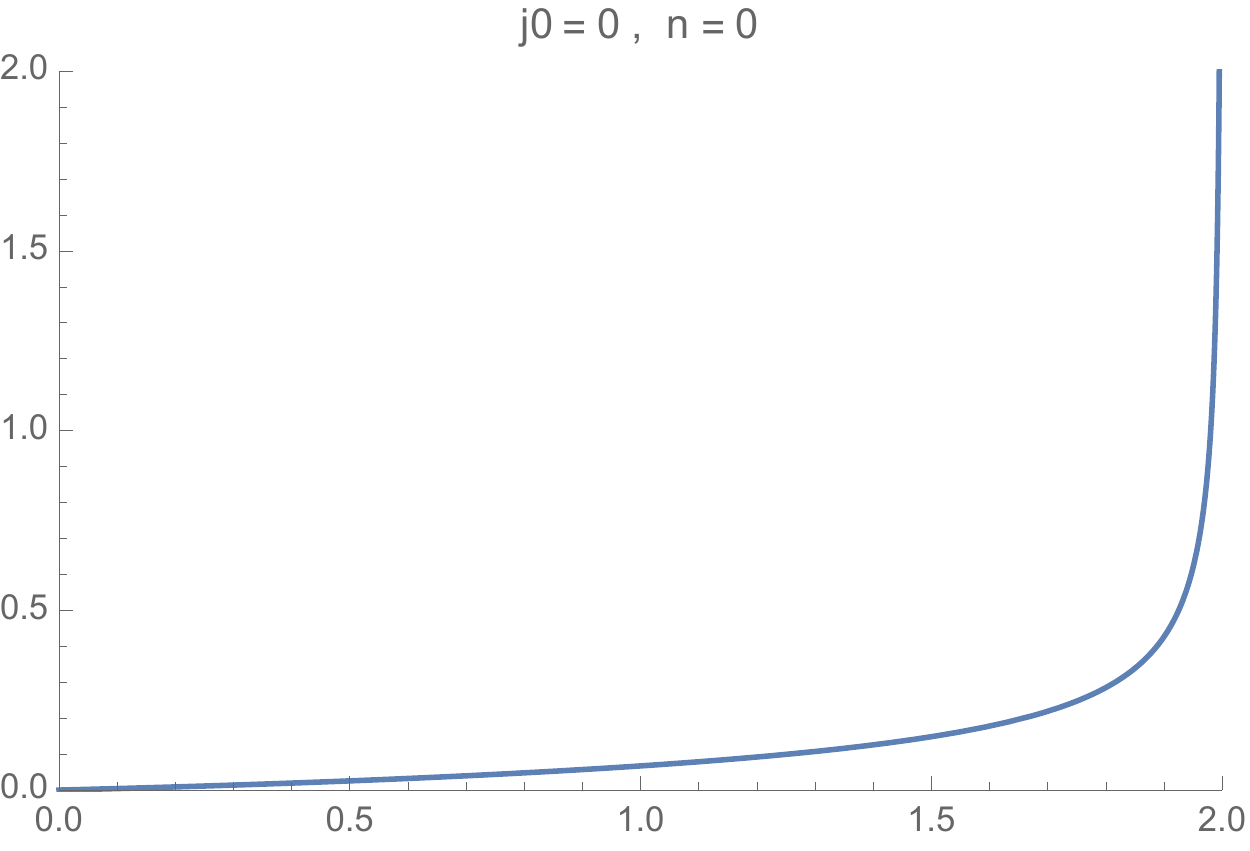}
\includegraphics[width=2in]{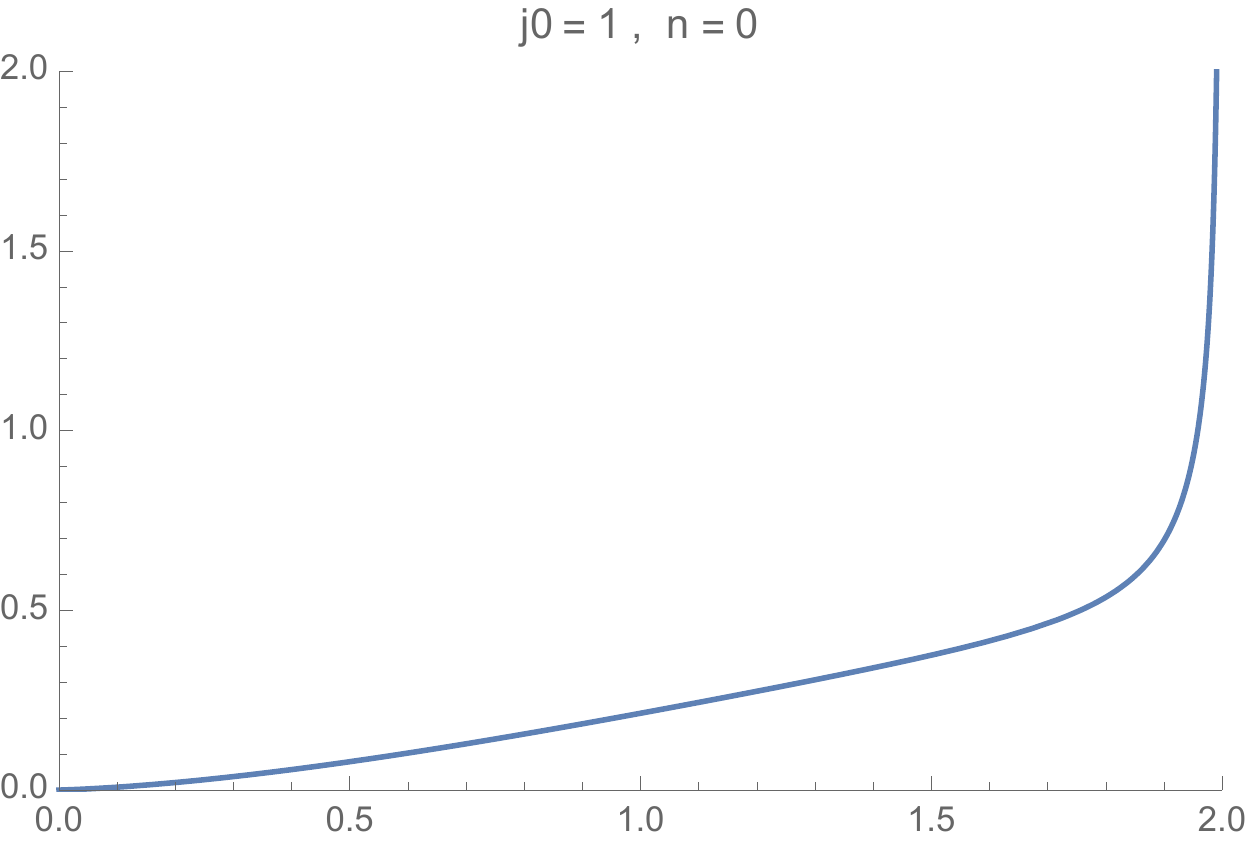}
\includegraphics[width=2in]{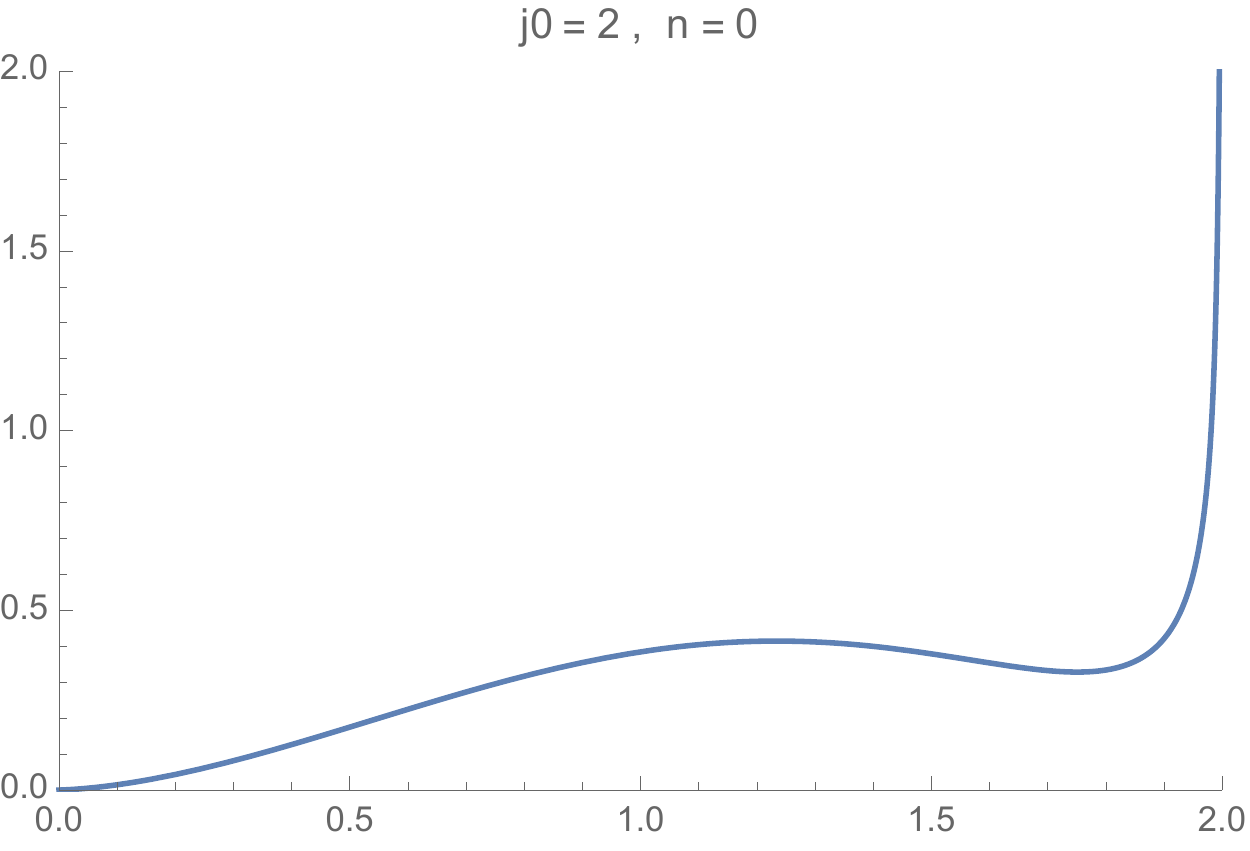}
\includegraphics[width=2in]{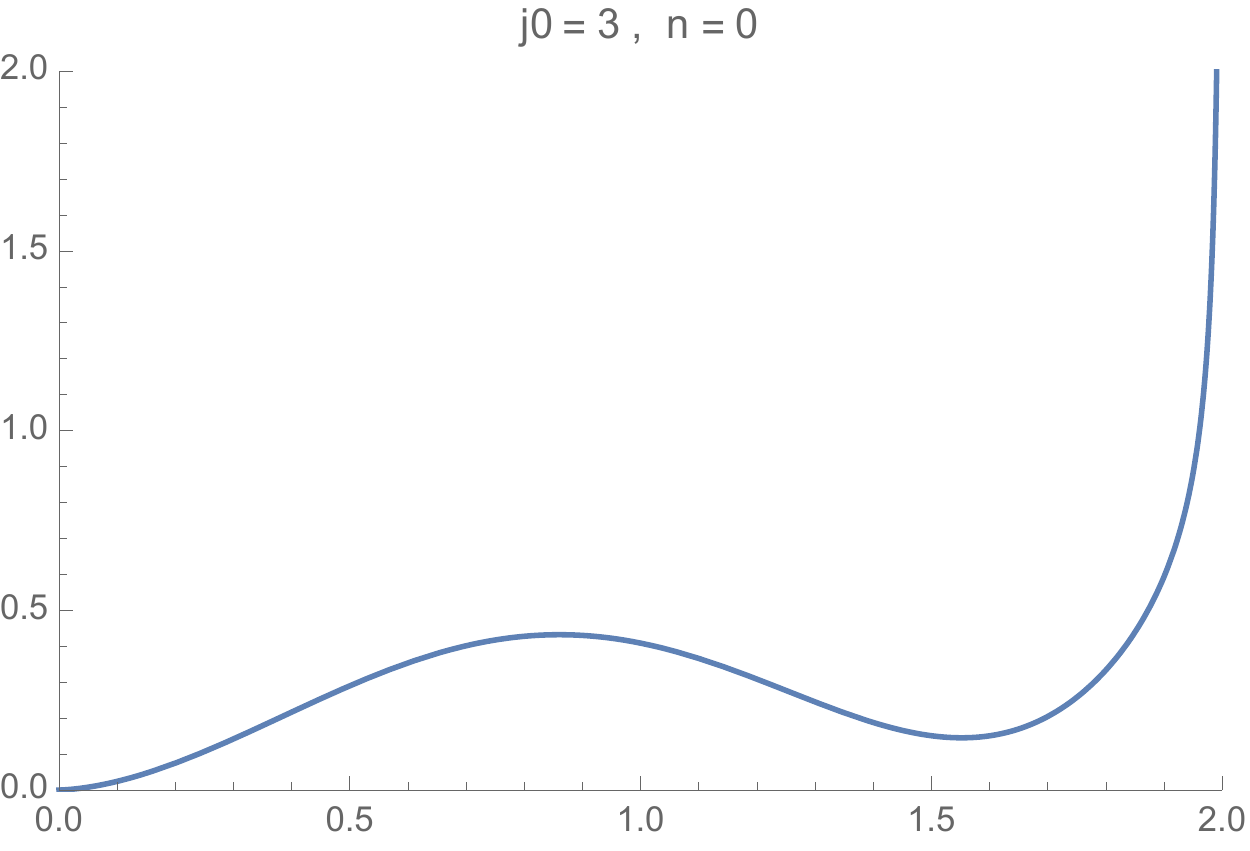}
\includegraphics[width=2in]{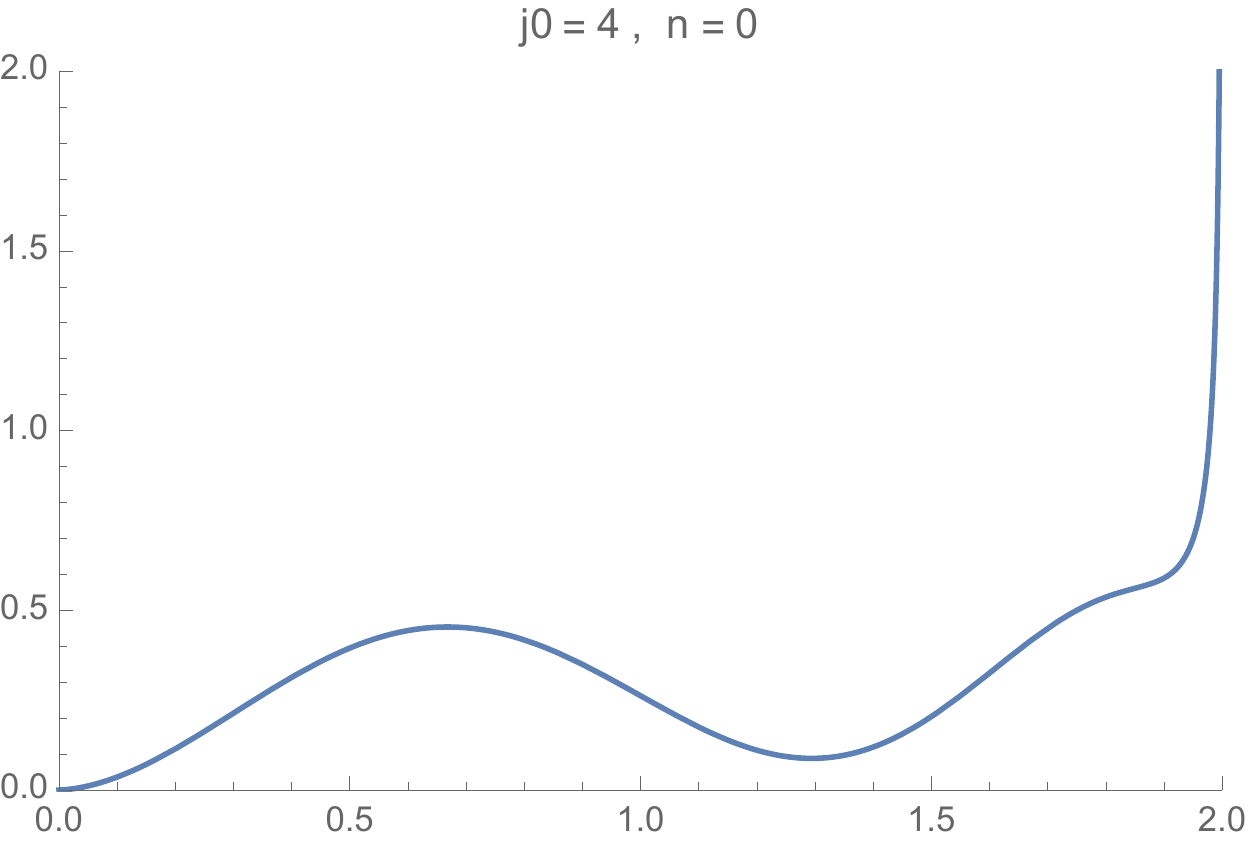}
\includegraphics[width=2in]{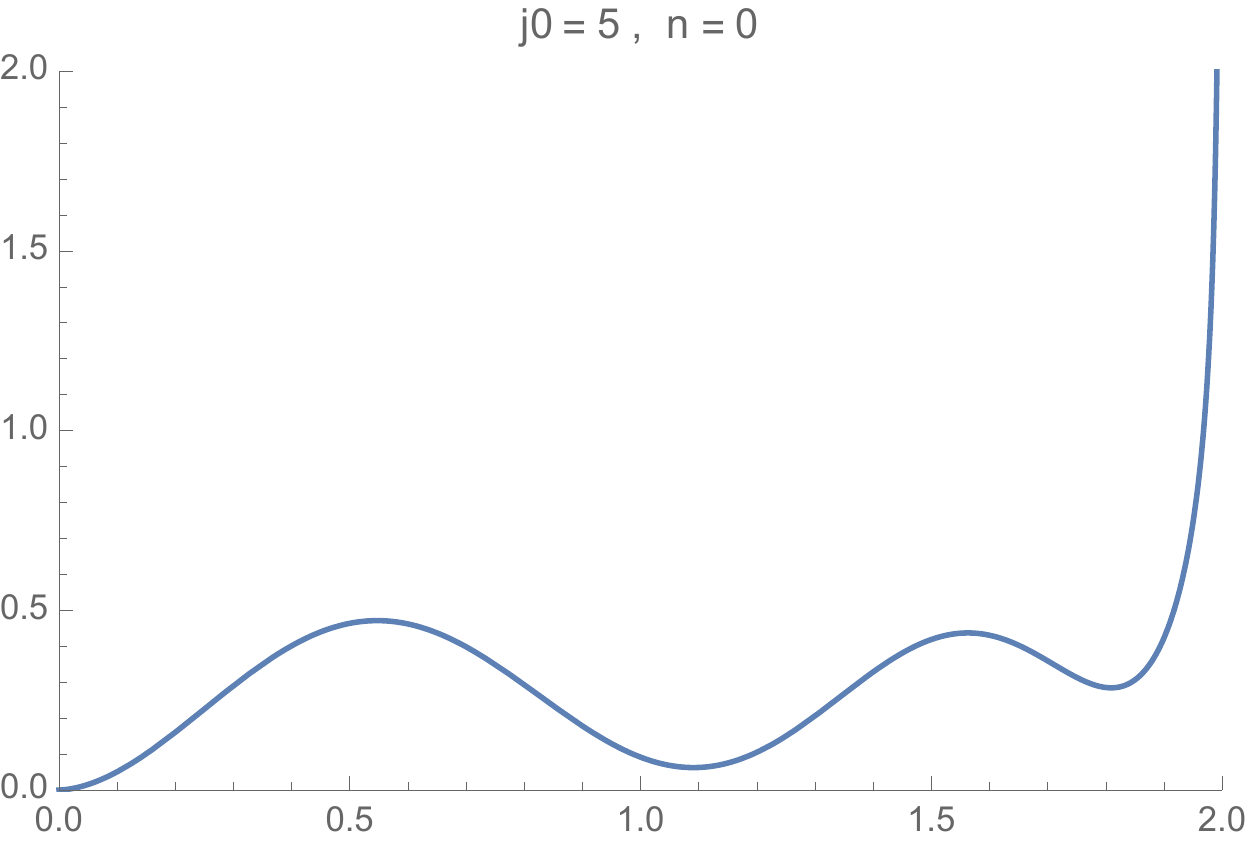}
\includegraphics[width=2in]{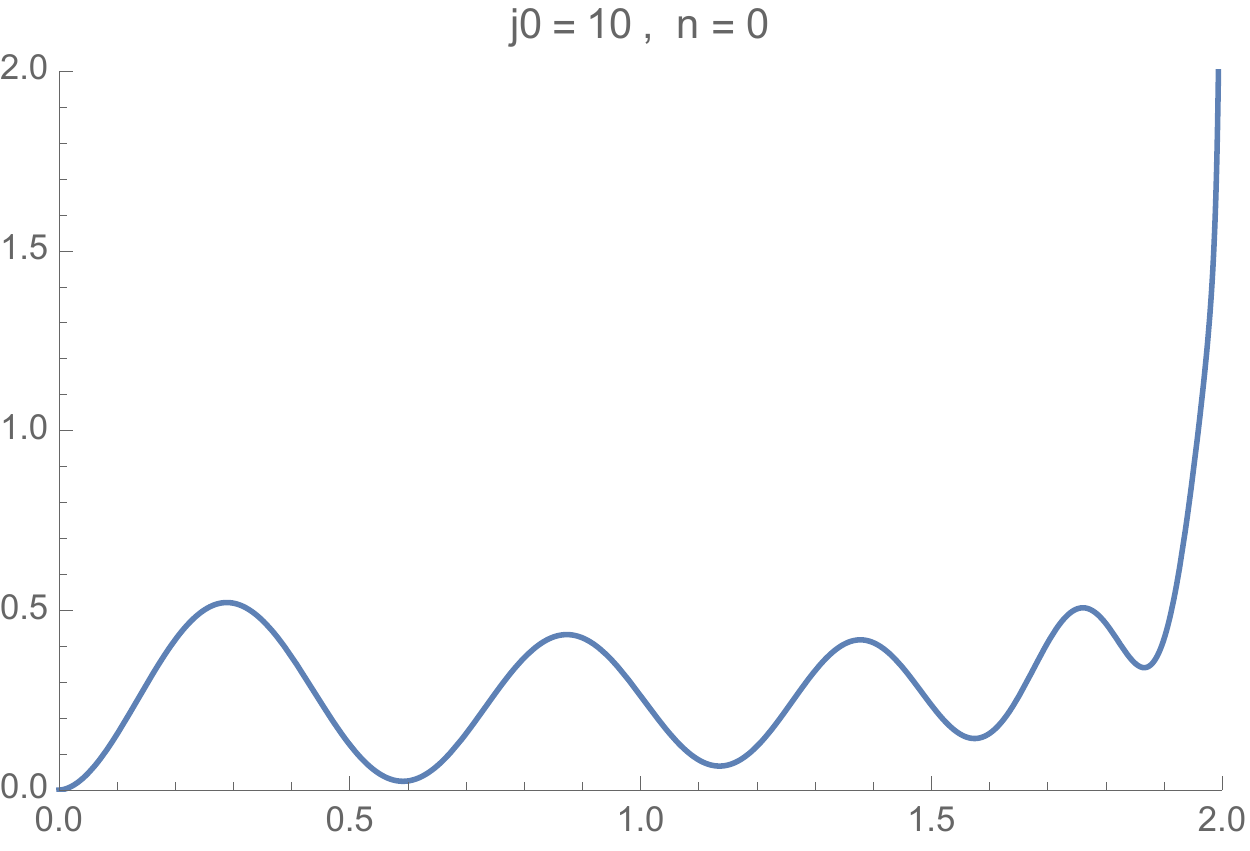}
\includegraphics[width=2in]{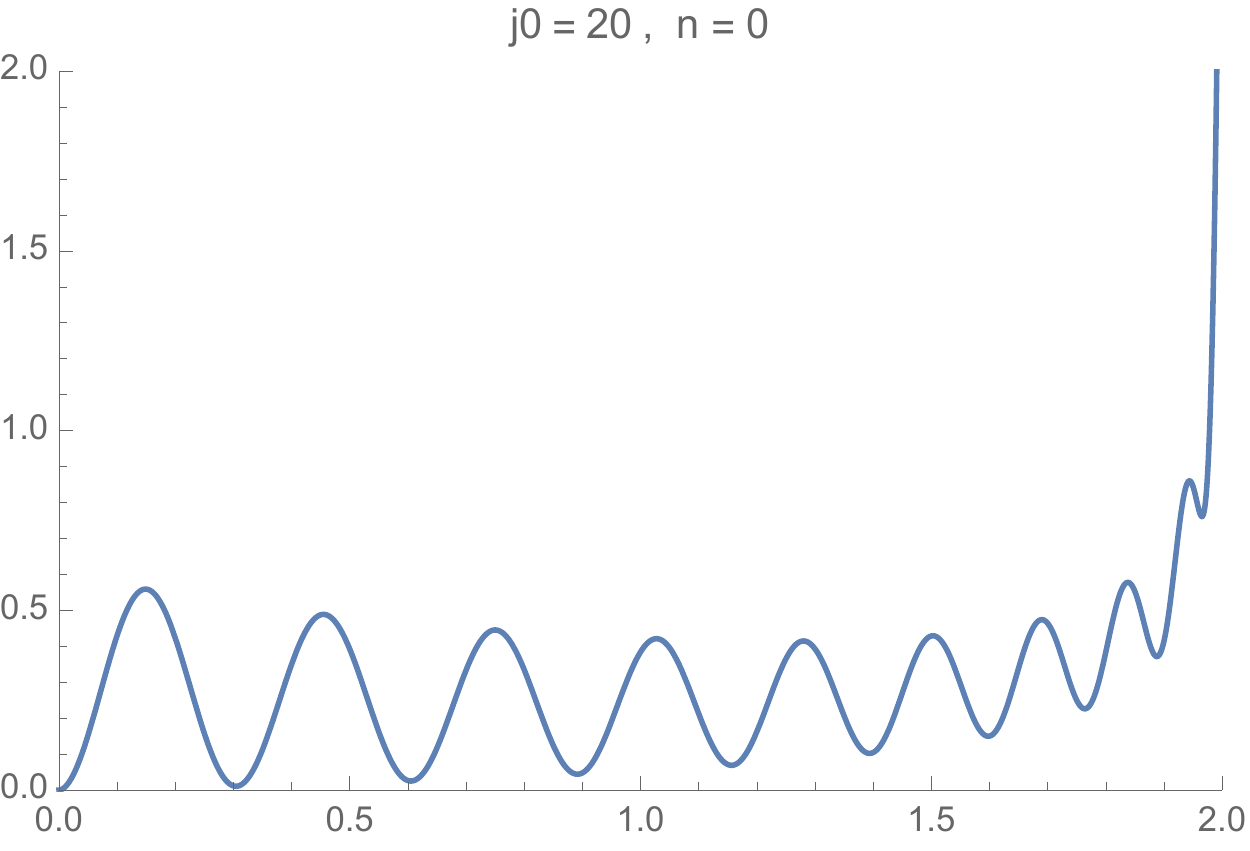}
\includegraphics[width=2in]{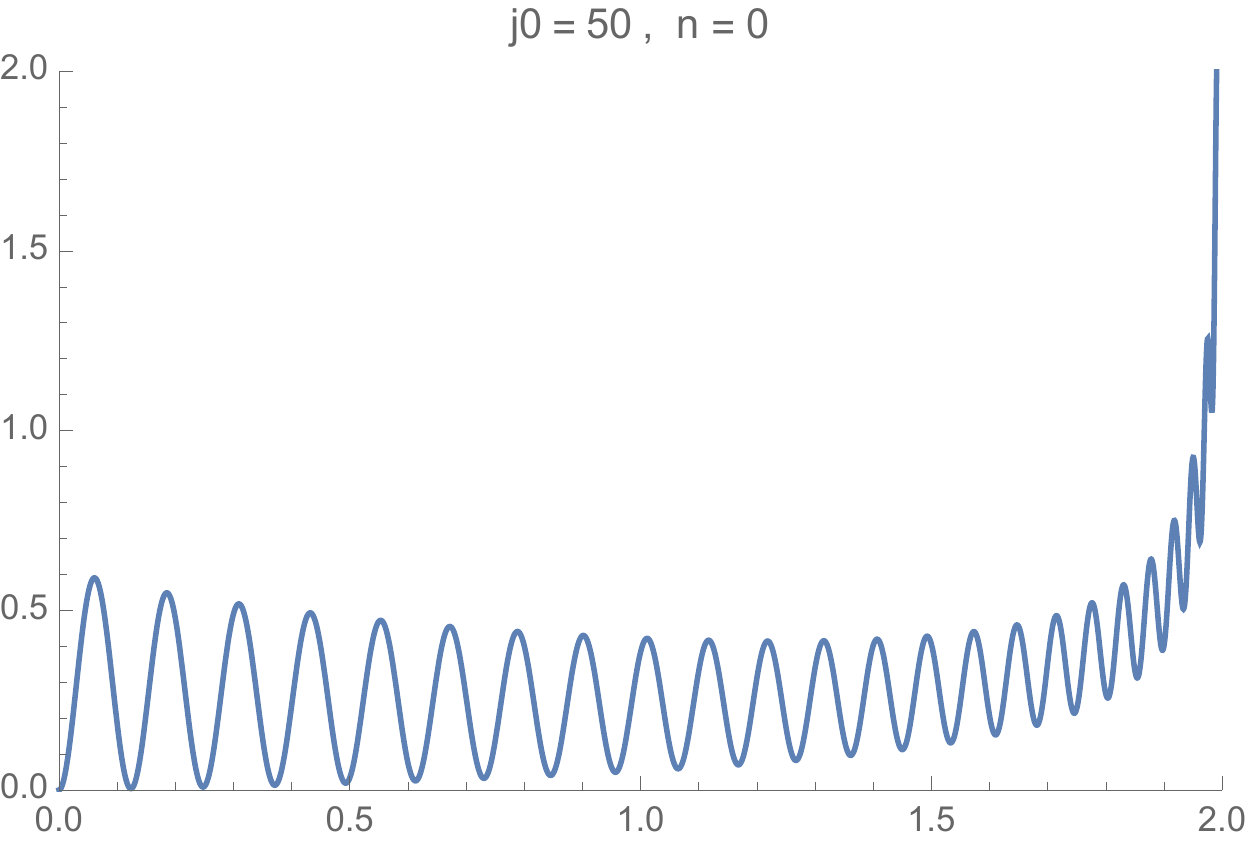}
\includegraphics[width=2in]{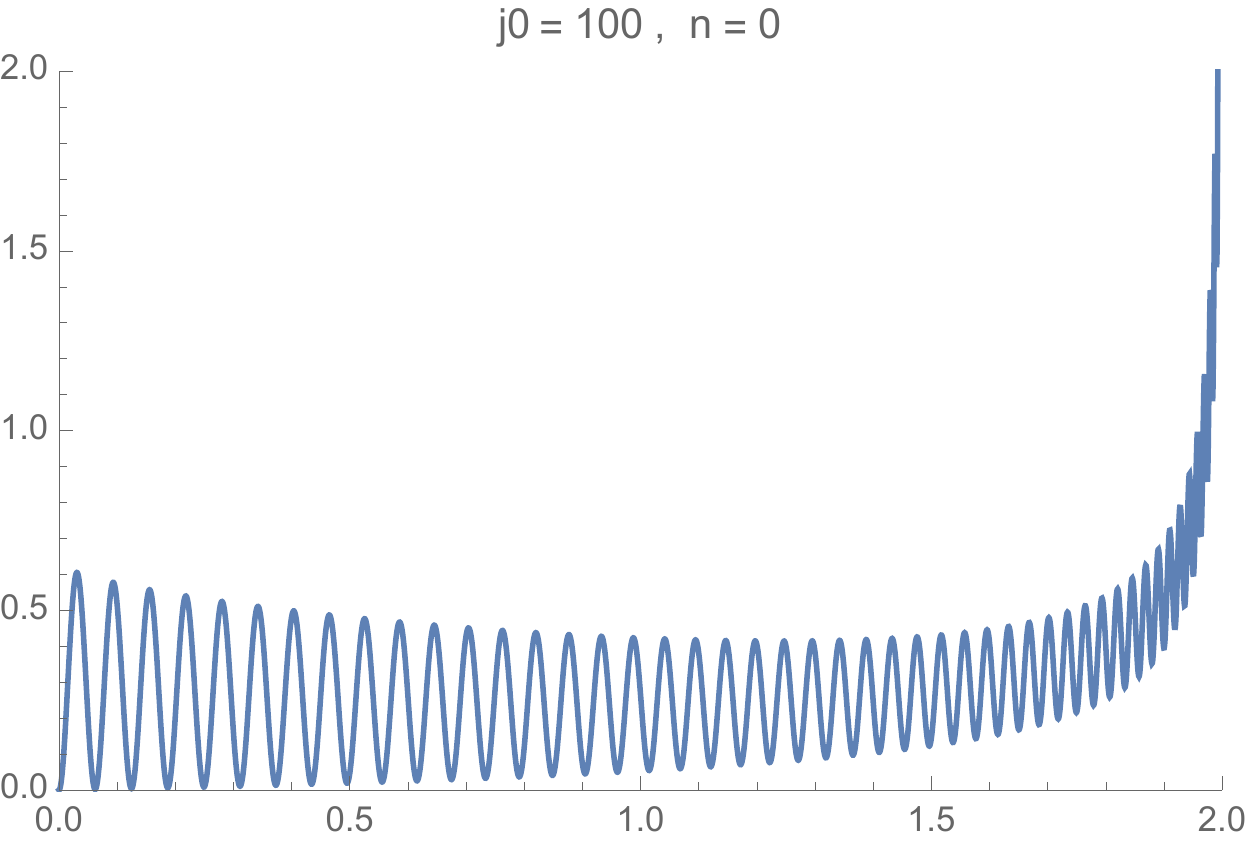}
\caption{The coarse-grained density probability profile on the initial tooth, $c(u;0,j_0)$, as a function of the rescaled position on the tooth $u=j/t$, for different values $j_0$ of the starting point.}
\label{profilesJ0}
\end{center}
\end{figure}

\begin{figure}[h!]
\begin{center}
\includegraphics[width=4in]{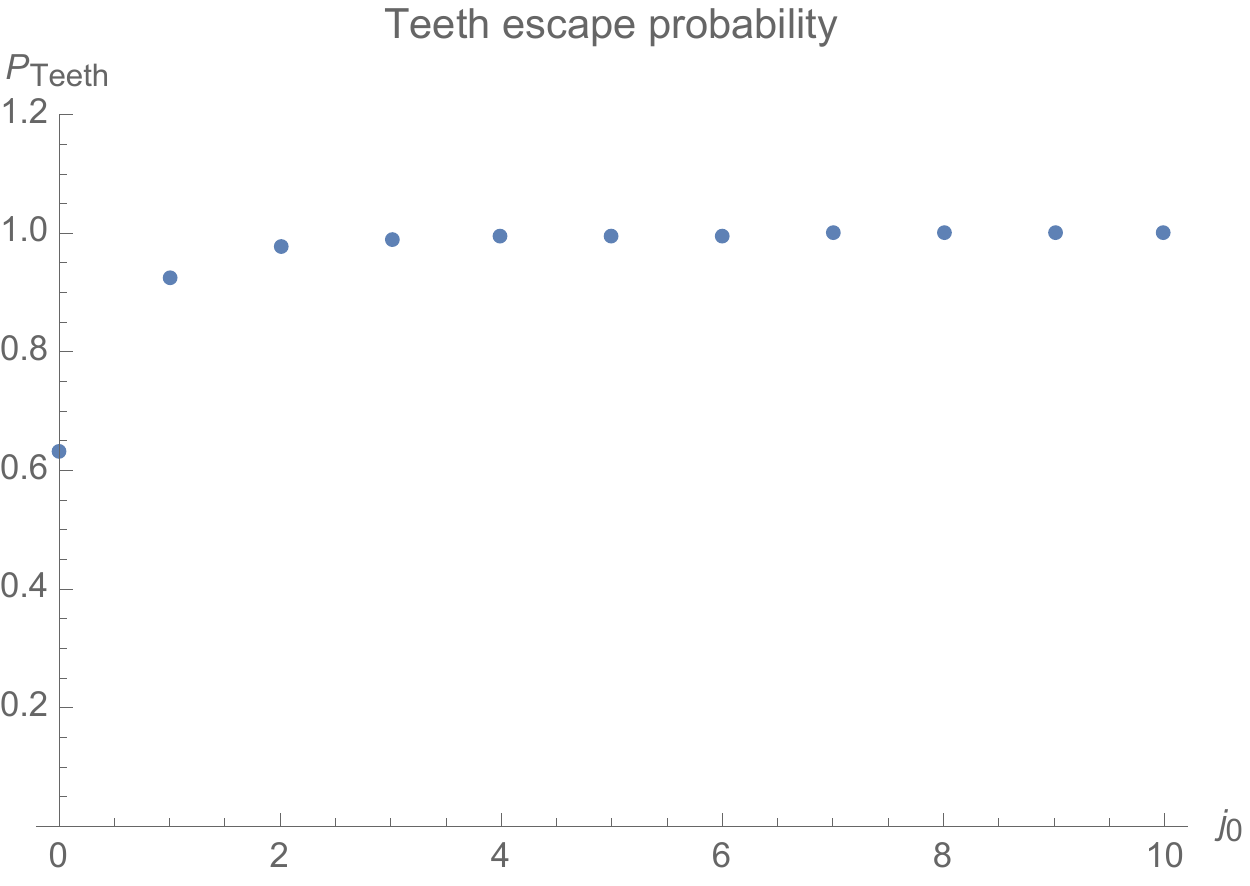}
\caption{Probability $P_{\mathrm{teeth}}(j_0)$ to be on teeth at a finite distance from the origin as $t\to\infty$, as a function of the initial position $j_0$. It saturates at  $P_{\mathrm{teeth}}=1$ as $j_0\to\infty$}
\label{prob_escape-teeth}
\end{center}
\end{figure}

\begin{figure}[h!]
\begin{center}
\includegraphics[width=4in]{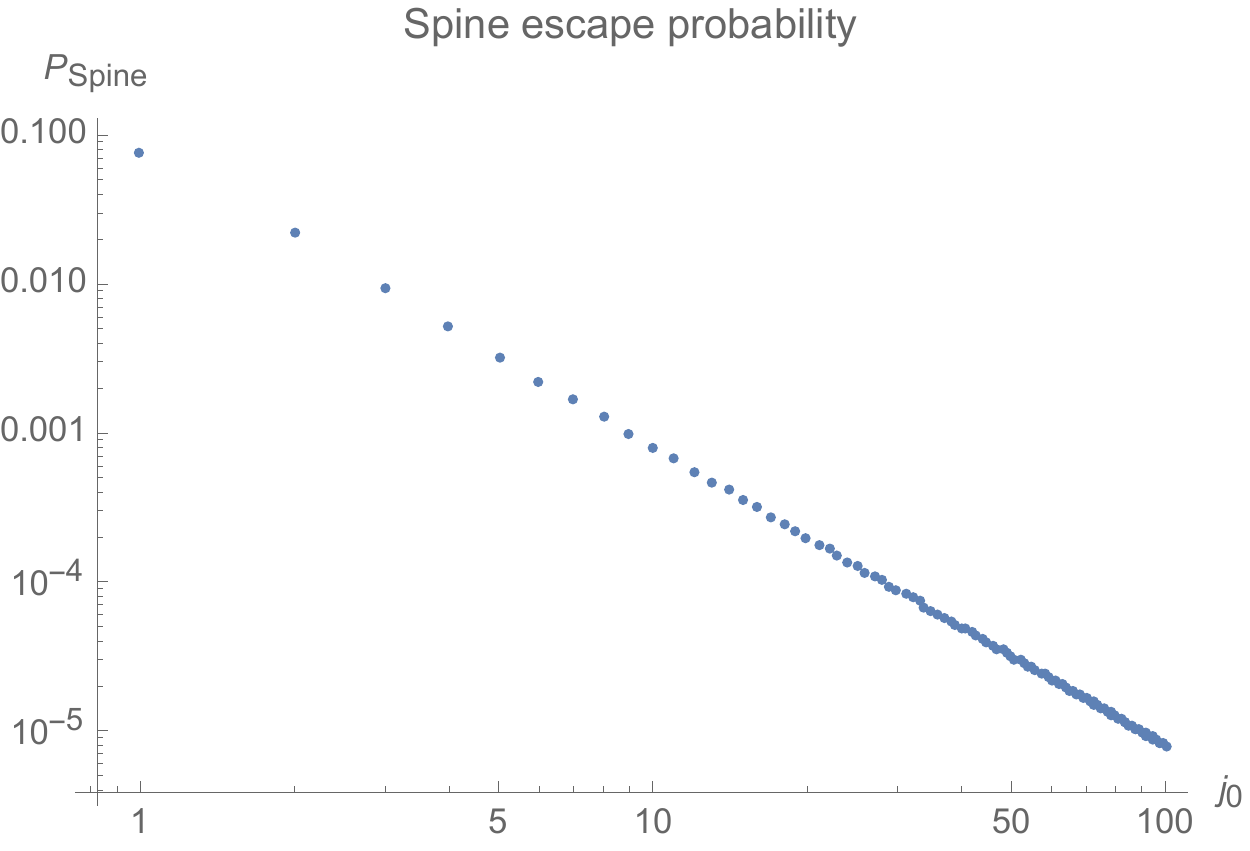}
\caption{Log-Log-plot of the probability $P_{\mathrm{spine}}(j_0)=1-P_{\mathrm{teeth}}(j_0)$ to be at a finite distance from the spine as $t\to\infty$, as a function of the initial position $j_0$. Numerical data suggests $P_{\mathrm{spine}}(j_0)\propto j_0^{-2}$.}
\label{prob_escape-spine}
\end{center}
\end{figure}

Out of the large $t$ asymptotics we can extract the coarse grained 
probability density profile along the initial tooth $n=0$, $c(u,j_0)$, defined as
in Sec.\ 4.3.2:
\begin{equation}
\label{cuj0}
c(u;j_0)= \text{``coarse-grained''} \lim_{t\to\infty}\  \left( t\ \Big | A_t(0,j_0;0,j=ut)\Big|^2 \right)
\end{equation}
More generally, we can consider the density profile along an arbitrary tooth at $n\neq 0$
\begin{equation}
\label{cunj0}
c(u;n,j_0)= \text{``coarse-grained''} \lim_{t\to\infty}\  \left(t\ \Big | A_t(0,j_0;n,j=ut)\Big |^2 \right)
\end{equation}
We can then study the total asymptotic probability to be on the tooth $n$ at large time
\begin{equation}
\label{ptnj0}
P_T(n,j_0)=\int_0^\infty du\ c(u;n,j_0).
\end{equation}
We do not give the details of the calculation, but describe the general features of the $j_0$-dependence of these probability profiles.
\begin{enumerate}
\item As for $j_0=0$, the profiles are non zero in the interval $[0,2]$, with a 
front at $u=2$, corresponding to the singularity of $(2-u)^{-1/2}$, 
and an Airy-like profile when looked at close to $u=2$ on a finer scale.
 
  \item If $n\neq 0$, the probability profile does not depend on $j_0$ and is identical to the one calculated previously when $j_0=0$, i.e.\ $c(u;n,j_0)=c(u,n)$.
   Therefore the asymptotic probability to be on tooth $n$ (the escape probability for tooth $n$) does not depend 
   on $j_0$, i.e.\
   $$n\neq 0\ \implies\ P_T(n,j_0)=P_T(n)\ .$$
   
  \item On the initial tooth $n=0$, the probability profile $c(u;0,j_0)$ depends in a non\-trivial way on $j_0$, 
  and also the asymptotic probability to be on the initial tooth $P_T(0,j_0)$. 
  
  \item The profile $c(u,0,j_0)$ exihibits interesting oscillations with $u$, which increase in frequency with $j_0$. 
  This depicted in Fig.~\ref{profilesJ0}.
  These macroscopic oscillations have a natural interpretation as quantum interference 
  between the ``right moving'' part of the wave-function
  which starts from $j=j_0$ and moves to $\infty$ and the ``bounced back'' part of the wave-function which starts as a left-moving wave from $j=j_0$ towards $j= 0$, hits the spine, so that part of it bounces back as a right moving wave function towards $j\to\infty$, and then interferes with the initial right moving part.
  
  \item The total asymptotic probability to be on the first tooth, $P_T(0,j_0)$, increases with $j_0$. 
  Moreover, the total asymptotic probability to be on a tooth at a finite distance from the initial one, 
  $$
  P_{\mathrm{teeth}}(j_0)=\sum_{n\in\mathbb{Z}} P_T(n,j_0)
  $$ 
  is found to increase very fast and to saturate to 1 when $j_0$ becomes large.
 This is depicted on Fig.~\ref{prob_escape-teeth}
 
 \item As a consequence, the probability to escape to $\infty$ along the spine, $$P_{\mathrm{spine}}(j_0)=1-P_{\mathrm{teeth}}(j_0)$$
  goes to zero as $j_0$ becomes large !
 This is depicted in Fig.~\ref{prob_escape-spine}.
 This is to be expected, since the propagation along the spine is carried by the states 
 $|\alpha,\gamma\rangle$ localized near the 
 spine, whose overlaps with the inital state $|n=0,j=j_0\rangle$ decay as $j_0$ becomes large.
The numerical data suggests that this probability $P_{\mathrm{spine}}(j_0)$ decreases as $j_0^{-2}$.
\end{enumerate}

Finally, let us note that the quantum walk when one starts from an initial state which is a coherent quantum superposition of states localized at different positions on the comb can be studied by the same method. It could lead to interesting interference patterns. 

\section{Discussion}
We have studied in some detail the behaviour of quantum walk on an infinite comb.
The analysis is of course much simplified due to the translational invariance along the spine 
so in particular the eigenfunctions of the Hamiltonian are Bloch waves.
The matrix elements of the time development operator have an explicit representation as a contour integral which can be studied by the steepest descent method.
The main result is that the quantum walk is ballistic both along the spine and into the teeth, but the wave function 
decays exponentially into the bulk of the combs.  Starting from a vertex on 
the spine, in the teeth and along the spine close to starting point, 
the quantum walk behaves qualitatively as quantum walk on the discrete line.
The quantum spectral dimension is 
$d_{\mathrm{qs}}=2$, and differs from the classical spectral dimension $d_\mathrm{s}=3/2$ of the infinite comb. We recall that the classical Hausdorff dimension of the infinite comb is $d_\mathrm{H}=2$. 

These different behaviours correspond to the different large time geometrical scaling of the region where the walker is most likely to be located.

\begin{figure}[h!]
\begin{center}
\includegraphics[width=3in]{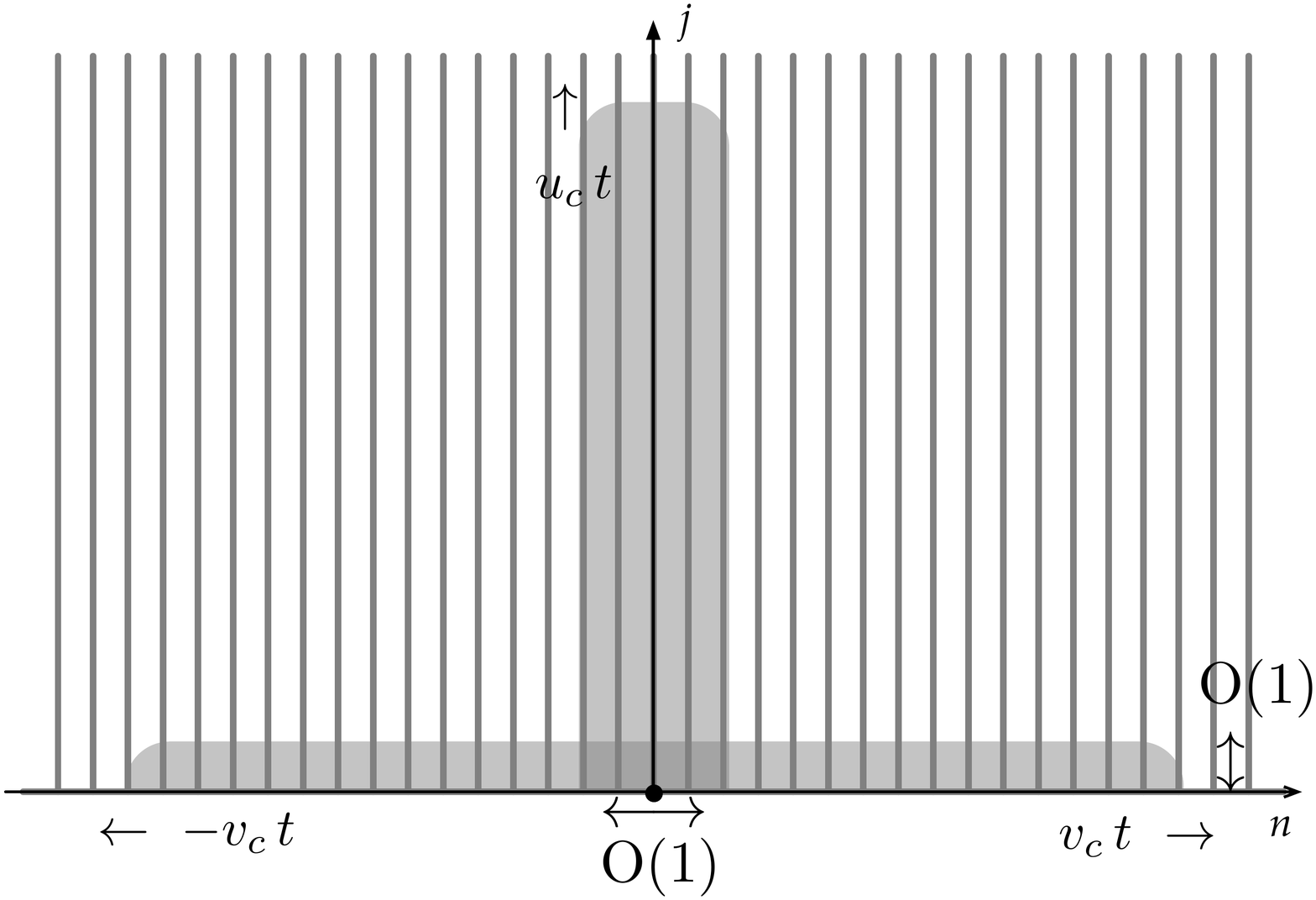}
\caption{Schematic representation of the propagation of the quantum walker on the infinite comb, starting from the origin on the spine.
The shaded region indicates where the quantum walker is most likely to be located as $t\to\infty$. The propagation takes place along a few  teeth at a finite distance from the starting point, as well as along the spine.}
\label{Comb1}
\end{center}
\end{figure}
\begin{figure}[h!]
\begin{center}
\includegraphics[width=3in]{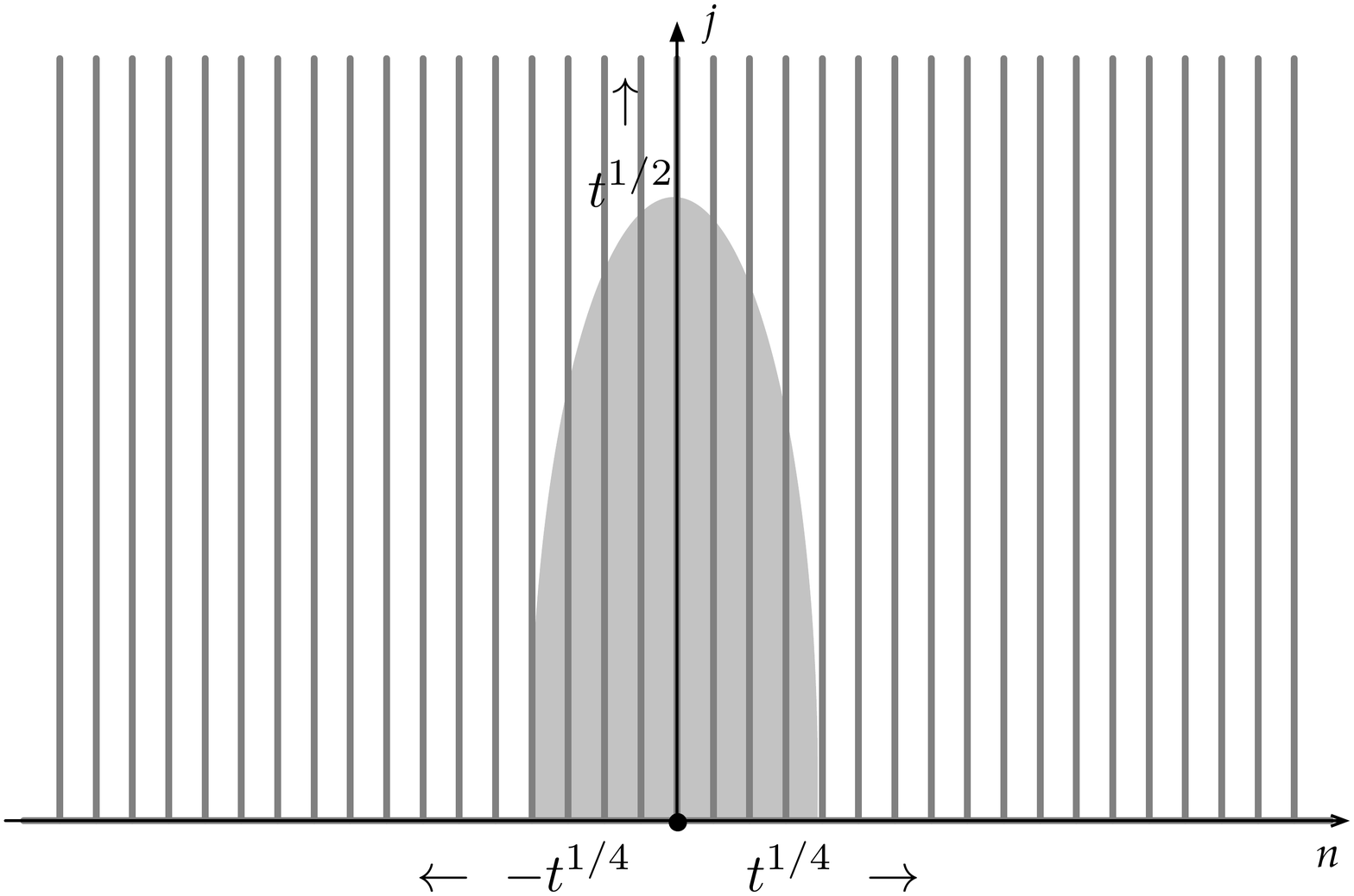}
\caption{Similar schematic representation of the classical random walker (classical diffusion on the infinite comb).}
\label{Comb3}
\end{center}
\end{figure}
\begin{figure}[h]
\begin{center}
\includegraphics[width=3in]{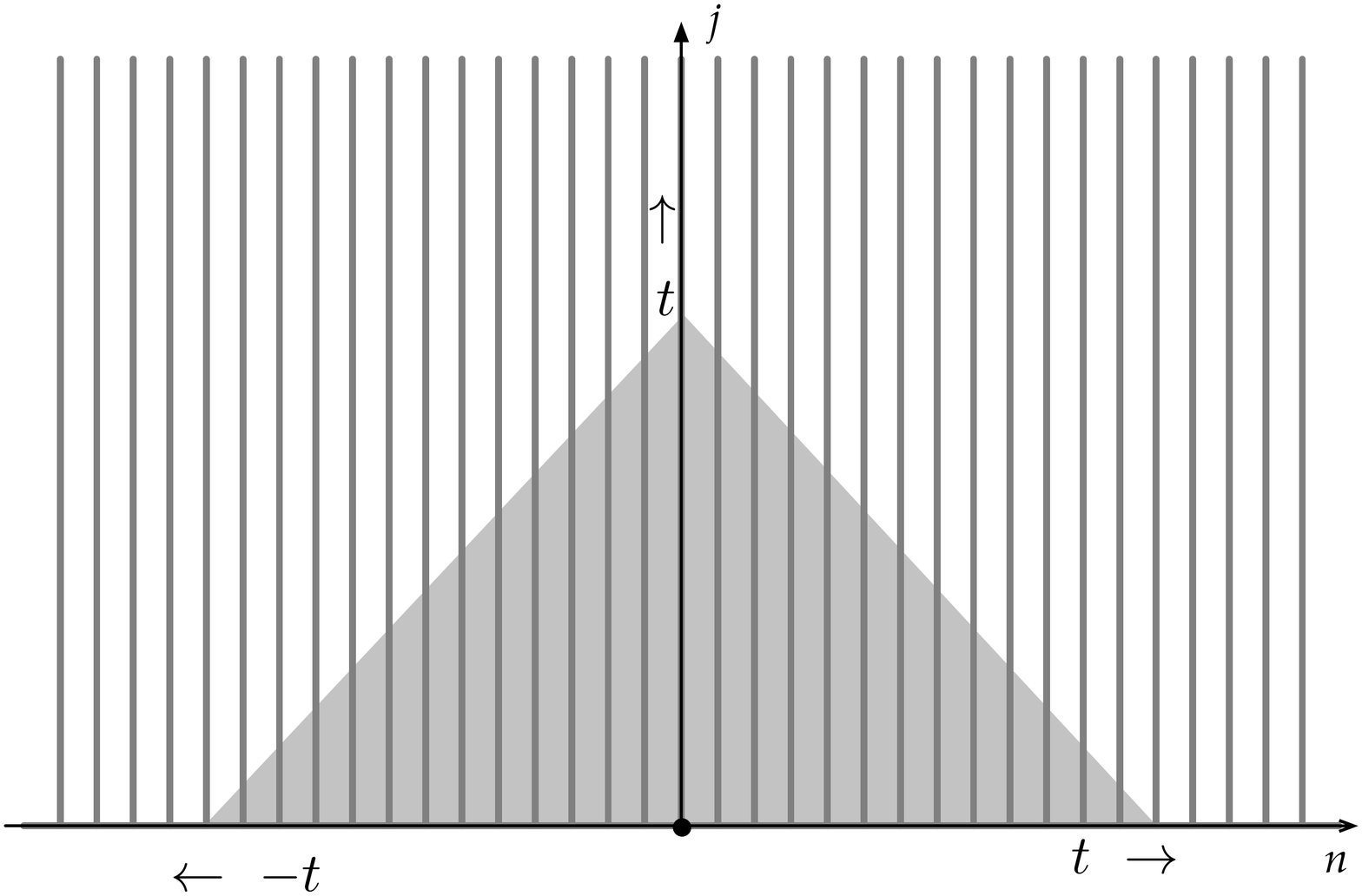}
\caption{Similar schematic representation of the classical geodesic straight walker (ballistic) on the infinite comb.}
\label{Comb2}
\end{center}
\end{figure}

The quantum walker moves with a maximal speed $v_c$ along the spine ($n$ direction) and a maximal speed $u_c$ along the tooth ($j$ direction, as depicted in Fig.~\ref{Comb1}.
The classical random walker (classical diffusion) is located in a region which scales as $t^{1/2}$ in the $j$ direction, and only as $t^{1/4}$ in the $n$ direction, as depicted in Fig.~\ref{Comb3}.
A classical ballistic walker (geodesic flow) is located in a region which scales as $t$ both in the $j$ direction and in the $n$ direction, as depicted in Fig.~\ref{Comb2}.

One can easily replace the spine by the higher dimensional lattice $\bbZ^d$ and find the spectrum and eigenfunctions of the Hamiltonian as done in this paper.   There are eigenfunctions which decay exponentially in the teeth and we expect  that the steepest descent analysis can be used to study the probability amplitudes in this case as well.

When the teeth are finite rather than infinite we have a different problem:
The walk cannot disappear into the teeth so the quantum walk is expected to be qualitatively as on the discrete line..

What we find interesting about the infinite comb studied in this paper 
is the fact that there are eigenfunctions of the Hamiltonian which are exponentially decaying in the teeth and 
concentrated on the spine.  
It would be of interest to find out whether it is a generic feature of infinite inhomogeneous graphs that some eigenfunctions are concentrated on some infinite subgraphs.

While it is generally believed that coined discrete time quantum walks behave qualitatively as continuous time 
quantum walks, it is not clear how to extend our analysis to the discrete time case.  The problem is that one needs to introduce
a new degree of freedom (the coin) and there is no simple Hamiltonian.  The time development operator acts in a much
more complicated way since both the state of the coin and the state of the walker change. 
It is nontrivial to find the eigenvalues and eigenfunctions of the time development operator.  However, this would be an interesting problem to study.

We are not  aware of any direct aplications of quantum walks on combs,  We view it as a toy model for studying such
walks on graphs which are slightly irregular.   Understanding 
simple cases is likely to pave the way for analysing quantum
walks on graphs which might model interesting physical systems.  There are 
recent experimental results on quantum walks
on structures close to the comb we have analysed, see e.g.\ \cite{exp1,exp2} and references therein.

The ultimate goal is to study quantum walk on infinite random graphs.  One could for example consider
combs with random teeth as has been done for classical random walk \cite{DJW}.  In this case one might encounter
the phenomenon of localization, which is found for quantum walk on the line with a random 
potential \cite{luck,localization1,localization2}, since it is natural to conjecture that random teeth could mimick the effects of a random potential.

Another direction is to address this problem on the so called quantum graphs, see e.g.\ \cite{BerKuc}, which are composed of continuous segments characterized by their individual length, instead of the discrete structures where the quantum walker hops from site to site.
One more interesting problem is the case of 
random planar graphs, and their $d>2$ extensions, which appear in the context of quantum gravity, see e.g.\ \cite{AmDuJo}.
\bigskip

\noindent
{\bf Acknowledgement.}  T.~J.\ is grateful for hospitality at IPhT, Saclay.
F.~D. is grateful for hospitality at Science Institute, U. of Iceland, Reykjavik. 
F.D. was partly funded by the ERC-SyG project, Recursive and Exact New Quantum Theory (ReNewQuantum) which received funding from the European Research Council (ERC) under the European Union's Horizon 2020 research and innovation programme under grant agreement No 810573.


\bigskip

\newpage

\noindent
{\bf \Huge Appendices}

\appendix
\section{Calculation of $N(\alpha ,\theta)$}\label{aCalcN}

%

Using the definition \rf{6} we calculate
$$
\br\theta ',\alpha '|\theta ,\alpha\kt = \sum_{n=-\infty}^{\infty}\sum_{j=0}^\infty \left[ \bar{A'}e^{-i\alpha'n-i\theta'j}+
\bar{B'}e^{-i\alpha'n+i\theta'j} \right] \Bigl[ Ae^{i\alpha n+i\theta j}+Be^{i\alpha n-i\theta j} \Bigr]
$$
where $A'$ and $B'$ are the coefficients defined in \rf{9} with $\alpha$ and $\theta$ replaced by their primed counterparts.
The sum over $n$ is trivial and we find
$$
\br\theta ',\alpha '|\theta ,\alpha\kt =2\pi\delta (\alpha-\alpha ')\sum_{j=0}^\infty \Bigl[ (y+e^{-i\theta '})(y+e^{i\theta} )
e^{i(\theta-\theta')j} -(\theta\rightarrow -\theta ) - (\theta'\rightarrow -\theta' ) +(\theta\leftrightarrow \theta')
\Bigr]
$$
We view the sum over $j$ as a distribution in $\theta$ and $\theta'$ and use the regularization
$$
\sum_{j=0}^\infty e^{i\phi j}=\lim_{\epsilon\downarrow 0}\sum_{j=0}^\infty e^{i\phi j -\epsilon j}
=\lim_{\epsilon\downarrow 0}\frac{1}{1-e^{i\phi -\epsilon}}.
$$
Then the sum over $j$ becomes
\beq{ss}
S(\theta,\theta') = \frac{(y+e^{-i\theta'})(y+e^{i\theta})}{1-e^{i(\theta -\theta')-\epsilon}}
 -(\theta\rightarrow -\theta ) - (\theta'\rightarrow -\theta' ) +(\theta\leftrightarrow \theta')
\eeq
and we drop writing $\lim_{\epsilon\downarrow 0}$.  Putting the four terms above on a common denominator we obtain
\beq{ll}
\frac{f(\theta,\theta')-f(\theta,-\theta')}{\left[ 1+\zeta^2-2\zeta \cos(\theta+\theta')\right]\left[1+\zeta^2-2\zeta \cos(\theta-\theta')\right]}
\eeq
where $\zeta=e^{-\epsilon}$ and
$$
\!\!\!\! f(\theta,\theta')= 2\!\left[ y^2(1-\zeta\cos (\theta -\theta'))+y(1-\zeta )(\cos\theta +\cos\theta')+\cos (\theta-\theta')-\zeta
\right] \!\left[ 1+\zeta^2 -2\zeta\cos(\theta+\theta')\right]
$$
If $\theta\neq\theta'$ then the denominator in \rf{ll} is bounded away from zero for all values of 
$\epsilon$ and
 we can let $\epsilon$ go to 0.  In that limit the two terms in the numerator cancel exactly.  It follows that the distribution $S$ defined in \rf{ss} has support at $\theta =\theta'$. 

For $\theta\approx \theta'$ we can write the denominator in \rf{ll} as
$$
\left[ 2-2\cos(\theta-\theta')+O(\epsilon )\right] \left[\epsilon^2 +(1+O(\epsilon ))(\theta-\theta')^2
+O((\theta-\theta')^3)\right].
$$ 
In order to identify the distribution \rf{ll} we need to expand the numerator in $\epsilon$.  If the first order term in $\epsilon$ is nonzero at $\theta=\theta'$ we obtain a $\delta$-function at $\theta=\theta'$ and this turns out to be the case.  A straightforward calculation shows that in the sense of distributions
$$
S(\theta,\theta')=2\pi(1+2y\cos\theta +y^2)\delta (\theta-\theta')
$$
which establishes \rf{11}.

\bigskip
\noindent
\section{ Completeness of the eigenfunctions}\label{aCompleteness}

We need to show that 
\beq{a21}
\int_0^{2\pi} d\alpha \int_0^\pi d\theta \,N^{-1}(\alpha,\theta)\bar{\phi}_{\alpha,\theta}(n,j)\phi_{\alpha,\theta}(n',j')
+\int_{\pi /2}^{3\pi /2} d\alpha \,\bar{\psi}_\alpha (n,j)
\psi_\alpha (n',j') =\delta_{nn'}\delta_{jj'}.
\eeq
Using 
\rf{6}, \rf{10} and
\beq{a22}
N(\alpha,\theta)= 4\pi^2 (y+e^{i\theta})(y+e^{-i\theta})
\eeq
the first integral can be written
$$
\frac{1}{4\pi^2}\int_0^{2\pi}d\alpha\int_0^\pi d\theta\, e^{i\alpha (n'-n)}
\left(e^{i\theta (j'-j)}+e^{i\theta (j-j')} -\frac{y+e^{-i\theta}} {y+e^{i\theta}}e^{-i\theta (j+j')} -\frac{e^{i\theta}} {y+e^{-i\theta}}e^{i\theta (j+j')}\right).
$$
The sum of the first two terms in the above integral is $\delta_{nn'}\delta_{jj'}$. 
Making the change of variable $z=e^{-i\theta}$ in the third term and $z=e^{i\theta}$ in the fourth term they combine into an integral 
around the unit circle which is easily evaluated.  The sum of the last two terms then becomes
\beq{a24}
-\frac{1}{2\pi}\int_{\pi/2}^{3\pi/2} d\alpha \,e^{i\alpha (n'-n)} (1-y^{-2})(-y)^{-(j+j')}.
\eeq
It is not hard to check that the second integral in \rf{a21} equals \rf{a24} with opposite sign.  



\section{General analysis of the saddle points in the $w$ complex plane as a function of the velocity parameters $u$ and $v$.}
\label{aSteepestPathUVPlane}
In this appendix we summarize and illustrate  the general features of the saddle points and of the steepest descent path in the $w$ complex place for general values of the two velocities $u$ and $v$ in the upper-right quadrant $u\ge 0$, $v\ge 0$. This might be interesting for readers interested in the full precise mathematical structure of the large $t$ asymptotics from the point of view of  WKB theory and of resurgence theory.

We already discussed these features in the $w$ plane in the original integral representation of the wave function $u=0, v=0$ (see Fig.~\ref{u=0_v=0}) and in the analysis of the large time behavior of the wave-function along the spine corresponding to the cases $u=0$ and $v>0$ (see Fig.~\ref{u=0_v<vc}, Fig.~\ref{u=0_v=vc} and Fig.~\ref{u=0_v>vc}).
A similar analysis can be extended in the general $u\ge 0$ $v\ge0$ case, but we have not attempted a full rigorous analytical study. We present here pictures obtained by solving numerically the 
algebraic saddle point equation and the algebraic equations for the steepest descent curves, and discuss the corresponding asymptotics.

\begin{figure}
\begin{center}
\includegraphics[width=2.5in]{{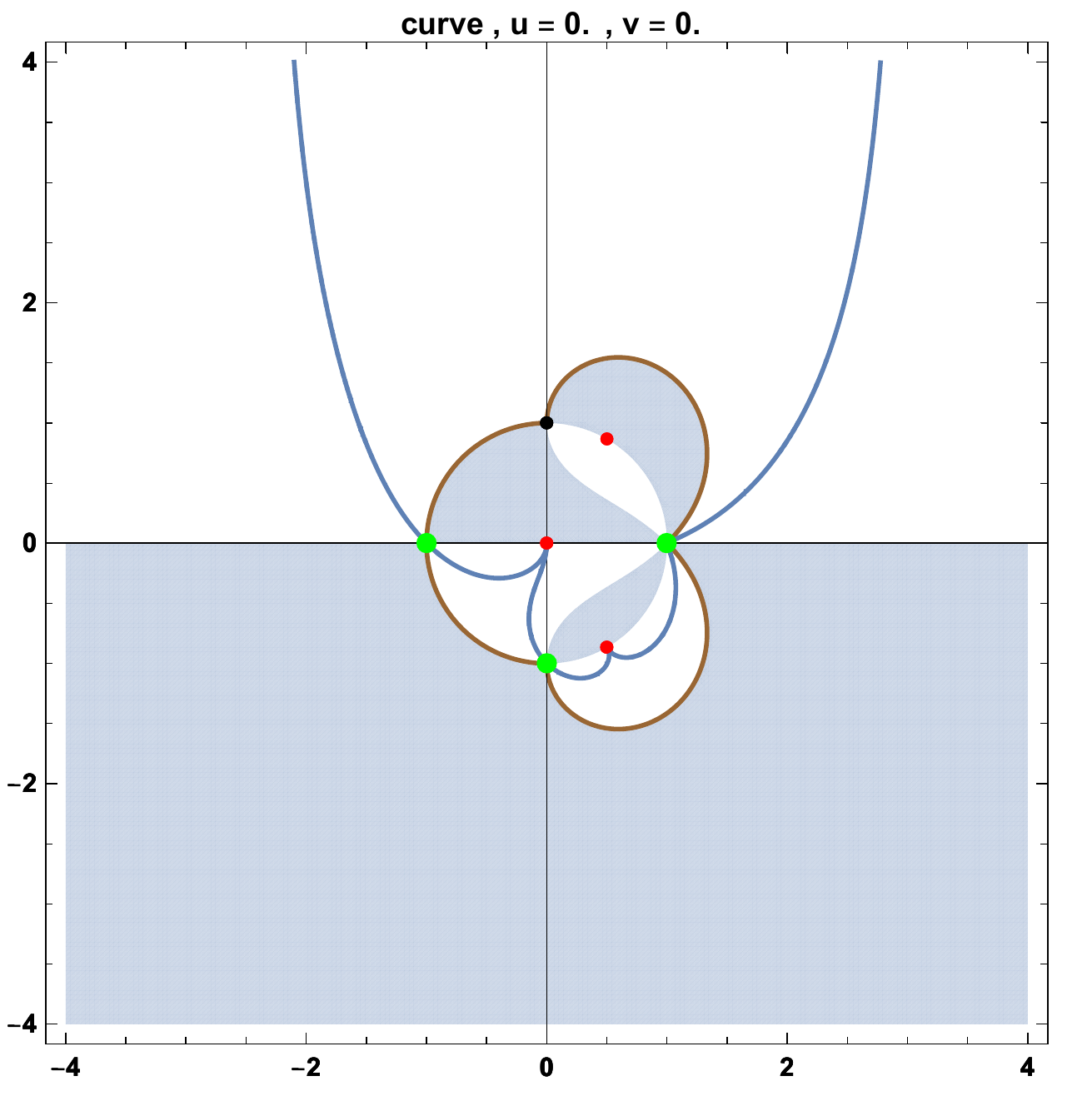}}
\includegraphics[width=2.5in]{{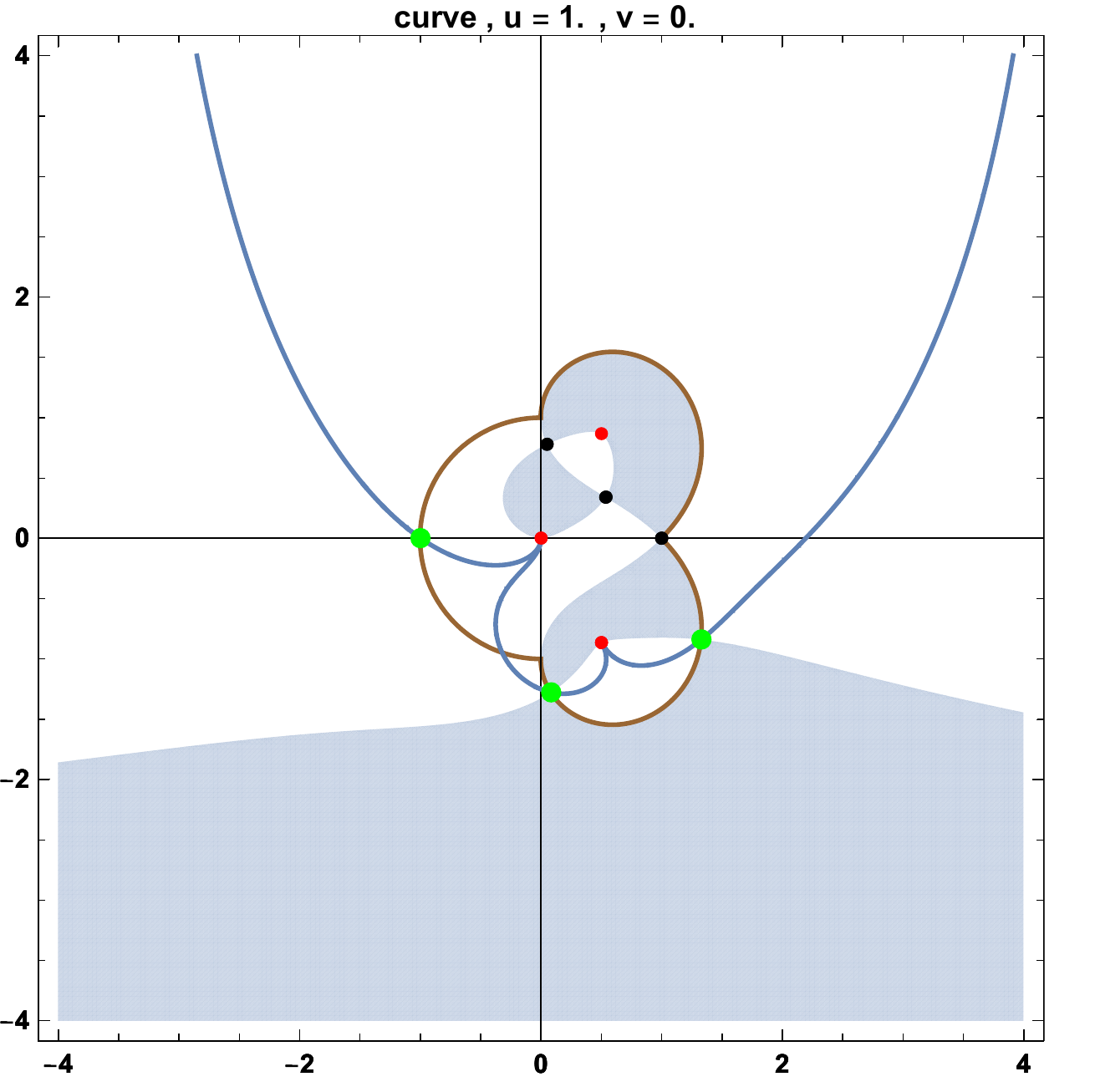}}
\includegraphics[width=2.5in]{{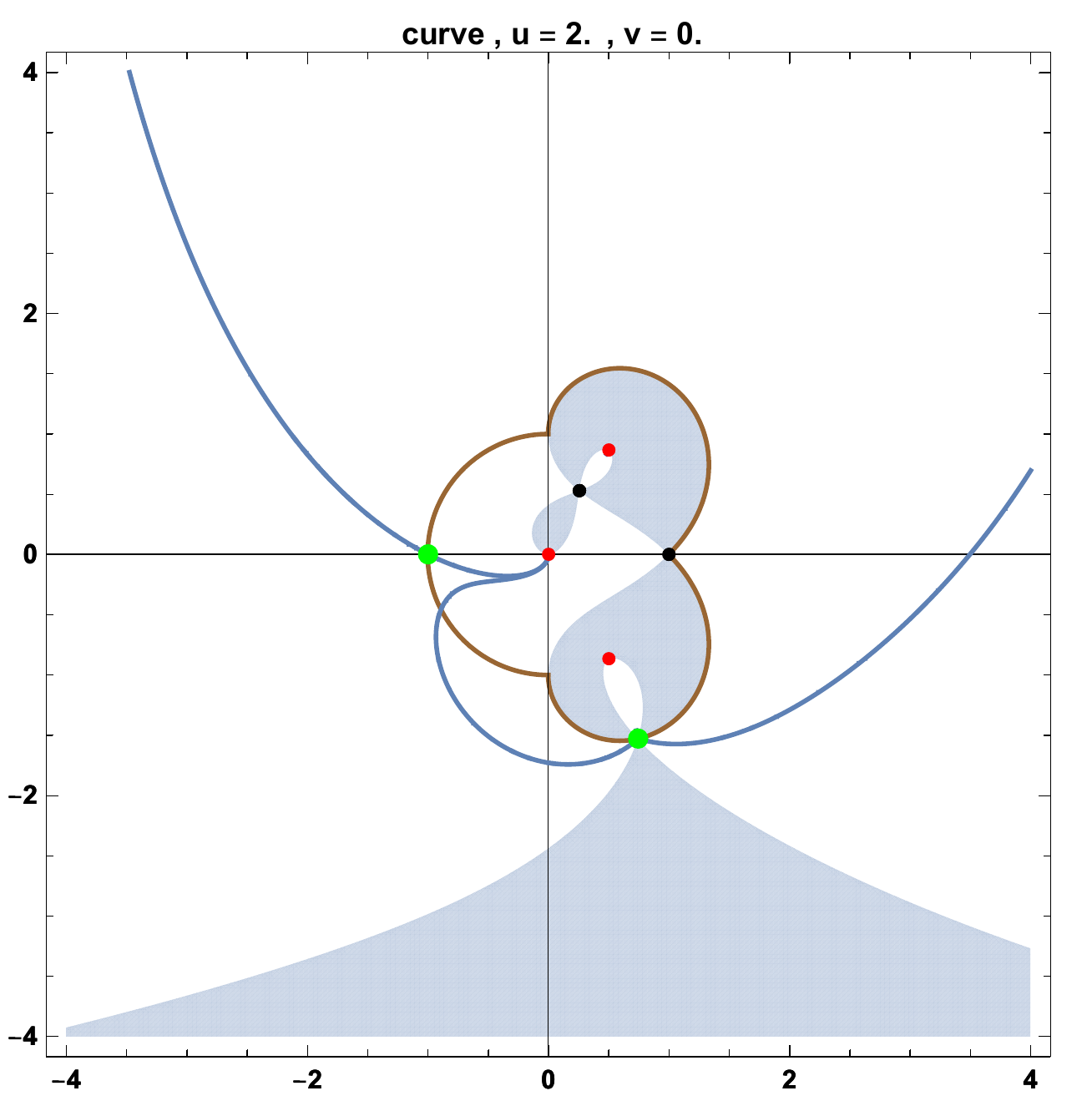}}
\includegraphics[width=2.5in]{{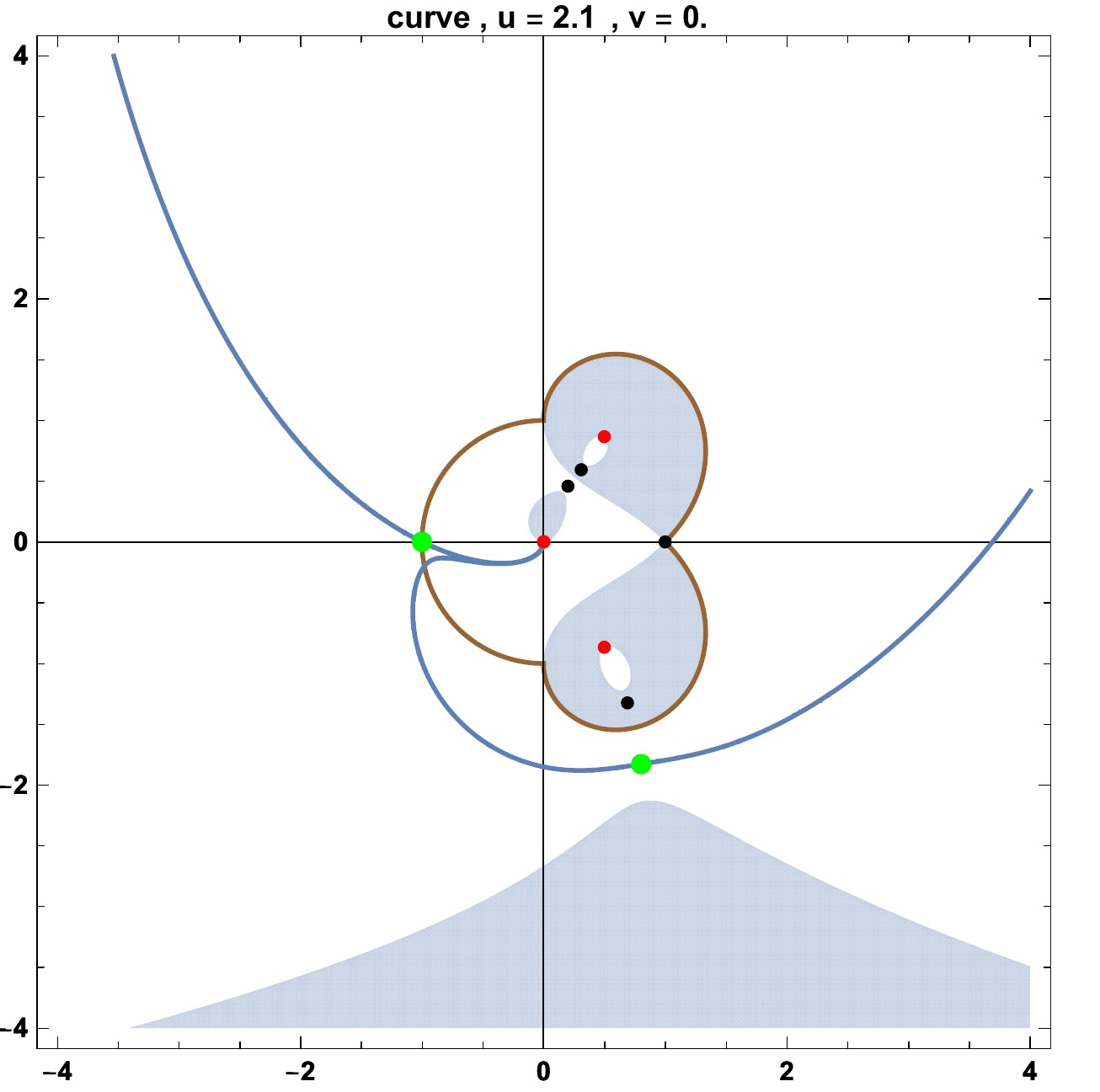}}
\includegraphics[width=2.5in]{{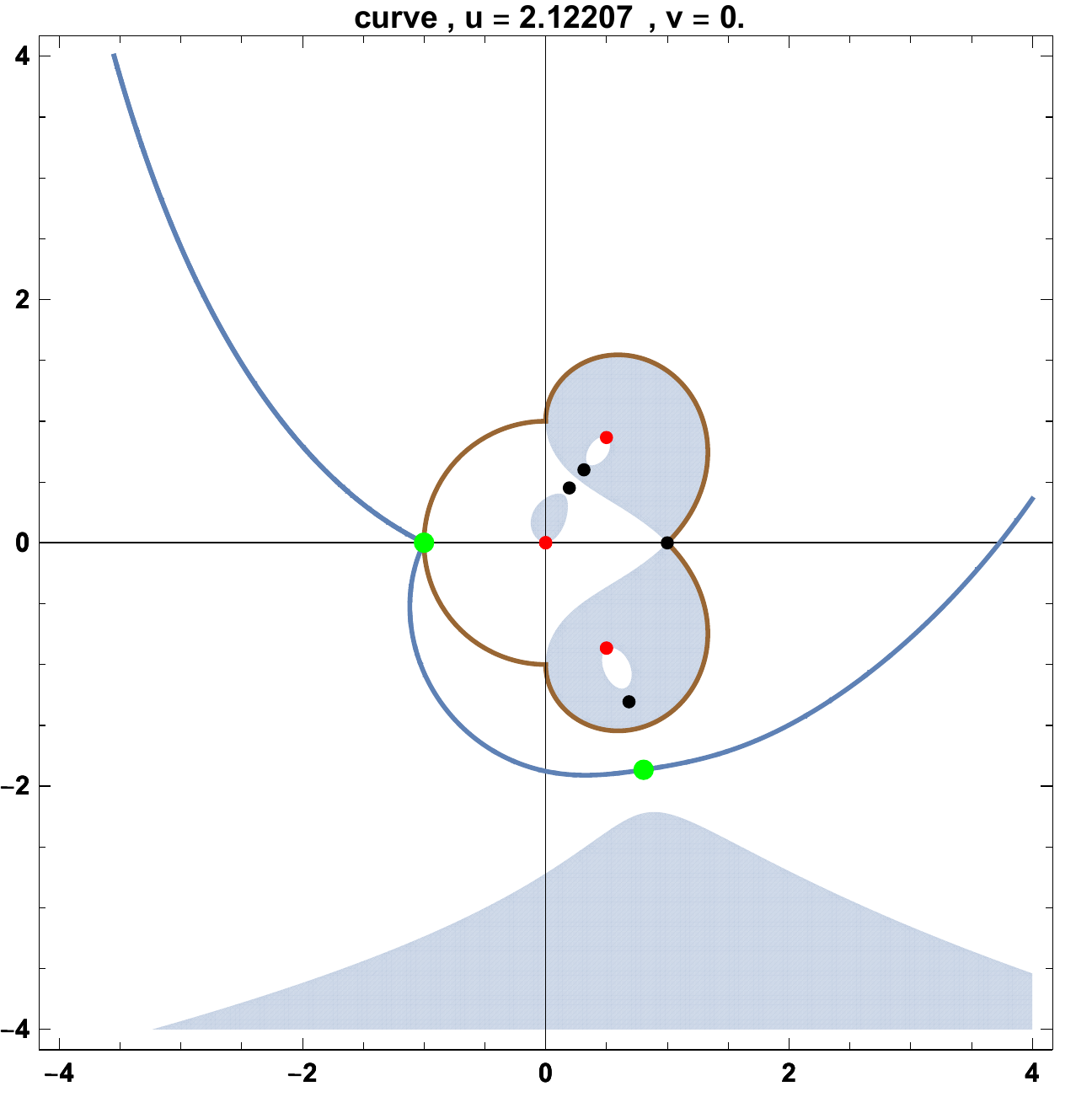}}
\includegraphics[width=2.5in]{{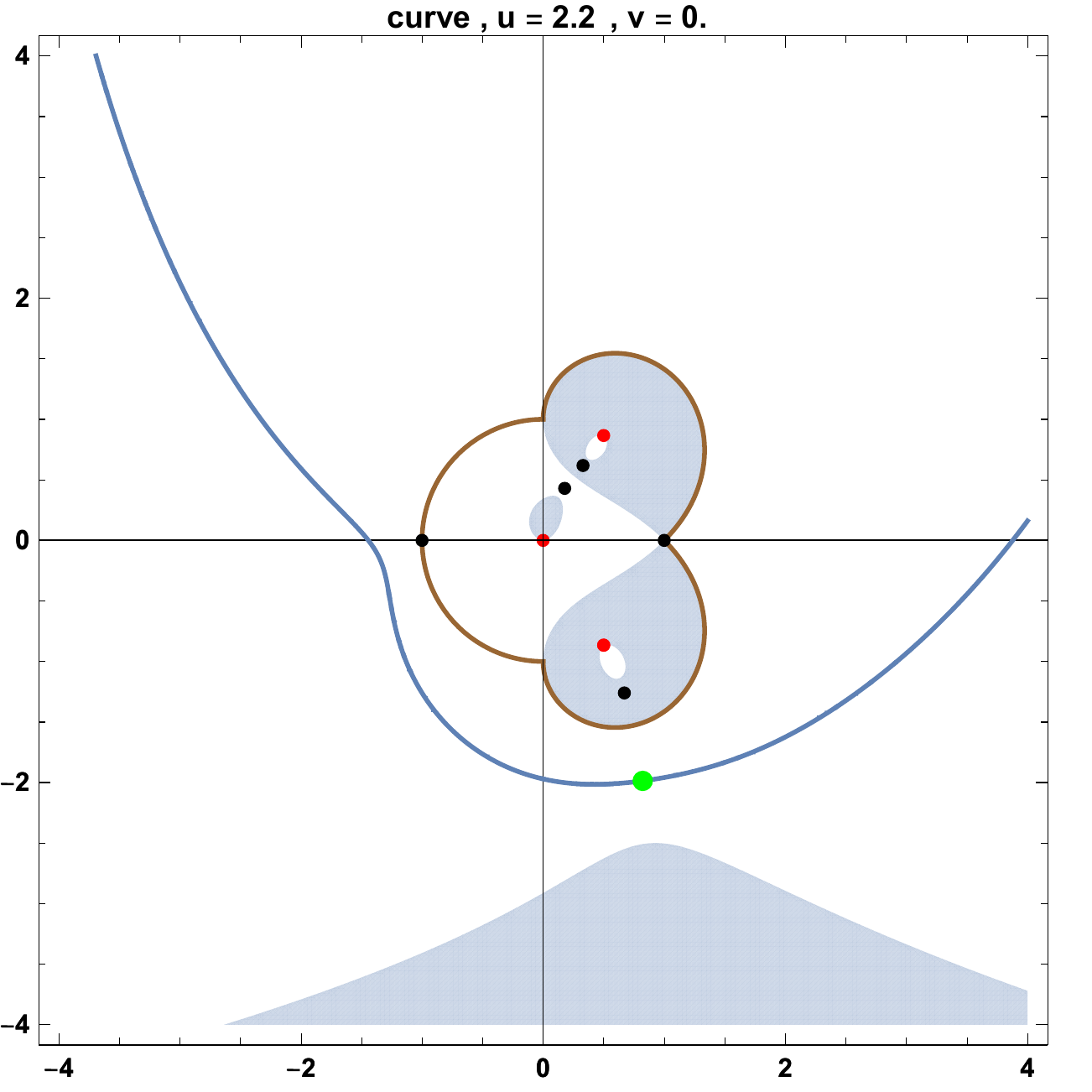}}
\caption{The steepest descent path and the saddle points in the $w$ plane for $v=0$ and various values of $u$, previously studied in the $z$ plane. The saddle points govern the large $t$ asymptotics along the teeth which are at a finite distance from the origin.}
\label{fv=0plane}
\end{center}
\end{figure} 
Firstly, the analysis for $v=0$ and $u>0$ that we did in the $z$-plane to study the large time behaviour of the wave function along the teeth at a finite distance from the origin can be repeated in the $w$-plane. On Fig.~\ref{fv=0plane} we illustrate the features for $u\ge 0$. As before, the white regions are the ``allowed'' regions where the real part of the potential $\mathrm{Re}(\mathbb{W}(w))<0$, while the grey regions are the ``forbiden'' regions where $\mathrm{Re}(\mathbb{W}(w))>0$.
The first picture is the $u=0$ case already discussed. The steppest descent path goes through the three relevant saddle points at $w_1=-1$, $w_2=-\imath$ and $w_3=1$.
The second picture illustrates the $0<u<u_c=2$ case. The path still goes through the three saddle points, but only $w_2$ and $w_3$ are relevant with  $\mathrm{Re}(\mathbb{W}(w_2))=\mathrm{Re}(\mathbb{W}(w_3))=0$ (oscillatory behaviour of the wave function) while $w_1$ becomes less relevant since $\mathrm{Re}(\mathbb{W}(w_1))<0$ (exponential decay).
The third picture is the critical case $u=u_c=2$. The $w_2$ and $w_3$ saddle points merge. This corresponds to the Airy-like front behaviour.
The forth picture describes the case when $u$ is slightly larger that $u_c$. The $w_2$ and $w_3$ saddle points split again. One of them $w_2$ is in the allowed region, the other one $w_3$ is in the forbidden region. The steepest descent path goes through $w_1$ and $w_2$ which are both in the allowed region.  This corresponds to the exponential decay of the wave function. It turns out that the most relevant (less decaying) saddle point is $w_2$ since $\mathrm{Re}(\mathbb{W}(w_1))<\mathrm{Re}(\mathbb{W}(w_2))<0$.
The fifth and sixth pictures show an interesting phenomenon. There is a second critical point at $u'_c=2.12207\ldots$ (numerical estimate) beyond which the steepest descent path goes only through $w_2$. The saddle point $w_1$ is not relevant anymore. This is a Stokes phenomenon, and $u'_c$ can be viewed as a Stokes point. The segment $\mathcal{L}_u=\{(u,v):\,0<u<u_c=2,\, v=0\}$ can be viewed as an anti-Stokes line.

\begin{figure}
\begin{center}
\includegraphics[width=13.cm]{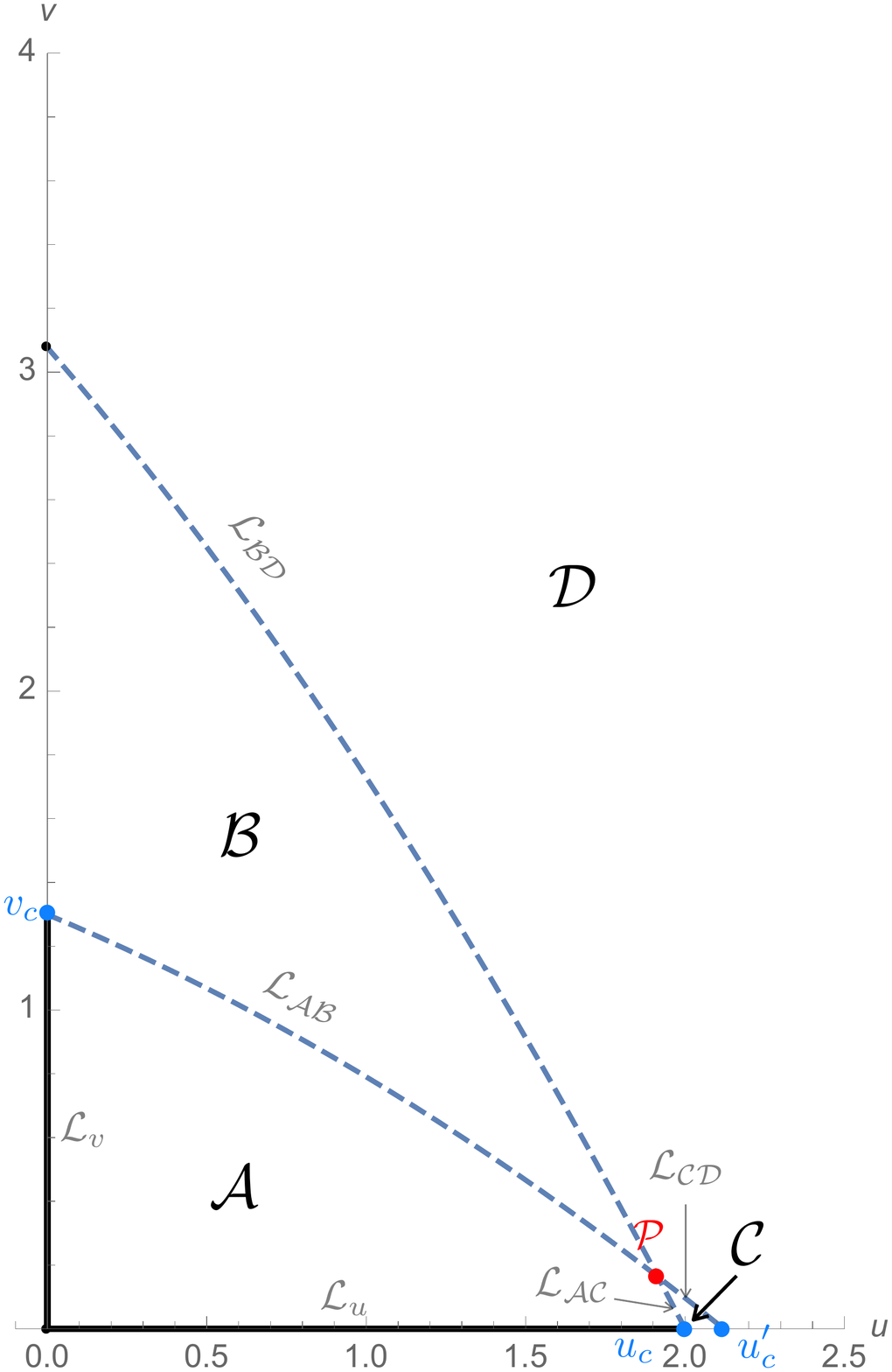}
\caption{The structure of the $(u,v)$ plane (we consider the first quadrant $u$ and $v>0$). The regions $\mathcal{A}$, $\mathcal{B}$, $\mathcal{C}$ and $\mathcal{D}$ correspond to different asymtotics and sub-asymptotics at large time, and are discussed in the text. The Stokes lines separating the regions correspond at changes in the sub-asymptotics.
The critical points are the critical tooth velocity $u_c$ and  crtical edge velocity $v_c$.}
\label{uv-plane-full}
\end{center}
\end{figure}
We now discuss the general quadrant $u\ge0,\,v\ge0$. Fig.~\ref{uv-plane-full} schematically depicts the structure of the quadrant.
Away from the anti-Stokes lines $\mathcal{L}_u$ and $\mathcal{L}_v=\{(u,v):\,u=0, \,0<v<v_c\}$ where the large $t$ behaviour is oscillatory, the rest of the quadrant corresponds to large $t$ exponential decay, controlled by the saddle points inside the allowed regions of the complex $w$ plane.
However, this region is partitionned into four subregions, labelled $\mathcal{A}$, $\mathcal{B}$, $\mathcal{C}$ and $\mathcal{C}$, separated by Stokes lines $\mathcal{L}_{\mathcal{AB}}$, $\mathcal{L}_{\mathcal{AC}}$, $\mathcal{L}_{\mathcal{CD}}$, $\mathcal{L}_{\mathcal{BD}}$, and meeting at a single point $\mathcal{P}$.

\begin{figure}
\begin{center}
\includegraphics[width=3.5in]{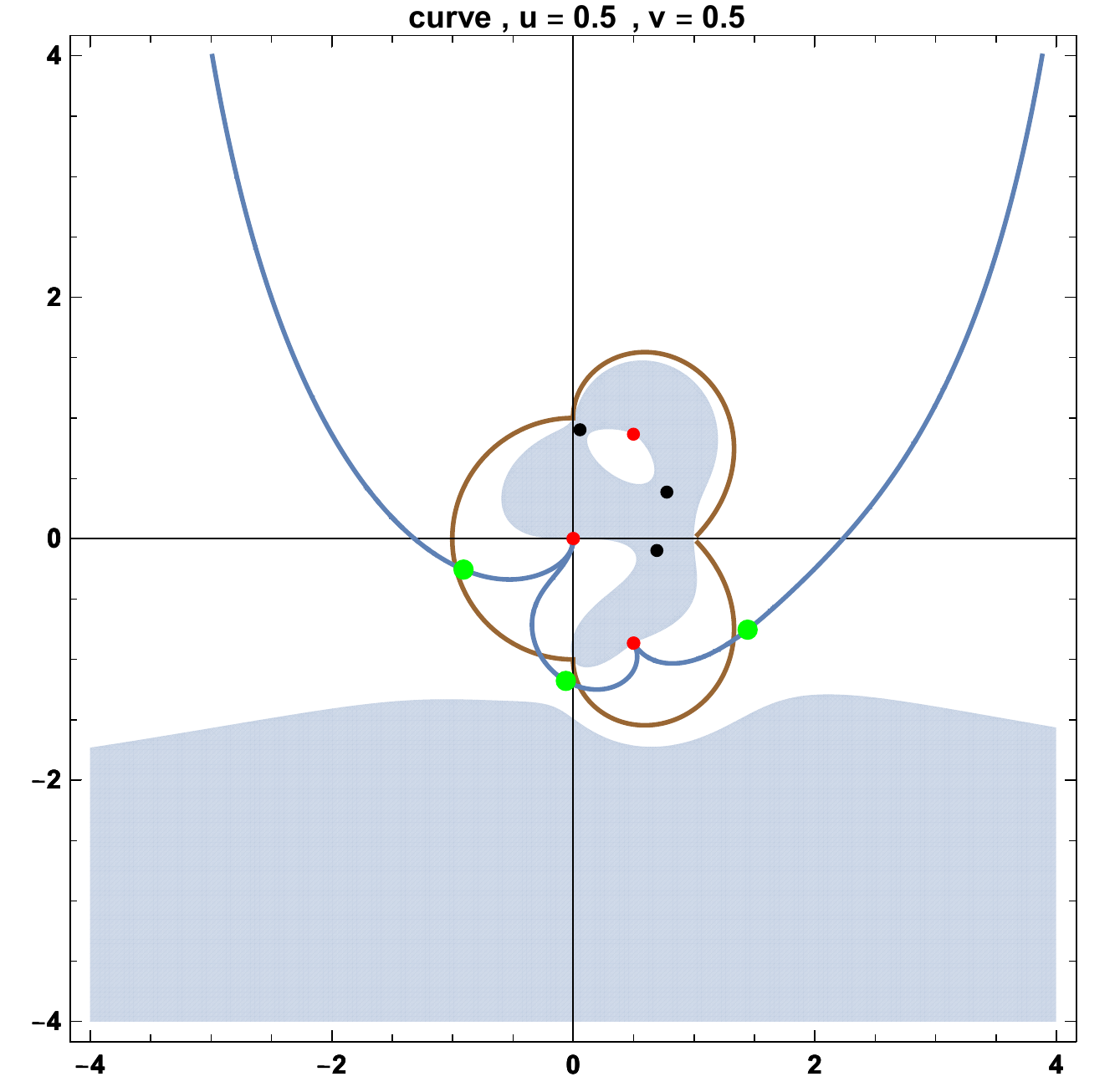}
\caption{Steepest descent path and saddle points in region $\mathcal{A}$}
\label{pathA}
\end{center}
\end{figure}

\begin{figure}
\begin{center}
\includegraphics[width=3.5in]{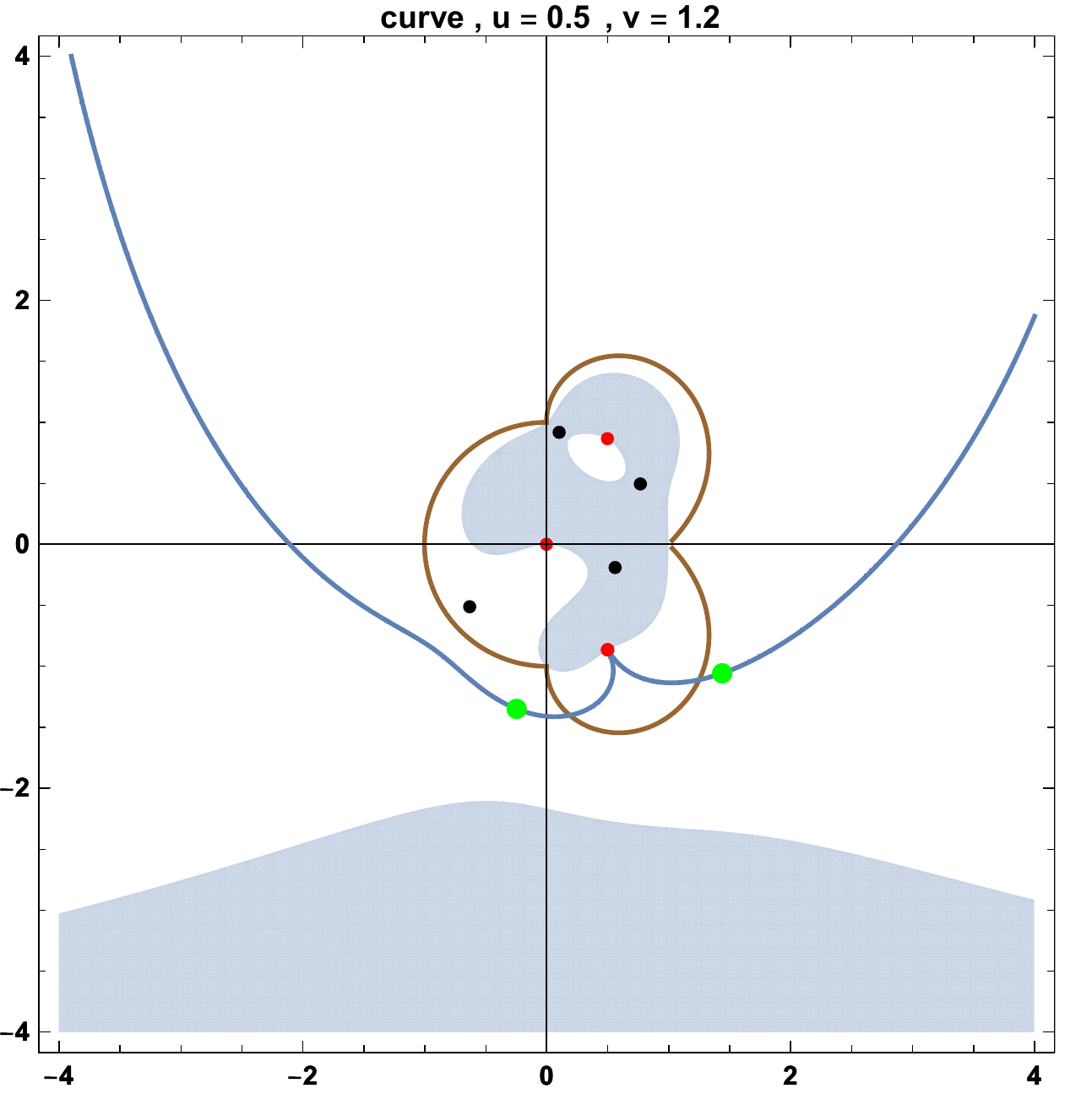}
\caption{Steepest descent path and saddle points in region $\mathcal{B}$}
\label{pathB}
\end{center}
\end{figure}

\begin{figure}
\begin{center}
\includegraphics[width=3.5in]{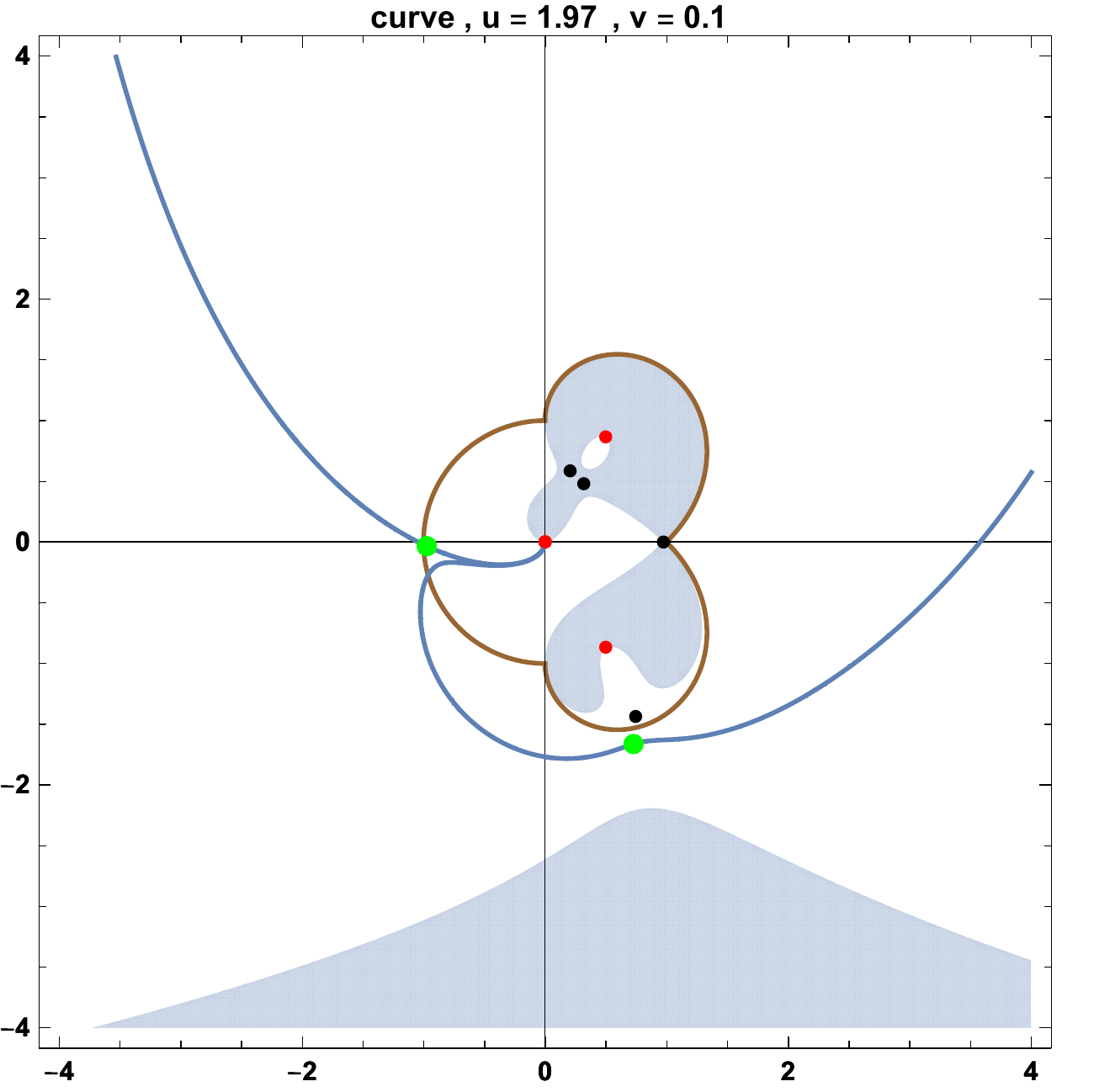}
\caption{Steepest descent path and saddle points in region $\mathcal{C}$}
\label{pathC}
\end{center}
\end{figure}

\begin{figure}
\begin{center}
\includegraphics[width=3.5in]{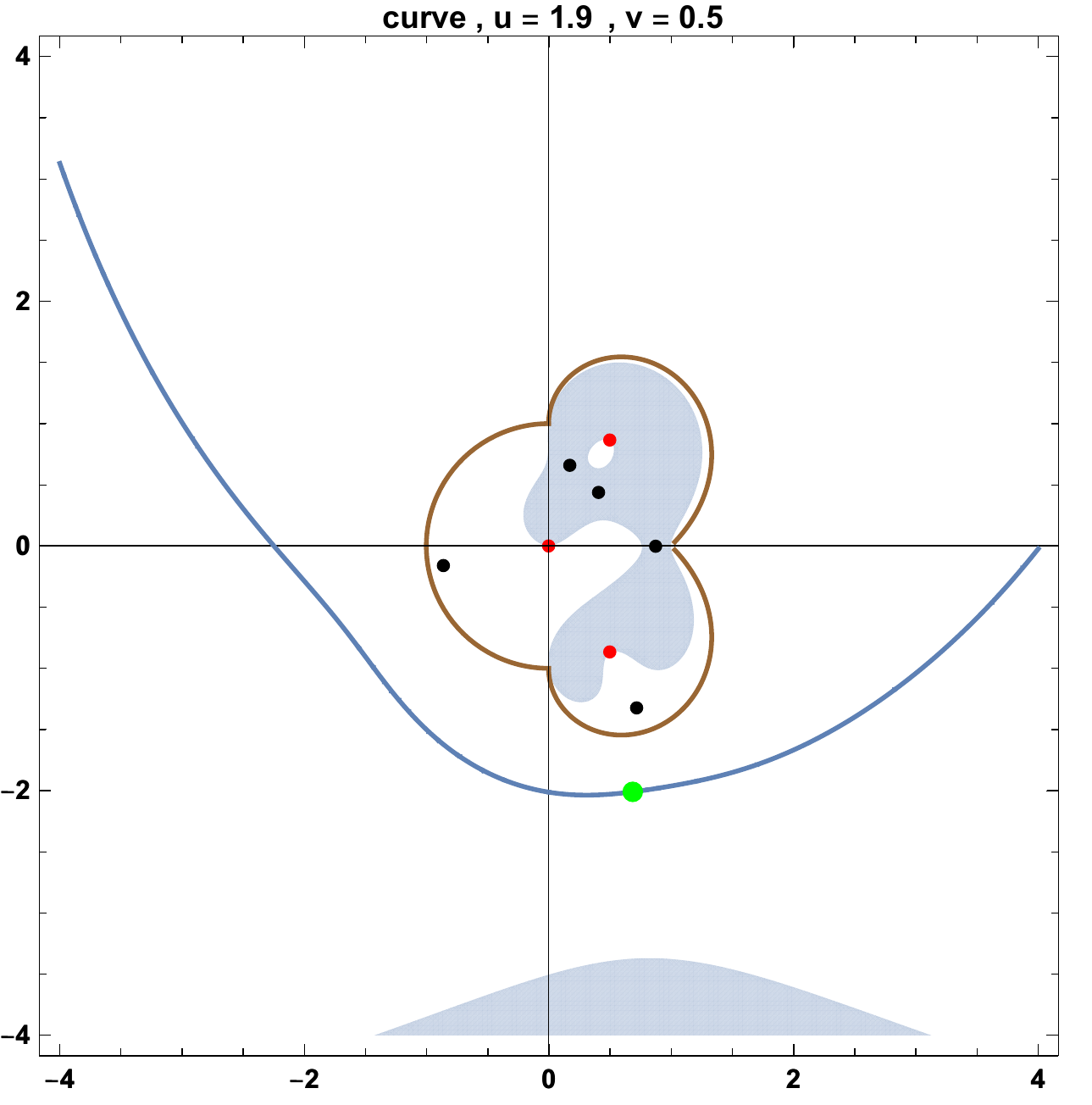}
\caption{Steepest descent path and saddle points in region $\mathcal{D}$}
\label{pathD}
\end{center}
\end{figure}

\begin{figure}
\begin{center}
\includegraphics[width=3.4in]{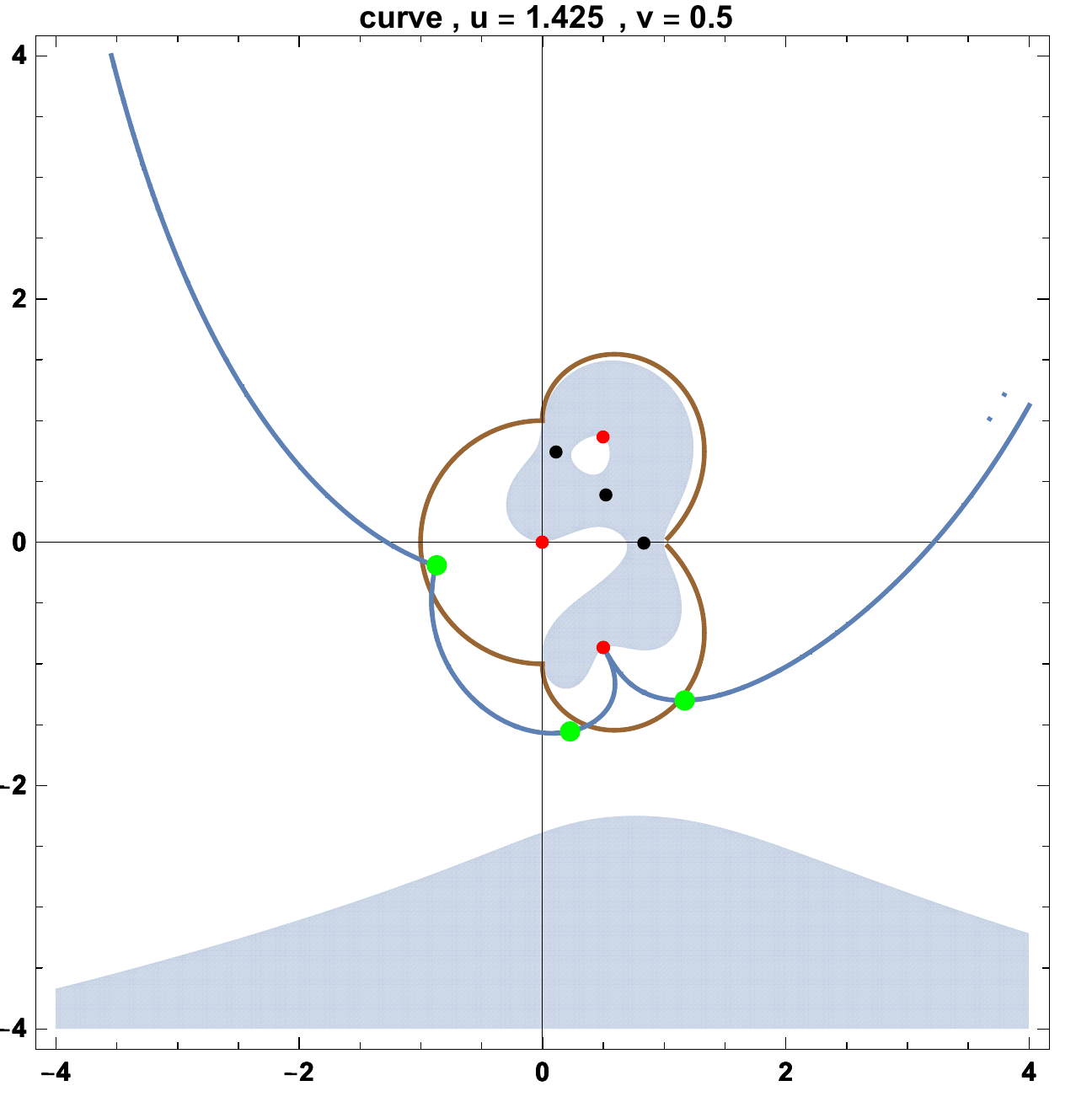}
\caption{Steepest descent path and saddle points on line $\mathcal{L_{AB}}$}
\label{pathAB}
\end{center}
\end{figure}

In the region $\mathcal{A}$, the steepest descent path goes through the three saddle points $w_1$, $w_2$ and $w_3$, and the dominant saddle point is $w_2$. An example is shown in Fig.~\ref{pathA}.
In the region $\mathcal{B}$, the steepest descent path goes through saddle points $w_2$ and $w_3$, and the dominant saddle point is still $w_2$, $w_1$ is not relevant anymore. An example is shown in Fig.~\ref{pathB}.
In the region $\mathcal{C}$, the steepest descent path goes through saddle points $w_1$ and $w_2$, and the dominant saddle point is still $w_2$, $w_3$ is not relevant anymore. An example is shown in Fig.~\ref{pathC}.
In the region $\mathcal{D}$, the steepest descent path goes only through the 
saddle point $w_2$ which is of course the dominant one, $w_1$ and $w_3$ are not relevant anymore. An example is shown in Fig.~\ref{pathD}.
An example of configuration on the Stokes line $\mathcal{L_{AB}}$ is given on Fig.~\ref{pathAB}.

\begin{figure}
\begin{center}
\includegraphics[width=3.5in]{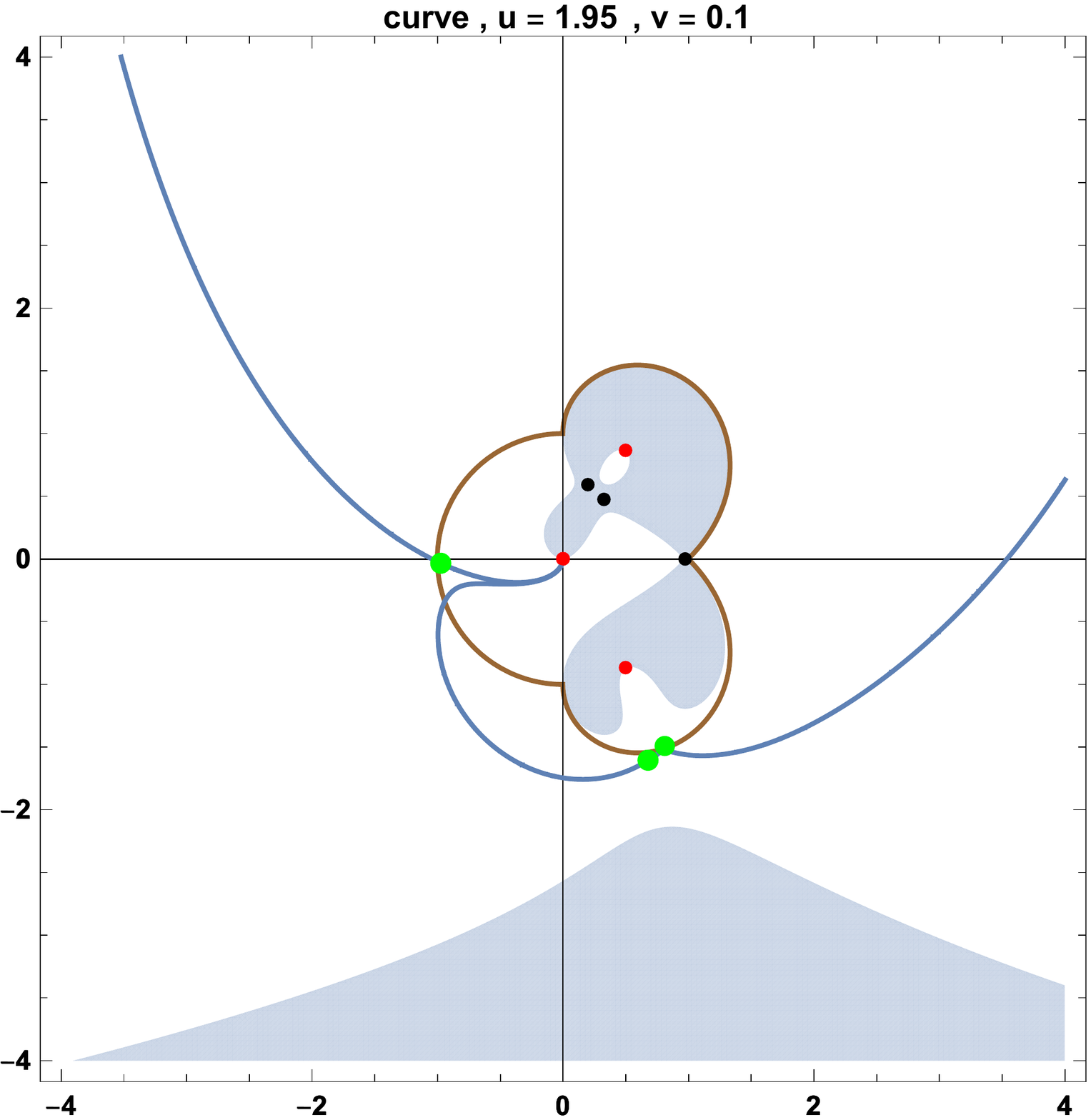}
\caption{Steepest descent path and saddle points on line $\mathcal{L_{AC}}$}
\label{pathAC}
\end{center}
\end{figure}

\begin{figure}
\begin{center}
\includegraphics[width=3.5in]{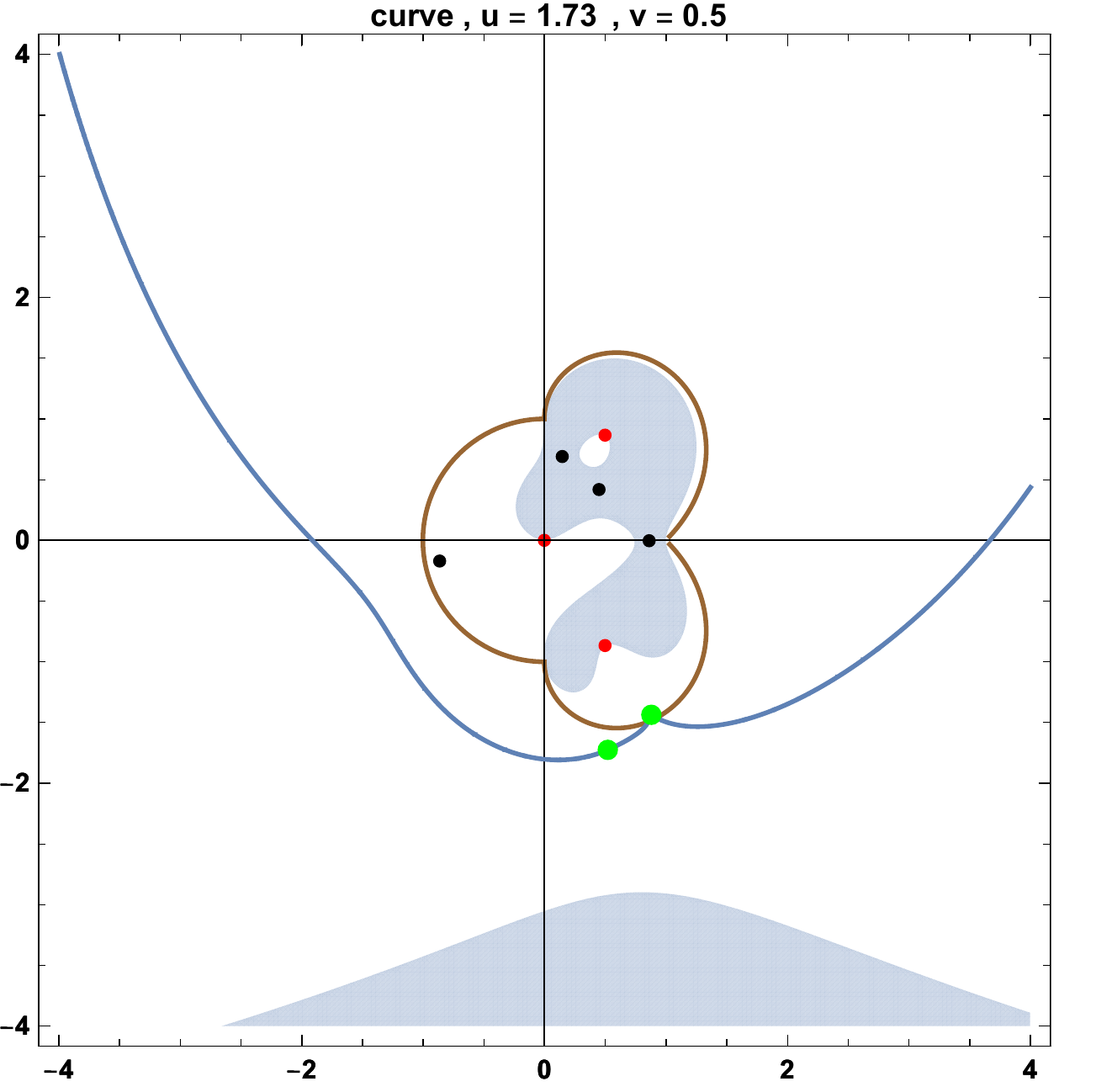}
\caption{Steepest descent path and saddle points on line $\mathcal{L_{BD}}$}
\label{pathBD}
\end{center}
\end{figure}

\begin{figure}
\begin{center}
\includegraphics[width=3.5in]{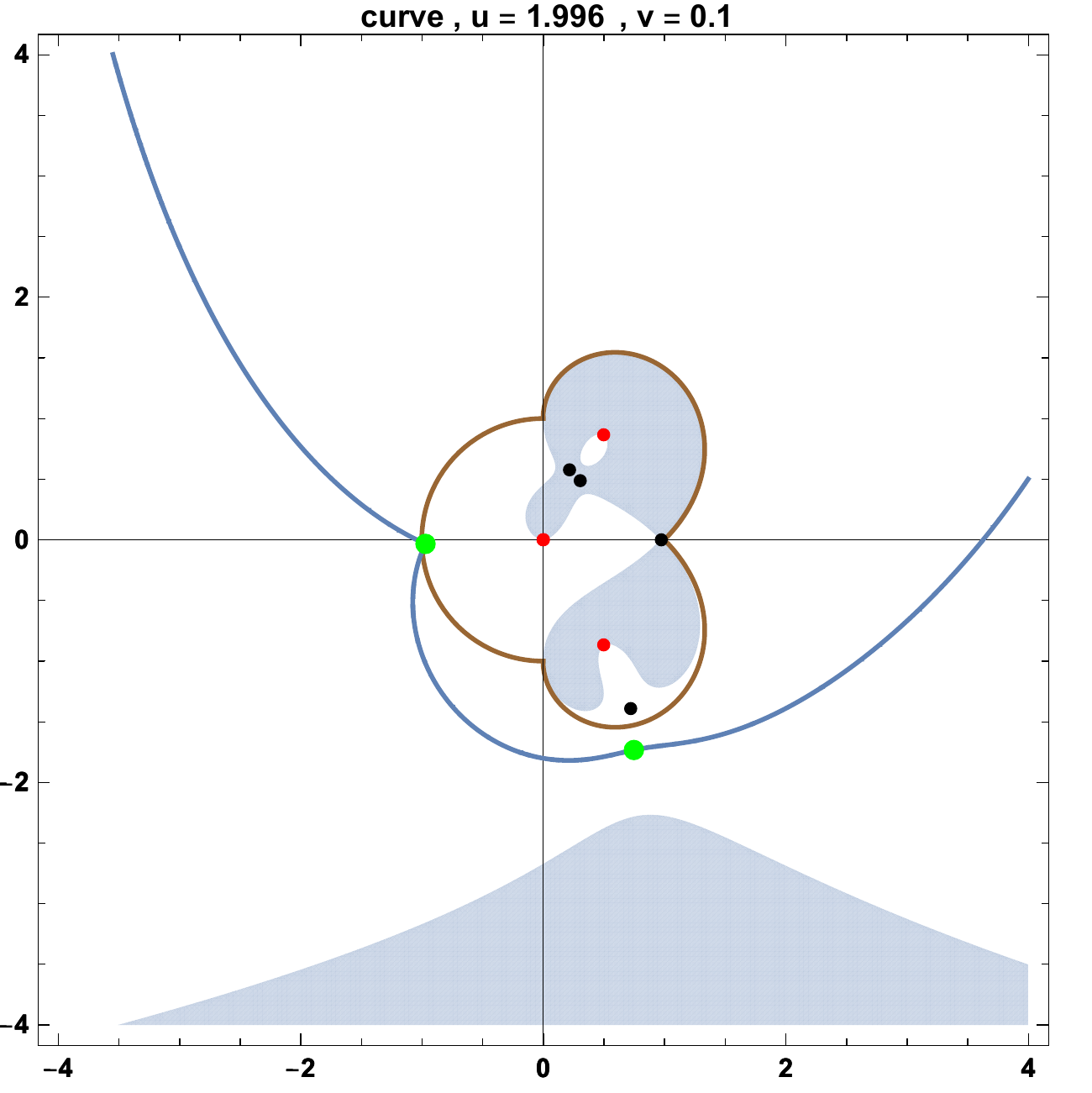}
\caption{Steepest descent path and saddle points on line $\mathcal{L_{CD}}$}
\label{pathCD}
\end{center}
\end{figure}

\begin{figure}
\begin{center}
\includegraphics[width=3.5in]{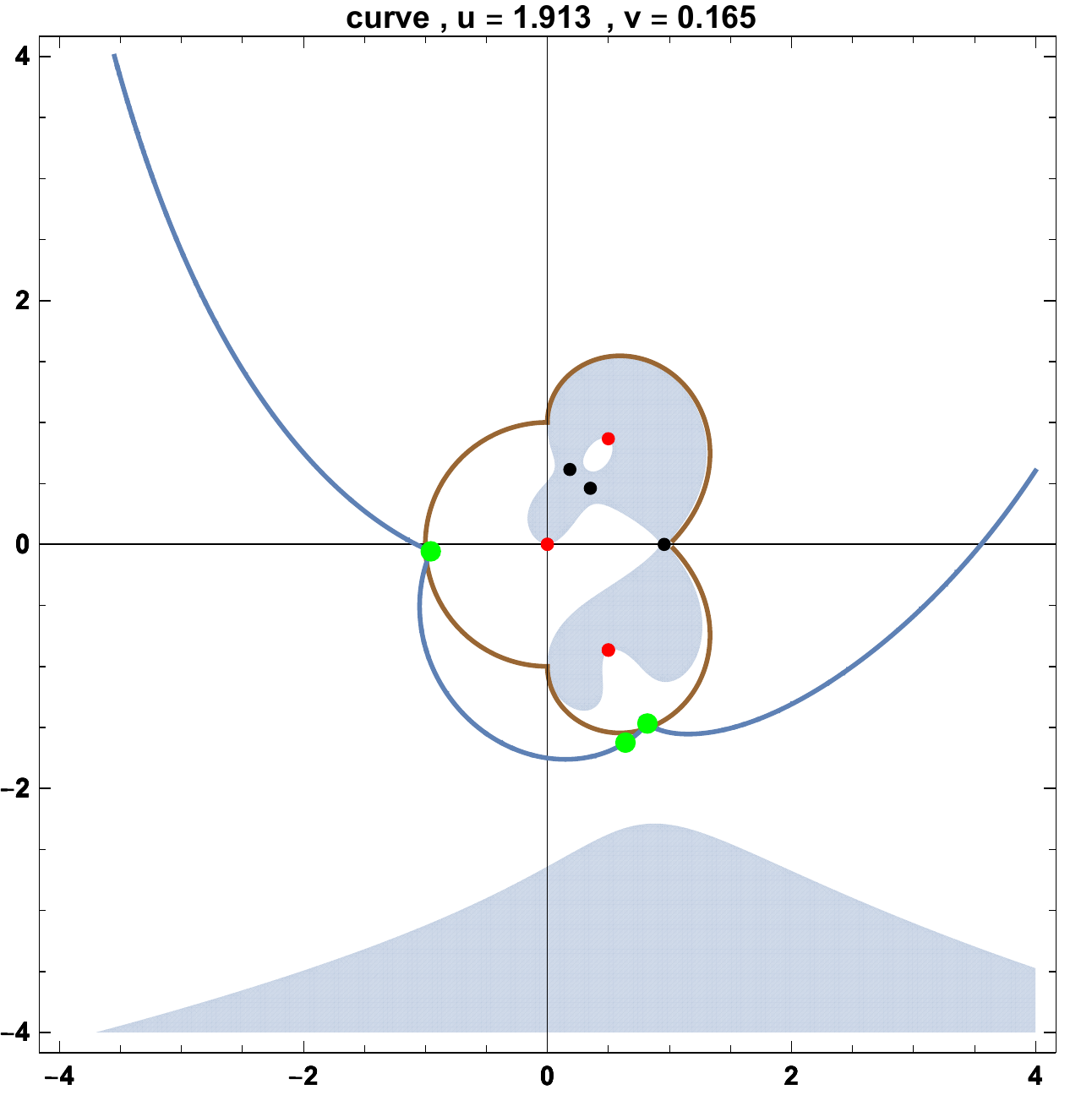}
\caption{Steepest descent path and saddle points in the $w$ plane at the point $\mathcal{P}$}
\label{pathABCD}
\end{center}
\end{figure}

An example of configuration on the Stokes line $\mathcal{L_{AC}}$ is given on Fig.~\ref{pathAC}.
An example of configuration on the Stokes line $\mathcal{L_{BD}}$ is given on Fig.~\ref{pathBD}.
An example of configuration on the Stokes line $\mathcal{L_{CD}}$ is given on Fig.~\ref{pathCD}.
Finally, the configuration on the special point $\mathcal{P}$ where the four Stokes lines meet is given on Fig.~\ref{pathABCD}.

We stress that the dominant sadlle point, which gives the dominant exponential decay term at large $t$ when $u>0$ and $v>0$ is always $w_2$, i.e.\ the deformation of the saddle point $\emath^{-\imath \pi/3}$ when $u=v=0$, which is defined unambiguously in the first quadrant, as long as one does not meet or turn around the critical points $\mathcal{P}_u=(u_c,0)=(2,0)$ or $\mathcal{P}_v=(0,v_c)=(0,\sqrt{27/16})$.

\begin{figure}[]
\begin{center}
\includegraphics[width=7.cm]{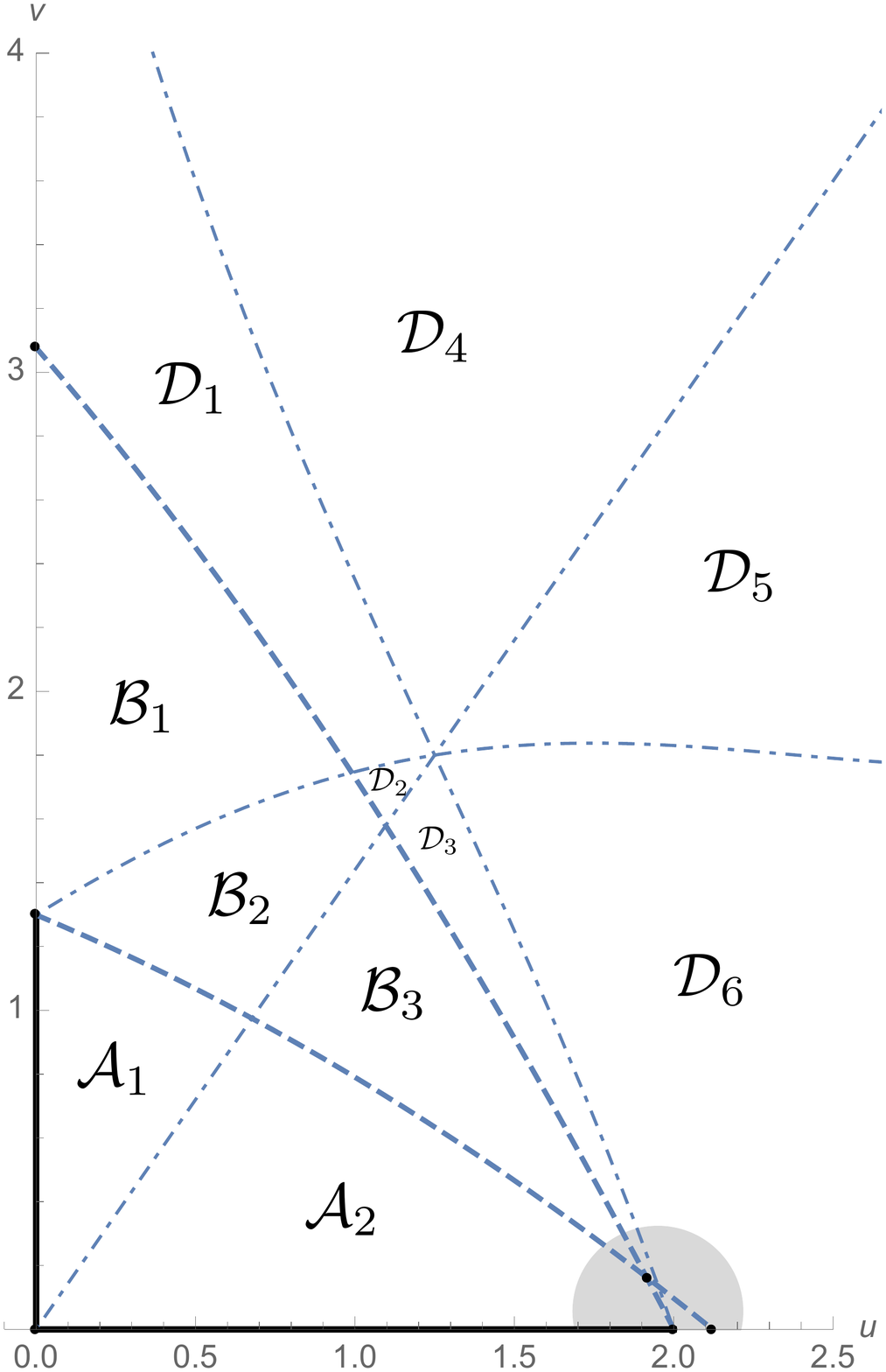}
\raisebox{0.0cm}{\includegraphics[width=7.cm]{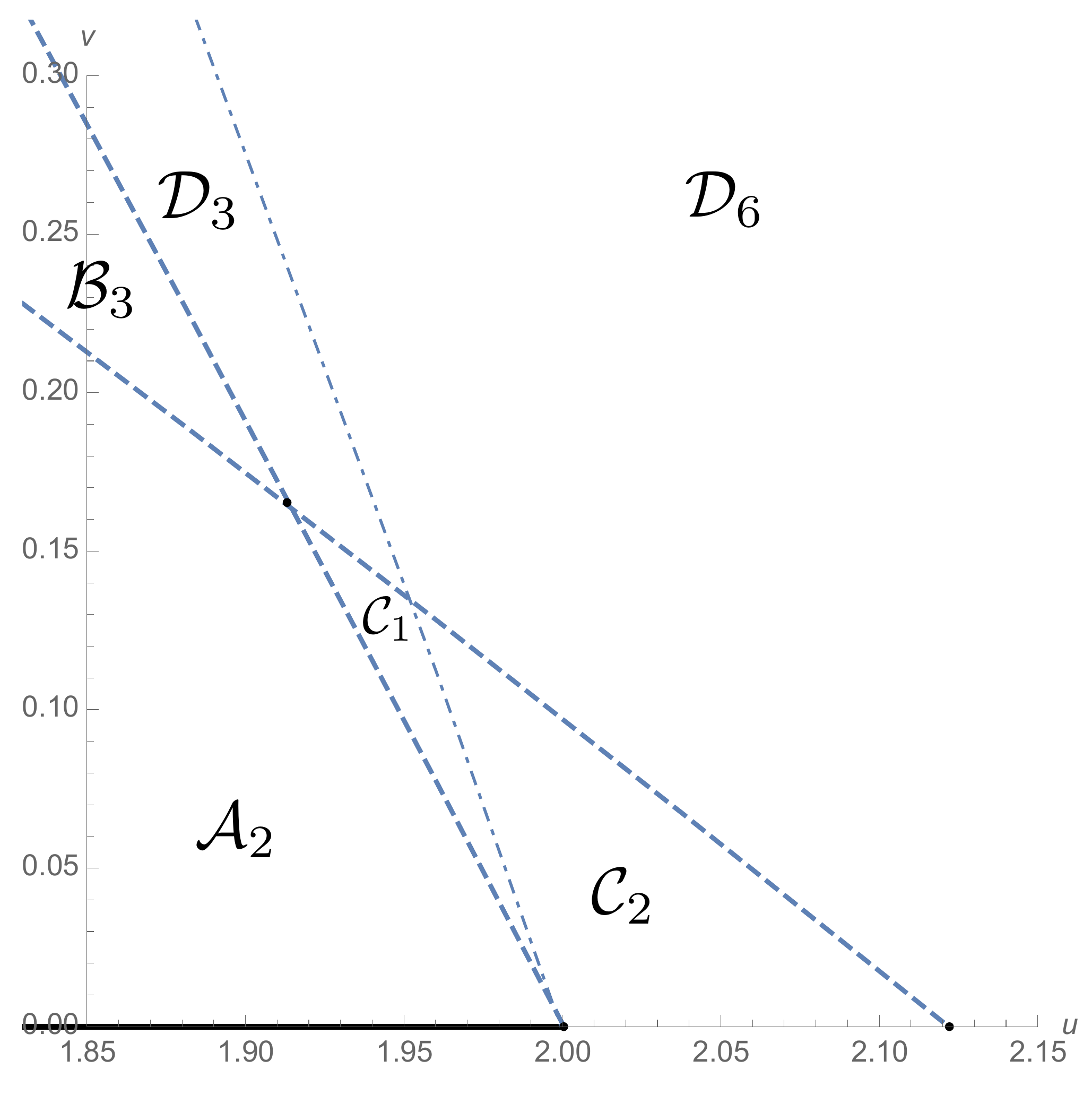}}
\caption{The detailed structure of the $(u,v)$ velocities space (first quadrant), with the Stokes lines (dashed) and the anti-Stokes lines (dot-dashed).
The rightmost picture is an enlarged view of the grey region near the critical $u_c$ point.}
\label{uv-plane-full-stks}
\end{center}
\end{figure}

To complete the discussion, it is possible to extend the study to the anti-Stokes lines, where two saddle points $w$ and $w'$ contribute at the same level to the large $t$ asymtotics. This happens if the modulus of the exponentially decreasing terms coming from the two saddle points have the same behaviour, but the terms have different oscillatory phases.
This therefore occurs when $\mathrm{Re}(\mathbb{W}(w)) =\mathrm{Re}(\mathbb{W}(w'))$ but $\mathrm{Im}(\mathbb{W}(w)) \neq \mathrm{Im}(\mathbb{W}(w'))$.
Let us denote this schematically by
\begin{equation}
\label{ }
w\thickapprox w'\qquad\iff\qquad \mathrm{Re}(\mathbb{W}(w)) =\mathrm{Re}(\mathbb{W}(w'))\qquad\text{(anti-Stokes)}
\end{equation}
and the condition that the contribution of $w$ is dominant w.r.t. that of $w'$ (if both are relevant) by
\begin{equation}
\label{ }
w\succ w'\qquad\text{iff}\qquad \mathrm{Re}(\mathbb{W}(w))> \mathrm{Re}(\mathbb{W}(w')).
\end{equation}
We note that the Stokes lines  for a pair of saddle points is defined by the Stokes condition
\begin{equation}
\label{ }
w\equiv w'\qquad\iff\qquad \mathrm{Im}(\mathbb{W}(w)) =\mathrm{Im}(\mathbb{W}(w'))\qquad\text{(Stokes).\hphantom{anti-}}
\end{equation}
On Fig.~\ref{uv-plane-full-stks} we depict for completeness the $u-v$ plane with its anti-Stokes lines in addition to the Stokes lines discussed above. The anti-Stokes line are the dot-dashed lines. Note that they meet at an anti-Stokes point where $w_1 \thickapprox w_2 \thickapprox w_3$. The domain $\mathcal{A}$ is separated into two subregions 
$\mathcal{A}_1$ ($\mathbf{w}_2 \succ \mathbf{w}_1\succ \mathbf{w}_3$), $\mathcal{A}_2$ ($\mathbf{w}_2\succ \mathbf{w}_3 \succ \mathbf{w}_1$); 
the domain $\mathcal{B}$ into three subregions 
$\mathcal{B}_1$ (${w}_1 \succ \mathbf{w}_2\succ \mathbf{w}_3$), $\mathcal{B}_2$ ($\mathbf{w}_2\succ w_1 \succ \mathbf{w}_3$), $\mathcal{B}_3$ ($\mathbf{w}_2\succ \mathbf{w}_3 \succ w_1$);
the domain $\mathcal{C}$ into two subregions 
$\mathcal{C}_1$ ($\mathbf{w}_2 \succ w_3\succ \mathbf{w}_1$), $\mathcal{C}_2$ ($w_3\succ w_\mathbf{2} \succ \mathbf{w}_1$);
and the domain $\mathcal{D}$ into six subregions
$\mathcal{D}_1$ ($w_1 \succ \mathbf{w}_2\succ w_3$), $\mathcal{D}_2$ ($\mathbf{w}_2\succ w_1 \succ w_3$), $\mathcal{D}_3$ ($\mathbf{w}_2\succ w_3 \succ w_1$), 
$\mathcal{D}_4$ ($w_3 \succ \mathbf{w}_2\succ w_1$), $\mathcal{D}_5$ ($w_3\succ w_1 \succ \mathbf{w}_2$), $\mathcal{D}_6$ ($w_1\succ w_3 \succ \mathbf{w}_2$).
The relevant saddle points (those picked by the steepest descent path) are denoted by bold letters.


\end{document}

%% file: eqmacros.tex
\def\void{}
\def\labelmark{}

\newenvironment{formula}[1]{\def\labelname{#1}
\ifx\void\labelname\def\junk{\begin{displaymath}}
\else\def\junk{\begin{equation}\label{\labelname}}\fi\junk}%
{\ifx\void\labelname\def\junk{\end{displaymath}}
\else\def\junk{\end{equation}}\fi\junk\labelmark\def\labelname{}}

{\ifx\void\labelname\def\junk{\end{array}\end{displaymath}}
\else\def\junk{\end{array}\right.\end{equation}}
\fi\junk\labelmark\def\labelname{}\def\junk{}
}

\newcommand{\beq}{\begin{formula}}
\newcommand{\eeq}{\end{formula}}
\newcommand{\beqv}{\begin{formula}{}}

%% file: totalmacroe.tex
\newcommand{\rf}[1]{(\ref{#1})}
\newcommand{\oh}{\frac{1}{2}}

\newcommand{\bea}{\begin{eqnarray}}
\newcommand{\eea}{\end{eqnarray}}
\newcommand{\beas}{\begin{eqnarray*}}
\newcommand{\eeas}{\end{eqnarray*}}
\newcommand{\beqs}{\begin{displaymath}}
\newcommand{\eeqs}{\end{displaymath}}

\newcommand{\br}{\langle}
\newcommand{\kt}{\rangle}

\newcommand{\vp}{\varphi}

\newcommand{\cH}{{\cal H}}

\newcommand{\ben}{\begin{equation}}
\newcommand{\een}{\end{equation}}

\newcommand{\bdm}{\begin{displaymath}}
\newcommand{\edm}{\end{displaymath}}

\newcommand{\bbZ}{{\mathbb Z}}